\documentclass[oldversion]{aa}
\usepackage{times,graphics}
\usepackage{lscape}
\usepackage{longtable}
\usepackage{epsfig} 
\usepackage{txfonts} 
\usepackage{natbib}
\usepackage{psfig}

\bibpunct{(}{)}{;}{a}{}{,}

\def\logg{$\log  \rm {g}$}
\def\teff{$T_{\rm eff}$}
\def\vsini{$v \sin i$}

\def\etal{{\em et~al}.~}
\def\kms{kms$^{-1}$}
\def\ie{{\em i.e}.~}
\def\eps{$\xi$}
\def\2{{\sc ii}}
\def\1{{\sc i}}
\def\3{{\sc iii}}
\def\4{{\sc iv}}
\def\hg{H$_{\gamma}$}
\def\hd{H$_{\delta}$}

\begin{document} \title{The VLT-FLAMES survey of massive stars: Evolution of
surface N abundances and effective temperature scales in the Galaxy and
Magellanic Clouds. \thanks{Based on observations at the European Southern
Observatory Very Large Telescope in programmes 68.D-0369 and 171.D-0237.}$^{, }
$\thanks{Tables~\ref{ablin}-\ref{ngc330g} are only available in electronic
form at http://www.edpsciences.org}}

   \author{C.Trundle \inst{1}, P.L. Dufton \inst{1}, I. Hunter \inst{1,2}, C.J. Evans \inst{3}, D.J. Lennon \inst{2}, S.J
   Smartt \inst{1}, R.S.I. Ryans \inst{1}} 

   \offprints{C.Trundle,~\email{c.trundle@qub.ac.uk.}}

   \authorrunning{C.~Trundle \etal}
   
   \titlerunning{VLT-Flames survey: \teff~scales \& N abundances of B-type stars.}
    
   \institute{Astronomy Research Centre, Department of Physics \& Astronomy,
   School of Mathematics \& Physics, The Queen's University of Belfast, Belfast,
   Northern Ireland, UK
             \and The Isaac Newton Group of Telescopes, Apartado de Correos 321, E-38700,
	          Santa Cruz de La Palma, Canary Islands, Spain
             \and  UK Astronomy Technology Centre, Royal Observatory, Blackford Hill, Edinburgh, EH9 3HJ, UK
}                 
   \date{}

   \abstract{We present an analysis of high resolution VLT-FLAMES spectra of 61
   B-type stars with relatively narrow-lined spectra located in 4 fields
   centered on the Milky Way clusters; NGC3293 \& NGC4755 and the Large and
   Small Magellanic cloud clusters; NGC2004 and NGC330. For each object a
   quantitative analysis was carried out using the non-LTE model atmosphere code
   TLUSTY; resulting in the determination of their atmospheric parameters and
   photospheric abundances of the dominant metal species (C, N, O, Mg, Si, Fe).
   The results are discussed in relation to our earlier work on 3 younger
   clusters in these galaxies; NGC6611, N11 and NGC346 paying particular
   attention to the nitrogen abundances which are an important probe of the role
   of rotation in the evolution of stars. This work along with that of the
   younger clusters provides a consistent dataset of abundances and atmospheric
   parameters for over 100 B-type stars in the three galaxies.  We provide
   effective temperature scales for B-type dwarfs in all three galaxies and for
   giants and supergiants in the SMC and LMC. In each galaxy a dependence on
   luminosity is found between the three classes with the unevolved dwarf
   objects having significantly higher effective temperatures. A metallicity
   dependence is present between the SMC and Galactic dwarf objects, and whilst
   the LMC stars are only slightly cooler than the SMC stars, they are
   significantly hotter than their Galactic counterparts. }

   \keywords{stars: atmospheres -- stars: early-type -- stars: B-type -- 
   stars: abundances - Magellanic Clouds - Galaxies: abundances  - open clusters
   and associations: individual: NGC3293, NGC4755, NGC2004, NGC330  -- stars: evolution }

   \maketitle

%-----------------------------------------------------------------------------------

\section{Introduction.}
\label{intro}

Rotation was introduced as a crucial dynamical process in understanding the
evolution of massive stars due to a number of discrepancies found between
stellar evolution models and observations \citep[]{mey97}. Amongst these were
the discovery of surface enrichments of helium and nitrogen in OBA-type
supergiants \citep[]{jj67,duf72,wal72,venn99} and helium excesses in fast
rotating O-type stars \citep[]{her92}, which provided observational evidence
that additional mixing mechanisms were required in the models. Important
breakthroughs by \citet[]{zah92} on the theory of the transport mechanisms
caused by rotation, and subsequent work in that area has allowed for the
introduction of rotation in the stellar evolution models \citep[]{mey00,heg00}.
Additionally metallicity may play an important role in the evolution of
rotational velocities in massive stars through the more compact structures and
lower mass-loss rates predicted at low metallicities \citep[]{mae01}.  Yet
discrepancies still remain as previous studies show significant surface
enrichments which require more efficient rotational mixing at earlier stages of
evolution than the models predict \citep{l06}.

In recent years there has been a strong motivation in observational astronomy to
study the correlation of rotational velocities of OB-type stars and their
surface composition together with understanding the roles of metallicity and the
density of the stellar environment. \citet{kel04} presented the first
extragalactic study of the distribution of rotational velocities of B-type
main-sequence stars. They showed that young cluster objects rotate more rapidly
than field objects, whilst LMC objects rotate faster than their Galactic
counterparts, highlighting the existence of a metallicity dependence. Confirming
the report by \citeauthor{kel04}, \citet{str05} found that BA-type stars close
to the ZAMS in the h and $\chi$ Persei clusters had projected rotational
velocities twice that of a similar aged field population.  \citet{Mar06}
investigated the projected rotational velocity distribution of both B-type and
Be stars in the Large Magellanic Cloud cluster, NGC2004, with the result that
the latter population are rotating faster in their initial ascent along the
main-sequence. Subsequently \citet{wolf07} have studied the role of the initial
density conditions of the star-forming regions on the rotational velocity
distributions in seven galactic clusters. They found that stars formed in low
density regions have a higher number of slow rotators in comparison to those
formed in high density clusters.
\begin{center}
\begin{table*}
\caption[ Observational details]
{\label{obs} Observational details of the telescope/instrument combinations used for
this paper. The second column presents the complete wavelength coverage of the data,
whilst the numbers in parentheses are the number of wavelength settings required to obtain this
coverage. The third and fourth columns display the mean signal-to-noise (S/N) ratio 
and resolution of the data.}  
\begin{flushleft}
\centering
\begin{tabular}{lccc} \hline \hline
Telescope/Instrument	& $\lambda$-range (\AA)	 & S/N & R \\	  
\hline   
\\
VLT/FLAMES	& 3850-4755, 6380-6620             (6) & 100-150 & 20000-30000\\  	
VLT/UVES$^{1}$  & 3750-5000, 5900-7700, 7750-9600  (3) & 40      & 20000\\
ESO2.2m/FEROS   & 3600-9300                        (1) & $>$100  & 48000\\
\hline
\multicolumn{4}{l}{$^{1}$UVES data only used for one object NGC330-124}
\end{tabular}
\end{flushleft}
\end{table*}   
\end{center}

To study the roles of rotation, mass-loss and metallicity on the evolution of 
massive stars, we have undertaken a high resolution spectroscopic survey of
approximately 750 OB stars towards seven young clusters in the Galaxy, LMC and
SMC (the VLT-FLAMES Survey of Massive Stars). The Galactic and Magellanic Cloud 
samples have been discussed in \citet[]{mwpaper, mcpaper} respectively. The O
stars in the sample were analysed by \citet[]{rmsmc, rmlmc}, who derive their
atmospheric \& wind parameters, helium abundances and rotational velocities. 
Helium enrichments were found to be present at the  surface of many of these
stars, implying significant rotational mixing.  However the models considered by
\citeauthor[]{rmsmc} still  underpredicted the degree of helium enrichment
observed. They also found that the more evolved objects rotated slower than the
unevolved stars and that within the population of unevolved stars there was an
excess of fast rotators in the SMC compared to Galactic objects. 

Analysis of the much larger sample of B-type stars is currently underway;
\citet{mwvsini} have derived the rotational velocities of all the Galactic
stars. They confirmed the result of \citeauthor{str05} that the cluster objects
rotate faster than their field counterparts and confirm predictions that the
higher-mass stars with strong stellar winds have lower rotational velocities
due to the loss of surface angular momentum. To understand the efficiency of
rotation in mixing chemically processed material from the interior of a star to
the photospheric layers, it is important to study the surface chemical
composition of these objects in conjunction with their rotational velocity
distribution. \citet[][hereafter Paper~IV]{young} have derived the atmospheric
parameters and surface composition of 50 narrow lined B-type stars in the
youngest of our target clusters (NGC6611, N11, NGC346). In this paper we
extend that analysis to the older clusters in the survey studying 61 narrow
lined stars in NGC3293, NGC4755, NGC2004 \& NGC330. As these stars have low
projected rotational velocities, a detailed atmospheric analysis can provide
highly accurate atmospheric parameters and surface composition, thus providing
the baseline metallicities of these seven regions and an insight into the
evolution of nitrogen as a function of environment. Additionally we will
provide effective temperature scales as a function of spectral type, luminosity
and metallicity for these narrow lined objects which can be applied to the fast
rotators, in which the blending of the lines makes it impossible to determine
effective temperatures directly from the spectra. These effective temperature
scales will have important applications in many areas of astrophysics such as
for comparison with stellar evolution models, determining cluster properties
and understanding the properties of ionising stars.
%-----------------------------------------------------------------------------------

\section{Observations.} 
\label{obstext} 

The spectroscopic data analysed in this paper are from an ESO large
programme using the Fibre Large Array Multi-Element Spectrograph
(FLAMES) on the VLT, primarily with the Giraffe spectrograph, but
also using the fibre-feed to UVES (Ultraviolet and Visual Echelle
Spectrograph). In addition spectra from the Fibre-Fed Extended Range
Optical Spectrograph (FEROS) and UVES (without FLAMES feed) were
obtained for a number of targets in the Galactic clusters. The former
had been omitted from the FLAMES setups as they were too bright,
whilst the UVES data had been obtained prior to the large survey. As
explained in Sect.~\ref{selection}, we have enforced certain criteria
to select the dataset for this analysis which left only one suitable
UVES target. The properties of the datasets are summarised in
Table~\ref{obs}, while the target selection, data reduction and
observational details of all the observations have been discussed in
\citet[][hereafter Paper~I]{mwpaper} and \citet[][hereafter
Paper~II]{mcpaper}. Their target identifications will be used
throughout this paper. Whilst the survey covers seven clusters in
three distinct metallicity regimes; Galactic (NGC6611, NGC4755,
NGC3293), LMC (N11, NGC2004) and SMC (NGC346, NGC330), this paper
will concentrate on the narrow lined stars (\ie those with small
projected rotational velocities) in the older clusters NGC4755,
NGC3293, NGC2004, and NGC330. 

\begin{figure*}
\begin{center}
\epsfig{file=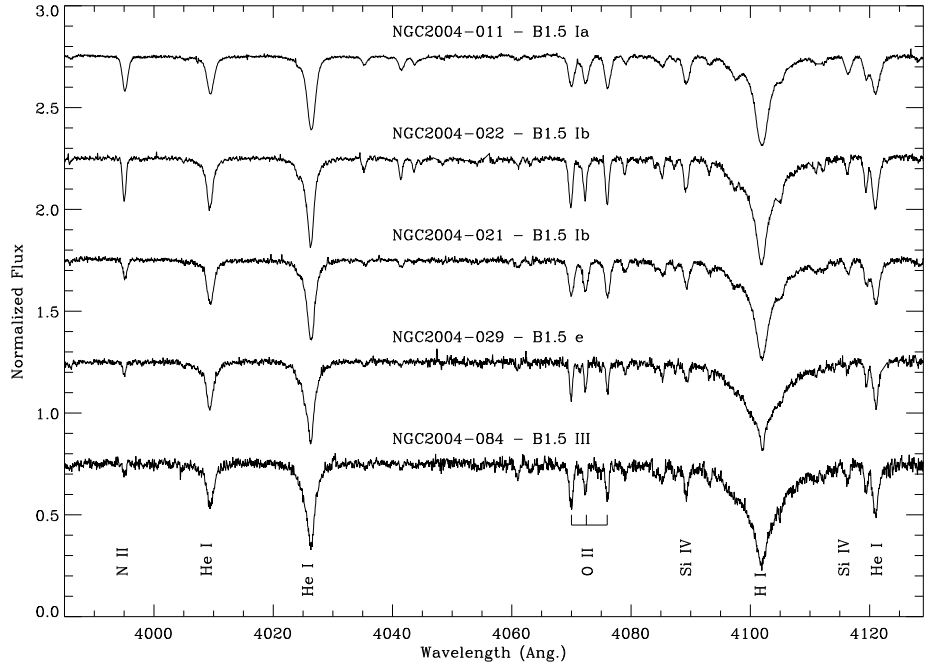, width=130mm, angle=0}
\caption[]
{ Examples of the FLAMES-Giraffe Spectra for B1.5 stars in NGC2004, additional
examples can be seen in Papers 1 \& 2. The spectra are shifted to rest wavelengths. The 
identified lines are: N {\sc ii} $\lambda$3995, He {\sc i} $\lambda\lambda$4009,
4026, 4120, O {\sc ii} $\lambda\lambda$4069, 4072,4076, Si {\sc
iv}$ \lambda\lambda$4089, 4116 and H {\sc i} $\delta$. Note the range in
line intensity of the N {\sc ii} line at 3995 \AA.
} 

\label{spectra}

\end{center}
\end{figure*}
\subsection{Selection of narrow lined stars.} 
\label{selection}

Our selection of objects follows closely the criteria set out in
Paper~IV. The main objective was to select the highest quality spectra
suitable for a reliable model atmosphere analysis. Fast rotators were
excluded because rotational-broadening blends the absorption lines,
thereby decreasing the accuracy with which equivalent widths can be
measured. The criteria applied were as follows:

\begin{itemize}

\item  Spectral types earlier than O9 were excluded as they are more
suited to analyses which utilise unified model atmosphere codes, and
can model the stronger stellar winds of these stars \citep[]{rmsmc,
rmlmc}.
\\
\item Any object whose spectrum was deemed to be contaminated by a
secondary object and for which the lines were not clearly separated
from those of the secondary, was omitted from the analysis.
\\
\item Only objects for which the effective temperature could be accurately
measured using the silicon ionisation equilibrium (viz. Si \3/Si \4~or Si \3/Si
 \2) were considered. 
\end{itemize}

In the case of the NGC2004 targets, the Si \3~lines at 4560 \AA, used for the
temperature determination, were observed in two wavelength settings. Therefore
in addition to the criteria listed above, if the measurement of the equivalent
widths from the two spectra did not agree to within 10\% the object was omitted
from this analysis.  

After applying the above criteria, we were left with 61 objects in total; 8
stars in NGC3293, 10 in NGC4755, 23 in NGC2004 and 20 in NGC330. These objects
are listed in Tables~\ref{galpar} \& \ref{mcpar}, whilst comments on two
objects which were considered during object selection, but that did not fulfill
all our criteria, are included in Appendix A.

\subsection{Data Reduction.}
\label{datared}

The FLAMES-Giraffe spectra were reduced using the Giraffe Base-Line
Reduction Software \citep[][girBLDRS]{ble03} as discussed in Paper~I \& II. An
inherent drawback in multi-fibre spectroscopy is the difficulty in sky
subtraction particularly when treating nebular emission. To deal with this,
typically 15 sky fibres were allocated in each FLAMES plate, those with
significant nebular emission were omitted prior to making a master sky
spectrum. The maximum variations in counts from the sky fibres across the
FLAMES plate were on the order of 10\%, which is comparable to the fibre
throughputs and would be difficult to disentangle from this effect. We carried
out significant testing of the sky subtraction. Initially the sky spectra were
smoothed but this did not remove very narrow absorption lines in the fainter
targets. Finally we used a master sky, which was scaled to the appropriate
fibre throughput and subtracted from all objects. Further to these steps and
those outlined in Papers~I \& II additional steps were required before the
spectra were suitable for analysis with the model atmosphere codes and these
are outlined here. \begin{center}
\begin{table}
\caption[Magellanic Cloud Observation details]
{\label{mcobs} Details of the Magellanic Cloud observations for FLAMES. Columns denoted by 
(a) give the number of exposures, where as columns denoted by (b) give the maximum
separation in days between exposures. Only single exposures were obtained for the
galactic objects and hence they are not included in this table.}  
\begin{flushleft}
\centering
\begin{tabular}{lccccc} \hline \hline
$\lambda$ Setting	& $\lambda_{c}$ (\AA)	&\multicolumn{2}{c}{ NGC2004}&\multicolumn{2}{c}{ NGC330} \\	  
        &                    	& (a) & (b) & (a) & (b)      	  \\
\hline   
\\
HR02	&	3958		&     6    &	6       &     9    &   	3        	  \\	  
HR03	&	4124		&     4    &	0       &     6    &	4     		  \\
HR04	&	4297		&     6    &	2       &     6    &	4     		  \\
HR05	&	4471		&     6    &	1       &     3    &	3     		  \\
HR06	&	4656		&     6    &	0       &     6    &	3     		  \\
HR14	&	6515		&     6    &	34      &     6    &	1     		  \\
\hline 
\end{tabular}
\end{flushleft}
\end{table}   
\end{center}

FLAMES observations were taken in six separate wavelength settings,
and in the case of the Magellanic Cloud fields, multiple exposures
were taken for each setting. These are summarized in Table~\ref{mcobs}
together with the maximum time separation of the individual exposures
for a given wavelength region; further information can be found in
Paper~I. As some of the wavelength settings were observed over an
extended period, (34 days for NGC2004 and 4 days for NGC330) careful
corrections for velocity shifts were required. For each wavelength
setting, each exposure was cross-correlated with the others,
identifying any radial velocity shifts. Stars were classified as {\it
possible} single-lined spectroscopic binaries if the mean radial
velocity of any two sets of exposures differed at the 3$\sigma$ level,
and are noted as such in Table~\ref{mcpar}. This method of
cross-correlation to determine if there were any radial velocity
shifts was dependent on sampling the binaries over a significant part
of their orbit. Hence it was of limited utility for some of the
clusters analysed here, and in particular for the NGC330 cluster where
it was unlikely to identify long period binaries. For the Galactic
cluster objects and the one UVES object, NGC330-124, no cross
correlation was possible as only one exposure was taken for each of
these objects.

A few objects were identified with significant radial velocity shifts
and are likely to be in binary systems, but a number of objects have
also been identified with very small shifts of $<$ 5 \kms. Objects with
similarly low velocity shifts were highlighted in Paper~IV. These
shifts may be significant but require further sampling in time for
corroboration and we simply label these objects as radial velocity
variables (see Table~\ref{mcpar}).

Once the spectra had been cross-correlated, the individual exposures were
combined and any cosmic rays were removed using the {\sc scombine} procedure in
{\sc iraf}\footnote{{\sc IRAF} is distributed by the National Optical Astronomy
Observatories, which are operated by the Association of Universities for
Research in Astronomy, Inc., under agreement with the National Science
Foundation.}. The combined spectra, and in the case of the Galactic stars the
single exposures, were then normalised and individual wavelength settings merged
using the spectral analysis package {\sc dipso} \citep[]{how94}. The spectra
from the four clusters were then inspected for 53 prominent metal lines, the
equivalent widths of these lines were measured if they were clearly visible and
unblended with neighbouring lines. The low order Balmer lines plus the neutral
helium lines in each star were then normalised for comparison with the
theoretical models. Additionally when observed, the singly ionised helium lines
at 4199, 4541 and 4686 \AA~were also normalised as they provide useful
supplementary checks on the effective temperature estimates. Figure~\ref{spectra}
displays some examples of the FLAMES-Giraffe spectra of B1.5 type stars in
NGC2004, additional examples can be seen in Papers 1 \& 2.

%-----------------------------------------------------------------------------------

\section{Spectral Analysis: tools and techniques.} 
\label{tools} 
\begin{center}
\begin{table*}
\caption[Atmospheric parameters of the Milky Way clusters]
{\label{galpar}  Atmospheric parameters for B-type stars in NGC3293 \& NGC4755 as
derived from non-LTE TLUSTY model atmospheres. The majority of the data comes from
FEROS however those taken with FLAMES are marked with $^{1}$. Identifications and
spectral classifications are taken from Paper I. Both the initial and corrected atmospheric
parameters are shown following the discussion in Sect.~\ref{par}. The
uncertainties in these parameters are typically 1000K for \teff, 0.20 dex for
\logg, 3-5 \kms~for \eps~and 5 \kms~for \vsini.  }
\begin{flushleft}
\centering
\begin{tabular}{llccccccccccc} \hline \hline
                  &                 &\multicolumn{3}{c}{Initial Parameters}& &\multicolumn{3}{c}{Corrected Parameters}& &  &  & \\
Star              & Sp.Typ          & \teff & \logg & \eps$_{\rm Si}$ &&\teff & \logg &  \eps$_{\rm Si}$ & &\vsini     & $M_{\star}$ &$\log$($L_{\star}$)  \\	
                  &                 &  (K)  &(cm$^{-2}$)& (\kms)& & (K)  &(cm$^{-2}$)& (\kms)& &(\kms)& ($M_{\odot}$)		  &($L_{\odot}$) \\
\hline   
                  &                 &               &         & & 	&	&	 & & 	   &	       &          \\
NGC3293-003       & B1   III	    & 20700 &  2.75 &	 15   & & 20500 &  2.75 &    13  & & 80    & 18$\pm$ 2 &  4.92 \\  %	    
NGC3293-004       & B1   III	    & 22700 &  3.13 &	 13   & & 22700 &  3.13 &    13  & & 105   & 17$\pm$ 1 &  4.76 \\  %	 
NGC3293-007       & B1   III	    & 22700 &  3.10 &	 12   & & 22600 &  3.10 &    11  & & 65    & 15$^{+2}_{-1}$ &  4.86 \\  %si ii/iii/iv crossed	
NGC3293-010       & B1   III	    & 21325 &  3.20 &	 10   & & 21450 &  3.20 &    11  & & 70    & 12$\pm$ 1 &  4.37 \\  %si ii 24250K 3.45	 
NGC3293-012       & B1   III	    & 21150 &  3.30 &	 10   & & 21500 &  3.30 &    11  & & 100   & 12$\pm$ 1 &  4.37 \\  %	
NGC3293-018       & B1   V	    & 23250 &  3.75 &	  3   & & 23450 &  3.75 &     5  & & 26    & 12$\pm$ 1 &  4.23 \\  %vt flattens off $<$0?      
NGC3293-026$^{1}$ & B2   III	    & 21700 &  3.65 &  $<$0   & & 22100 &  3.65 &     2  & & 30    &  9$\pm$ 1 &  3.83 \\  %$<0$?      
NGC3293-043$^{1}$ & B3   V	    & 19500 &  4.05 &	$<$0  & & 19500 &  4.05 &  $<$0  & & 14    &  7$\pm$ 1 &  3.32 \\  %
%NGC3293-062      & B3   V	    & 17550 &  3.95 &	  5   & & 17550 &  3.95 &     5  & & 24    &	       &       \\  %uncertain vt 5 adopted si3 ew range small  
                  &                 &               &         & & 	&	&	 & & 	   &	       &       \\
\hline 
                  &                 &               &         & & 	&	&	 & & 	   &	       &       \\
NGC4755-002       & B3   Ia	    & 15950 &  2.20 &	  19  & & 15950 &  2.20 &     18 & & 70    & 22$\pm$ 1 &  5.15 \\  %	
NGC4755-003       & B2   III	    & 17600 &  2.50 &	  17  & & 17700 &  2.50 &     15 & & 38    & 19$^{+1}_{-2}$ &  4.97 \\  %	  
NGC4755-004       & B1.5 Ib	    & 19400 &  2.60 &	  18  & & 19550 &  2.60 &     17 & & 75    & 19$^{+1}_{-2}$ &  5.00 \\  %si ii/iii/iv crossed      
%NGC4755-005      & B2   III	    & ------&  ---- &	  --- & & ------&  ---- &      --& & ---   &	       &       \\  %SB2 include in paper.      
NGC4755-006       & B1   III	    & 19000 &  2.85 &	  14  & & 19900 &  2.95 &     17 & &100   & 11$\pm$ 1 &  4.36 \\  %O\&Si abundances strange	
%NGC4755-011      & B1.5 V	    & 23300 &  4.05 &	$<$0  & & 23300 &  4.05 &   $<$0 & & 22    &	       &       \\  %sIii 23800 4.10  suspected SB2 abundance strange leave out.   
NGC4755-015       & B1   V	    & 21800 &  3.65 &	   2  & & 22400 &  3.70 &      5 & & 48    & 10$\pm$ 1 &  3.98 \\  %vt=2 or $<$2  
NGC4755-017       & B1.5 V	    & 20500 &  3.90 &	   6  & & 20400 &  3.90 &      3 & & 75    &  9$\pm$ 1 &  3.83 \\  %log g difficult 
NGC4755-020$^{1}$ & B2   V	    & 21800 &  3.95 &	   3  & & 21700 &  3.95 &      1 & & 120   &  9$\pm$ 1 &  3.78 \\  %
NGC4755-033$^{1}$ & B3   V	    & 18000 &  3.90 &	  10  & & 17300 &  3.85 &      6 & & 75    &  6$\pm$ 1 &  3.11 \\  %si3 v.weak
NGC4755-040$^{1}$ & B2.5 V	    & 18250 &  4.00 &	$<$0  & & 18900 &  4.10 &      2 & & 65    &  6$\pm$ 1 &  3.25 \\  %si3 v.weak
NGC4755-048$^{1}$ & B3 V	    & 18200 &  3.95 &	   6  & & 17800 &  3.95 &      4 & & 55    &  6$\pm$ 1 &  2.98 \\  %si3 v.weak
%NGC4755-061      & B5 V	    & 	    &       &	      & & 	&	&	 & & 	   &	       &       \\  %si3 v.weak
\\
\hline
\multicolumn{13}{l}{$^{1}$Spectra from FLAMES with Giraffe spectrograph.}
\end{tabular}											
\end{flushleft} 										
\end{table*}   
\end{center}

Our analysis follows the methodology presented in Paper~IV, as we
have strived to provide a consistent analysis of the entire FLAMES
dataset. This is important as later in this paper the implications
from the results of both the young and old clusters will be discussed
together. Due to the similarities with Paper~IV the details of the
spectral analysis will not be reiterated here but we will simply
provide a summary. 

The spectra were analysed with the Queen's University Belfast (QUB) B-type star
grid \citep[]{rya03, duf05}, which was generated using the non-LTE model
atmosphere code TLUSTY and line formation code SYNSPEC. For Hydrogen Lyman
and Balmer lines the broadening tables of \citet[]{vid73} were used, whilst for
the higher members of the spectral series the approach described by
\citet[]{hub94} was applied. Further details can be found in the SYNSPEC user
manual. The QUB grid has been created specifically for B-type stars of all
luminosity classes, covering the effective temperatures (\teff), surface
gravities (\logg) and microturbulent velocities ($\xi$) appropriate for late O
to late B-type stars (\teff: 35-12 kK, steps of 2.5 kK; \logg: 4.5 down to the
Eddington limit, steps of 0.25; $\xi$: 0, 5, 10, 15, 20 and 30 \kms). The grid
covers a range of metallicities appropriate to the Milky Way, LMC and SMC. In
addition, for each of these metallicity grids the light elements are varied
around their normal abundances by +0.8, +0.4, -0.4 and -0.8. The atmospheric
parameters along with the photospheric abundances were determined for each star
by interpolation within this grid via QUB IDL routines \citep[]{rya03}. The
energy levels and oscillator strengths relating to the transitions for the metal
lines considered in this work are available online at
http://star.pst.qub.ac.uk/~pld/line\_identifications.html. \\

%-----------------------------------------------------------------------------------

\section{Stellar parameters}
\label{par}
%\begin{center}
\begin{table*}
\caption[Atmospheric parameters of the MC clusters]
{\label{mcpar}  Atmospheric parameters for B-type stars in NGC2004 \& NGC330 as derived
from non-LTE TLUSTY models.  Identifications and spectral classifications are taken from Paper II.
 The uncertainties are as described in Table~\ref{galpar}.}  
\begin{flushleft}
\centering
\begin{tabular}{llccccccccccc} \hline \hline
            &               &\multicolumn{3}{c}{Initial Parameters}& &\multicolumn{3}{c}{Corrected Parameters}& &  &  & \\
Star        & Sp.Typ        & \teff & \logg &  \eps$_{\rm Si}$ & &\teff & \logg& \eps$_{\rm Si}$ & & \vsini    & $M_{\star}$/$M_{\odot}$ &$\log$($L_{\star}$/$L_{\odot}$)  \\	
                 &                 &  (K)  &(cm$^{-2}$)& (\kms)& & (K)  &(cm$^{-2}$)& (\kms)& &(\kms)& ($M_{\odot}$)		  &($L_{\odot}$) \\
\hline   
\\
            &               &        &       &        & &	&	&	 &&	 &	       &      \\ %	     \\
NGC2004-003$^{R5}$ & B5   Ia& 14450  &  2.10 &    14  & &14450  &  2.10 &    15  &&  42  & 20$^{+2}_{-1}$   & 5.10 \\ % binary 4km/s \\ 
NGC2004-005$^{R5}$ & B8   Ia& 12600  &  1.90 &    24: & &12390  &  1.90 &    12  &&  31  & 18 $\pm$ 2  & 4.93 \\ % binary 3km/s \\ 
NGC2004-007 & B8   Ia       & 12560  &  2.00 &    29: & &12250  &  2.00 &    10  &&  25  & 17 $\pm$ 2  & 4.88 \\ %		     \\ 
NGC2004-010 & B2.5 Iab      & 17050  &  2.40 &    16  & &17160  &  2.40 &    14  &&  45  & 19 $\pm$ 2  & 5.02 \\ %		     \\ 
NGC2004-011 & B1.5 Ia       & 21300  &  2.75 &    14  & &21250  &  2.75 &    13  &&  62  & 24 $\pm$ 2  & 5.22 \\ % binary    \\
NGC2004-012 & B1.5 Iab      & 21270  &  2.87 &    12  & &21270  &  2.87 &    12  &&  47  & 18 $\pm$ 2  & 4.92 \\ %		     \\
NGC2004-014 & B3   Ib       & 17660  &  2.85 &    14  & &17800  &  2.85 &    10  &&  20  & 15 $^{+2}_{-1}$ & 4.72 \\ %  	     \\ 
NGC2004-021 & B1.5 Ib       & 21400  &  3.00 &    12  & &21450  &  3.00 &    14  &&  59  & 16 $^{+2}_{-1}$ & 4.82 \\ %  	     \\
NGC2004-022 & B1.5 Ib       & 21700  &  3.15 &    10  & &21780  &  3.15 &    11  &&  42  & 16 $^{+2}_{-1}$ & 4.79 \\ % korn:23450,3.34,14 \\
NGC2004-026$^{R}$ & B2   II & 22700  &  3.65 &     1  & &22900  &  3.65 &     0  &&  19  & 14 $\pm$ 1  & 4.68 \\ % binary 40km/s\\ 
NGC2004-029$^{R}$ & B1.5 e  & 23100  &  3.50 &     1  & &23100  &  3.50 &     1  &&  30  & 14 $\pm$ 1  & 4.65 \\ % Binary 15-4    \\ 
NGC2004-036 & B1.5 III      & 22200  &  3.35 &     5  & &22870  &  3.35 &     7  &&  42  & 13 $\pm$ 1  & 4.58 \\ %		     \\ 
NGC2004-042 & B2.5 III      & 20930  &  3.45 &     3  & &20980  &  3.45 &     2  &&  42  & 12 $\pm$ 1  & 4.45 \\ %		     \\ 
NGC2004-046 & B1.5 III      & 25770  &  3.80 &  $<$0  & &26090  &  3.85 &     2  &&  32  & 15 $\pm$ 1  & 4.62 \\ % V or B1III? SiIV 56m\AA \\ 
NGC2004-053 & B0.2Ve        & 32000  &  4.15 &     3  & &31500  &  4.15 &     6  &&   7  & 18 $\pm$ 1  & 4.77 \\ % heII4541 - 500K hotter	 \\ 
NGC2004-061 & B2   III      & 21090  &  3.35 &  $<$0  & &20990  &  3.35 &     1  &&  40  & 11 $\pm$ 1  & 4.31 \\ %		     \\ 
%NGC2004-062 & B0.2V         & 31760  &  4.38 &  $<$0  & &31670  &  4.38 &     1  &&  106 & 19 $\pm$ 1  & 4.72 \\ % heII4541 good pos  hotter	 \\ 
NGC2004-064 & B0.7-B1 III   & 25700  &  3.70 &     3  & &25900  &  3.70 &     6  &&  28  & 13 $\pm$ 1  & 4.48 \\ %		     \\ 
NGC2004-070 & B0.7-B1 III   & 27200  &  3.90 &  $<$0  & &27400  &  3.90 &     4  &&  46  & 14 $\pm$ 1  & 4.51 \\ %		     \\ 
NGC2004-084 & B1.5 III      & 27170  &  4.00 &  $<$0  & &27395  &  4.00 &     3  &&  36  & 14 $\pm$ 1  & 4.46 \\ % (V) or B1III? SiIV ew=68m\AA \\ 
NGC2004-090 & O9.5 III      & 31750  &  4.05 &     3  & &32500  &  4.10 &  $<$0  &&  16  & 17 $\pm$ 1  & 4.64 \\ % heII4541 -33kK,logg=4.15	    \\ 
NGC2004-091$^{R5}$& B1.5 III& 26600  &  4.05 &     1  & &26520  &  4.05 &     0  &&  40  & 13 $\pm$ 1  & 4.42 \\ % binary  2	   \\  
NGC2004-108$^{R}$& B2.5 III & 22600  &  4.00 &  $<$0  & &22600  &  4.00 &  $<$0  &&  13  & 10 $\pm$ 1  & 4.21 \\ % binary?	\\ 
NGC2004-119 & B2   III      & 23210  &  3.75 &  $<$0  & &23210  &  3.75 &  $<$0  &&  15  & 10 $\pm$ 1  & 4.15 \\ %		     \\ 
\\
\hline
\\
NGC330-002  & B3   Ib	    & 14500  &  2.15 &    20  & &14590  &  2.15 &    16  &&  14  & 15 $\pm$ 2  & 4.73 \\ %	      \\
NGC330-003  & B2   Ib	    & 17250  &  2.25 &    15  & &17210  &  2.25 &    20  &&  49  & 16$^{+2}_{-1}$  & 4.84 \\ %	      \\
NGC330-004  & B2.5 Ib	    & 17000  &  2.30 &    16  & &17000  &  2.30 &    16  &&  36  & 15 $\pm$ 1  & 4.77 \\ %	      \\
NGC330-005  & B5   Ib	    & 13700  &  2.25 &     8  & &13700  &  2.25 &     8  &&  16  & 13 $\pm$ 2  & 4.54 \\ %	      \\
NGC330-009  & B5   Ib	    & 14000  &  2.45 &    10  & &13940  &  2.45 &     6  &&  29  & 12 $\pm$ 2  & 4.41 \\ %	      \\
NGC330-010  & B5   Ib 	    & 14900  &  2.60 &     9  & &14820  &  2.60 &     4  &&   0  & 12 $\pm$ 1  & 4.40 \\ %	    \\
NGC330-014  & B1.5 Ib	    & 20000  &  2.75 &    15  & &20130  &  2.75 &    18  &&  81  & 14 $\pm$ 1  & 4.64 \\ %	      \\
NGC330-016  & B5:  II	    & 14300  &  2.60 &    10  & &14220  &  2.60 &     5  &&  40  & 10 $\pm$ 1  & 4.20 \\ %	    \\
NGC330-017  & B2   II	    & 22000  &  3.35 &  $<$0  & &22000  &  3.35 &  $<$0  &&  14  & 14 $\pm$ 1  & 4.62 \\ %	      \\
NGC330-018  & B3   II	    & 18000  &  2.95 &     5  & &18000  &  2.95 &     5  &&  46  & 12 $\pm$ 1  & 4.41 \\ %	    \\
NGC330-020  & B3   II	    & 16400  &  2.85 &     2  & &16720  &  2.85 &     5  &&  44  & 11 $\pm$ 1  & 4.31 \\ %	      \\
NGC330-022  & B3   II	    & 18450  &  3.00 &     7  & &18860  &  3.00 &     1  &&  23  & 12 $\pm$ 1  & 4.38 \\ %	      \\
NGC330-026  & B2.5 II	    & 22500  &  3.40 &  $<$0  & &22500  &  3.40 &  $<$0  &&  71  & 12 $\pm$ 1  & 4.46 \\ %	    \\
NGC330-027  & B1   V	    & 22000  &  3.20 &     6  & &22040  &  3.20 &     7  &&  80  & 12 $\pm$ 1  & 4.42 \\ %	      \\
NGC330-032  & B0.5 V	    & 29700  &  4.15 &  $<$0  & &29700  &  4.15 &  $<$0  &&  17  & 16 $\pm$ 1  & 4.63 \\ %	      \\
NGC330-042  & B2   II	    & 25650  &  3.75 &     3  & &25450  &  3.75 &     1  &&  26  & 12 $\pm$ 1  & 4.34 \\ %	      \\
NGC330-047  & B1   V	    & 26700  &  4.05 &     0  & &26700  &  4.05 &     0  &&  28  & 12 $\pm$ 1  & 4.30 \\ %	      \\
NGC330-074  & B0   V	    & 32300  &  4.20 &     2  & &32020  &  4.20 &     4  &&  29  & 15 $\pm$ 1  & 4.31 \\ % poss variable vr	 \\
NGC330-114  & B2   III	    & 23800  &  3.90 &     3  & &23800  &  3.90 &     4  &&  17  &  9 $\pm$ 1  & 3.79 \\ %	      \\
NGC330-124  & B0.2 V   & 31150  &  4.25 &  $<$0  & &30980  &  4.25 &     2  &&  95  & 15 $\pm$ 1  & 4.38 \\ %	      \\
\hline
\multicolumn{13}{l}{$^{R}$ Radial velocity variations detected at the 3$\sigma$
level, these objects are candidates for binaries, see Sect.~\ref{datared}.  }\\
\multicolumn{13}{l}{`R5' indicates that the radial velocity variation is less than 5\kms.}
\end{tabular}											
\end{flushleft} 										
\end{table*}   
%\end{center}

A static stellar atmosphere is characterised by four parameters: \teff, \logg,
$\xi$ and metallicity (Z). These parameters are interdependent and hence are
determined through an iterative procedure, that assumes an appropriate
metallicity (which depends on the cluster/galaxy) and estimates of the
atmospheric parameters based on the spectral type of the star. By choosing
suitable initial estimates of the stars properties, one can significantly
reduce the number of iterations required. The stellar parameters of our
targets, are presented in Tables~\ref{galpar} \& \ref{mcpar}.

{\bf Effective temperatures} were determined using the silicon ionisation
balance \ie that the abundance estimates derived from the Si \3 lines (4552,
4567 \& 4574 \AA) agree with that from the Si \4 line (4116 \AA) for hot
objects, or those from the Si \2 lines (4128 \& 4131 \AA) for cooler objects. 
Several of the B1 \& B2 objects in the galactic cluster, NGC3293, had  all
three ionisation stages present in their spectra. However the temperatures
determined from the Si {\sc iii}/\4 and Si {\sc iii}/\2  ionisation stages
differed, with the latter generally requiring higher temperatures. For
NGC3293-007 where the Si \4 line is relatively strong and well observed this
difference is only 200 K, but for NGC3293-010,-018, \& -026 the differences are
$\sim$ 2000 to 2500 K.  As the Si \2 spectrum is the weakest of the three
ionisation stages the \teff~estimated from the Si \3/\4 lines have been adopted
for these stars. NGC4755-004 also has the three silicon ionisation stages
present in its spectra and for this object the estimates were in excellent
agreement. 

For the hotter objects (with spectral types earlier than B1),  the He \2 4541
and 4199 \AA\ lines were used as an additional check on the temperature.  The
estimates from the two elements normally agreed to within 500K with the He
lines implying slightly higher temperature.  For one star, NGC2004-090, a
significant discrepancy was found with the estimate from the Si lines being
31750 K, whilst the He \2 lines implied a temperature of 33000 K. However
the Si \3 lines are relatively weak in this spectrum increasing the uncertainty
in this effective temperature estimate.  

We believe that our effective temperature estimates should have an uncertainty
of typically $\pm$1000 K. However there are a number of other  objects for
which larger error estimates are appropriate, due to either the Si lines from
one of the ionisation stages being very weak or the difficulty in constraining
the microturbulence. In these cases, errors of up to 2000 K have been
adopted when estimating uncertainties in the derived abundances.

{\bf Surface gravities} were determined by comparing theoretical spectra with
the observed profiles of the hydrogen Balmer lines,  \hg~and \hd. This was
achieved using automated procedures developed in IDL, to fit models within the
TLUSTY grids using $\chi^{2}$ techniques. To increase the accuracy of our 
estimates, a higher resolution TLUSTY grid has been generated in gravity space,
with steps of 0.1 dex in \logg~from 4.5 dex down to the Eddington limit. The
estimates derived from the two hydrogen lines normally agree to within 0.1 dex,
with any differences mainly arising from  errors in the normalisation of the
observed spectra. 

{\bf Microturbulences} have been derived from the Si \3 lines 4552, 4567 \& 4574
\AA, by ensuring that the abundance estimates were consistent (\ie a slope of
zero is obtained in a plot of equivalent widths against abundance estimates).
For a number of objects in our sample, and those presented in Paper~IV,  a
microturbulence of 0 \kms~has been adopted.  Unfortunately this did not  produce
a slope of zero in the equivalent-width versus abundance-estimate diagram but  
was the value of microturbulence which brought the slope the closest to zero
(these values are denoted in Tables~\ref{galpar} \& \ref{mcpar} with $<$ 0
\kms). The cause for this discrepancy is unclear and has been discussed in
detail in Paper~IV. The uncertainties in the adopted microturbulence depend on
the accuracy of the measured equivalent widths with typical errors of 3-5 \kms.
Uncertainties of 5 \kms~are only required for those objects with  large
microturbulences (\ie $\xi$~ $>$ 10 \kms),  whose derived silicon (and indeed
other) abundances are less sensitive  to the value adopted. The microturbulence
can also be estimated from other species and this can lead to values that are
generally consistent with our adopted uncertainties. Again a detailed discussion
of this can be found in Paper~IV and will not be repeated here. 

In Paper~IV, a microturbulence for each object which provides a silicon
abundance equal to the median of that from all targets in the cluster
was also considered. This significantly reduces the scatter in the
abundances derived for elements (excluding nitrogen) within a
cluster.  The changes in the microturbulence are normally consistent
with the errors discussed above and in most cases the effect of these
changes on the other atmospheric parameters are minor, typically less
than 200 K for \teff~and negligible for \logg. However for a small
number of objects the changes in microturbulence had a significant
effect on the other atmospheric parameters due to their 
interdependence. Hence, we decided to carry out our abundance analysis 
in three steps: 

\begin{enumerate}  \item Using the microturbulence determined from
the Si \3 lines we derived the stellar parameters and surface
abundances as described above (see Table~\ref{ablin}).

\item For each star in a cluster, the microturbulence was varied until
the abundance estimate from the Si \3 lines was that of the median 
silicon abundance of the cluster. The abundances of other elements 
were then recalculated with this new value of microturbulence 
(see Table~\ref{fixvtab}).

\item Finally, since the ionisation balance and microturbulence are
reliant on the silicon lines, the \teff~and \logg~were reiterated,
where necessary, for the new value of microturbulence, whilst
maintaining the median silicon abundance of the cluster. This
required, on occasions, an additional reiteration of the
microturbulence due to the interdependency of the parameters. The
other abundances were then recalculated with these parameters resulting in
the abundances presented in Table~\ref{fixall}.   
\end{enumerate} 

\noindent The estimates of the atmospheric parameters from step one and three
are listed in Tables~\ref{galpar} \& \ref{mcpar} as initial and corrected
parameters, respectively. Nine out of the sixty-one targets analysed in
this work have microturbulent velocities in the range of 15-18 \kms. These
velocities are typical of the sound speeds in NLTE model atmospheres at the
line formation depths of the metal lines considered in the abundance analysis
of these stars. (The sound speeds of such models have been discussed by
\citet[]{mcE98} and the reader is referred there for further details.) Assuming
that microturbulence represents a true microscopic velocity field in the
atmospheres and as this turbulent velocity is a significant
fraction of the sound speed in these nine stars, one would expect this to
result in the formation of shocks. This casts some doubt on the validity of
applying static atmospheres rather than hydrodynamical atmospheres in the
interpretation of these stars. However this should not affect the main results
of this paper, in particular the large range in nitrogen abundances observed in
these stars. Whilst these nine objects are amongst those with the highest
nitrogen abundances they are not the only objects with significant enrichments
of nitrogen in their atmospheres.

In order to fit theoretical spectra to the observed data, any
additional broadening  of the spectral lines due to, for example,
rotation  or macroturbulence must be included. Here we have assumed
that rotational broadening will be the dominant mechanism and have
estimated its magnitude from the line profiles of the He \1 4026 \AA
~line for objects where the projected rotational velocity (\vsini)
was greater than 50 \kms~and from the Si \3 lines where it was less than 50
\kms. Details of the methodology have been presented in Paper~IV \&
\citet[][hereafter Paper V]{hun07} and the estimates are listed in 
Tables~\ref{galpar} \&~\ref{mcpar}

In the case of dwarfs and giants these estimates can be safely  considered to
be a measurement of the projected rotational velocity as the instrumental
broadening and microturbulence have been taken into account. However for most
of the supergiants this excess broadening is likely to be a convolution of
rotational broadening and other broadening mechanisms as has been discussed by
\citet{rya02},  \citet{duf06}, \citet{sim06} \& \citet{sim07}. In Paper~V, a
Fourier method \citep{sim07} has been applied to these  supergiants to
deconvolve the rotational broadening from other mechanisms and more realistic 
estimates of the projected rotational velocities were obtained. However
we emphasize that the values quoted here have only been used to account  for
additional broadening in the line profile when comparing observation and
theory, for which purpose we believe them to be adequate.

{\bf Luminosities and masses} were estimated for each object and are 
presented in Tables~\ref{galpar} \& \ref{mcpar}. Luminosities for all the Milky Way
cluster targets in our survey have been presented in Paper~III, but here we have
recalculated them following the same technique, based on the new, more accurate
atmospheric parameters. For the Magellanic Cloud stars, the
luminosities of each object were determined by assuming a constant reddening
towards each cluster, Bolometric correction from the empirical solutions of
\citet[]{vac96} and \citet[]{bal94}, and the apparent magnitudes presented in
Paper~II. For the LMC a standard Galactic extinction law of $A_{\rm V}$ =
3.1E(B-V) was used whilst for NGC330 we took $A_{\rm V}$ = 2.72E(B-V) from
\citet[]{bou85}. We adopt an E(B-V) of 0.09 for NGC2004 \citep[]{sag91} and
NGC330 \citep[]{l97}. The distance moduli (DM) adopted were 18.91 and 18.56 for
 NGC330 and NGC2004, respectively \citep[]{hil05, gie05}.  The masses were then
derived by plotting luminosity and temperature for each object on a
Hertzsprung-Russell (HR) diagram and interpolating between stellar evolutionary
tracks of varying masses.  The evolutionary tracks adopted are from
\citet[]{mey94} together with those of \citet[]{sch92} for the Milky Way
clusters, and \citet[]{sch93} and \citet[]{char93} for NGC2004 and NGC330,
respectively. Quoted uncertainties in the derived masses assumed an uncertainty
of 0.1 dex in $\log$ $L_{\star}$/$L_{\odot}$ and a negligible error from the
uncertainty in the effective temperature estimates.

%-----------------------------------------------------------------------------------

\section{Photospheric Abundances.}
\label{abundances}

This Paper is concerned with analyzing the photospheric abundance patterns,
specifically of C, N, and O, of the selected targets. Therefore we have used the
atmospheric parameters, discussed in the previous sections, to derive absolute
abundances of these stars by interpolating between models with the same
atmospheric parameters but differing light element abundances. Using the
equivalent widths measured for each line in the spectra arising from C, N, O,
Mg, Si and Fe we have derived  abundance estimates for each line of a given
species (see Tables~\ref{ngc3293a} - \ref{ngc330g}), and from these the mean
abundances in a given star were determined and are presented in
Tables~\ref{ablin}, \ref{fixvtab}, \& \ref{fixall}. Abundances of C, N, O, Mg \&
Si were determined using the non-LTE model atmospheres and line formation
calculations. As will be discussed later in this section, the Fe abundances were
also derived using non-LTE model atmosphere structures but with LTE line
formation calculations. Table~\ref{ablin} presents the abundances derived using
the initial atmospheric parameters from Tables~\ref{galpar} \& \ref{mcpar}.

\subsection{Errors in abundances.} 
\label{aberror} 

There are several factors which contribute to the uncertainties in the
mean abundances derived in the manner described above, including the
atomic data adopted in the model atmosphere code, errors in EW
measurements from normalisation problems or blending with other lines and
errors in the stellar parameters. The former two will be decreased for
species for which there are many lines, whereas the latter arises mainly
from the interdependence of the parameters. The random uncertainty arising
from the atomic data and observational data are accounted for in the
standard error in the mean of the abundances. The systematic errors which
arise from the uncertainties in the atmospheric parameters were then
accounted for by changing these in turn by their relevant uncertainties
and re-determining the abundance estimates. The random and systematic
errors (from each parameter) were then summed in quadrature to give the
uncertainties listed in Tables~\ref{ablin}, \ref{fixvtab}, \&
\ref{fixall}.  In the case of species for which only one or two lines
(viz. Si \4 and Mg \2) were observed,  we have adopted the random error of
a better-sampled species (viz. oxygen).  Since we have not explicitly
accounted for the interdependence of the stellar parameters, specifically
\teff~and \logg, the uncertainties may be slightly overestimated.

\subsection{Effect of Microturbulence on abundances.}

As discussed in Section~\ref{par}, for a number of objects it was difficult to
obtain a microturbulence from the silicon lines and so the silicon abundance in
each star was fixed to the mean cluster abundance (NGC3293: [Si/H]=7.45;
NGC4755: [Si/H]=7.41; NGC2004: [Si/H]=7.21; NGC330: [Si/H]=6.81) by adjusting
the microturbulence. The abundances derived using the initial \teff~and
\logg~but with the microturbulence fixed to give the desired Si abundance
($\xi$$_{\rm ave}$) are presented in Table~\ref{fixvtab}.  A comparison with
Table~\ref{ablin} shows that the Si \3 abundances are now more consistent within
the clusters. Nevertheless, in comparing the Si \3 abundances with those from
the other Si ionisation stages one can see that the ionisation balance is no
longer maintained. Table~\ref{fixall} presents the abundances derived with the
new microturbulence and appropriate \teff~and \logg~(\ie using the corrected
atmospheric parameters from Tables~\ref{galpar} \& \ref{mcpar}) to maintain the
mean cluster Si abundance in each object, whilst also maintaining the ionisation
balance.  By fixing the microturbulence the scatter in Si abundances is greatly
reduced as those objects with [Si/H] $>$ [Si/H]$_{\rm cluster}$ can be easily
brought into agreement with the [Si/H]$_{\rm cluster}$, by increasing $\xi$~(see
Tables~\ref{fixvtab} \& \ref{fixall}).  This procedure only slightly reduces
the scatter in abundances from the other elements (by less than 0.03 dex),
except for the mean oxygen abundance. For those targets with [Si/H]
significantly ($\sim$ 0.2 dex) above the mean cluster [Si/H], their oxygen
abundances were also quite high (see Table~\ref{ablin}). By increasing the
microturbulence in these objects, their silicon and oxygen abundances are
reduced putting them into better agreement with the mean values of the cluster
(see Table~\ref{fixall}).  For some objects, where the silicon abundance was
significantly lower than the desired cluster abundance ([Si/H] $\leq$
[Si/H]$_{\rm cluster}$), it was not possible to increase the silicon abundance
as the microturbulence was close to/or at zero and would need to be lowered
further still (see for example NGC2004: \#91, \#108, \#119).

\subsection{Abundances of individual species.}
\label{individ_ab}
Before discussing the final results it is worth mentioning a few general 
points on the mean abundances for the individual elements:

{\bf Carbon} is a problematic species in B-type stars, a result of  the C
\2 lines being very sensitive to non-LTE effects. In the spectra of our
targets the strongest carbon lines are at 3921, 4267, 6578 \& 6582 \AA.
The carbon model atom in TLUSTY fails to reproduce consistent abundances
from these 4 lines, with the carbon abundance estimated from the 4267 \AA\
line normally found to be lower than for the other three lines. 

Recently \citet[]{c2a,c2b} have constructed a new comprehensive non-LTE carbon
model atom and have shown that for six slowly-rotating early B-type stars their
model can produce consistent carbon abundances from 21 C \2 lines (including
those mentioned above) in the visible spectrum. To investigate the offsets
between the C abundances derived with TLUSTY in this work and those derived by
\citeauthor[]{c2a} using their C \2 model atom and a hybrid approach to the
non-LTE line formation, we have analysed one of their stars in the same way as
our targets. Using the same equivalent widths and spectra as \citeauthor[]{c2a},
we analysed HR3468 a Galactic B1.5 III star. In Table~\ref{carbon} we present
the results of this comparison, the \teff~and \logg~estimated in the two
analyses are in good agreement. This is reassuring as there were a number of
differences in the analyses viz. (1) \citeauthor[]{c2a} estimated the \teff~from
the C \2/C \3 ionisation equilibrium and (2) they derived the microturbulence
from the 17 carbon lines. This last point is the reason for the differences in
the values of  microturbulence in Table~\ref{carbon}. The difference in
abundances from the two analyses are quite significant, and vary from line to
line. The C \2 4267 \AA\ line differs the most by 0.5 dex, this difference is
reduced to a factor of 2.5 if the carbon abundances are derived using the lower
microturbulence of 5 \kms~in the TLUSTY analysis.

\begin{table}
\begin{center}
\caption[Comparison carbon model atoms.]
{\label{carbon} Results of analysis of HR3468.  The first column are the TLUSTY
atmospheric parameters and abundances obtained following the procedures described in Sects.~\ref{obstext}
\& \ref{par}, the third column are the TLUSTY results using the parameters from
\citet{c2b}, and the fourth column are those from \citeauthor{c2b}.
\citeauthor{c2b} estimate the \teff~from the C \2/C \3 ionisation equilibrium and derive the
microturbulence from the carbon lines not the silicon lines, as done in the
TLUSTY analysis.  Abundances are presented as [X]=12 + $\log$([X/H]) in units of dex.}  
\begin{tabular}{lccc} \hline \hline
         &\multicolumn{3}{c}{HR 3468}      \\
         &\multicolumn{2}{c}{Tlusty}    &\citeauthor{c2a} \\
 \hline \\
\teff   (K)         &    22800        &	22900	     &  22900 \\
\logg   (cms$^{-2}$)&     3.55        &	 3.60	     &  3.60  \\
\eps    (\kms)      &~~~~~~~10~~~(Si) &5        &  ~~~~~~~5~~~~(C) \\
     	 &                 &		     &        \\
C \2 3921& 7.92            & 8.14	     &  8.34  \\
C \2 4267& 7.83            & 7.95	     &  8.33  \\
C \2 6578& 8.05            & 8.37	     &  8.40  \\
C \2 6582& 7.96            & 8.17	     &  8.40  \\
\\ \hline			  
\end{tabular}			  
\end{center}			  
\end{table}

Two factors prevent us from adopting these offsets and applying them to the
carbon abundances derived here using TLUSTY. Firstly, the carbon abundances
derived by \citeauthor[]{c2a} were estimated using a profile fitting technique
and not the curve-of-growth technique applied here. Profile fitting is very
reliant on the \vsini~values adopted and results in uncertainties of up to 0.15
dex. More importantly, a comparison with only one object does not allow us to get
a clear picture of the offsets in the carbon abundances derived throughout the
entire parameter range covered by our targets.  As each of the carbon lines
behaves differently and since the C \2 4267 \AA\ line is the only line
detectable throughout our spectral range, we have taken the absolute abundance
of this line to represent the carbon abundance in these stars. In
Tables~\ref{ablin}, \ref{fixvtab}, \& \ref{fixall} the absolute carbon
abundances are those derived from the C \2 4267 \AA\ line without any applied
offsets but with the caveat that the absolute value is likely to be {\it
significantly} lower than the true value and should only be used differentially.
The abundances estimated from the other carbon lines are presented in
Tables~\ref{ngc3293a} -\ref{ngc330g} for comparison.

{\bf Nitrogen} abundances in B-type stars generally span a large range as a
result of being processed in the CNO-cycle and this element being   present in
the photosphere through some, much debated, mechanism. The range of nitrogen
abundances in our targets are again large: NGC3293: 7.45- 7.66 dex, NGC4755:
7.43 - 8.18 dex, NGC2004: 6.81 - 8.16 dex, NGC330: 6.76 - 7.83 dex. The spread
in the nitrogen abundances of the stars in the Galactic cluster NGC3293, is
smaller that of NGC4755 (also a Galactic cluster), but the former has no
supergiants present in the analysed data. The three objects in NGC4755 with
significant nitrogen enrichments are the three most massive and luminous
Galactic objects in our sample. In each of the other clusters, it is the
supergiants which have the highest nitrogen abundances, although there is a
significant range in nitrogen abundances derived from the main sequence and
dwarf objects. The nitrogen abundances will be discussed further in
Sect.~\ref{discuss}.

{\bf Oxygen} has several strong features in the spectra of B-type stars and as
such its mean abundance is very dependent on the microturbulence adopted for a
given star. It has been previously noted that microturbulent velocities when
derived from a range of oxygen lines are generally higher than that derived from
the Si~\3 4560 \AA\ multiplet \citep[]{vra00,tru02}. However in Paper~IV it was
shown that by selecting the lines from a single oxygen multiplet,  the
microturbulence estimated from both oxygen and silicon were in better agreement
(in some cases to within 1 \kms). The oxygen abundances are reasonably
consistent within a cluster but a comparison of Tables~\ref{ablin},
\ref{fixvtab}, \& \ref{fixall} reveals the strong dependence of oxygen on the
microturbulence. For example, the oxygen abundances are lower than the mean
cluster abundances for a number of objects where the microturbulence adopted was
0 but for which an even lower value, if realistic, could have been adopted
(NGC3293-043, NGC2004-108, NGC330-017 \& -026). The two B8Ia objects in NGC2004
(\#005 \& \#007) have large oxygen abundances compared to the cluster mean, by
almost a factor of 3. Although the \teff~and \logg~derived for these objects are
reasonable, the oxygen abundances of these objects are highly dependent on the
microturbulence, varying from 8.56 \& 8.46 dex to 8.84 dex in both objects.
These objects are also very luminous lying close to the edge of the grid and
have a very weak oxygen spectrum so should be treated with caution. 

{\bf Magnesium} like silicon is an $\alpha$-processed element and so should
follow the same trend as silicon. There is only one strong line which
can be seen over our wavelength range, that is the doublet at Mg \2 4481 \AA. The
derived Mg abundances are normally in very good agreement throughout the
clusters. 

{\bf Silicon} is one of the main diagnostic elements. As a result of its
sensitivity to both temperature and microturbulence, we tend to see a large range
in silicon abundances (Table~\ref{ablin}). However as discussed above we have
fixed the Si abundance in each star to reflect the median abundance of the
cluster, where possible.  Those objects which have $<$0 in the microturbulence
column for the corrected parameters in Tables~\ref{galpar} \&~\ref{mcpar} are
those for which we could not obtain the desired Si abundance by changing the
microturbulence. For these objects a zero microturbulence was adopted as this
provided the closest possible Si abundance to that desired.  

During our analysis we have encountered a problem with the Si \2 spectrum, where
the Si abundances from the two Si \2 lines at 4128 \& 4131~\AA\ differ on
average by 0.13 dex with the former resulting in a higher abundance. We expect
that this is related to the oscillator strengths (log gf) included in TLUSTY
calculations which have a ratio of 0.66. If the two lines were to follow LS
coupling we would expect the ratio to be approximately 0.70.

\begin{landscape}
\vspace{-5mm}
\begin{table}
\setcounter{table}{5}
%\begin{center}
\caption[Abundances of NGC3293, NGC4755, NGC2004 \& NGC330 stars. Fixing all atmospheric parameters]
{\label{fixall}   Absolute Abundances of NGC3293, NGC4755, NGC2004 \& NGC330 stars. Presented are the mean of the absolute abundances for each species 
studied, obtained using the final corrected atmospheric parameters from Tables~\ref{galpar} \& ~\ref{mcpar}. Those objects with $<$0 in the microturbulence
column are those for which the microturbulence could not be lowered any further to obtain a Si abundance close to or at the mean silicon abundance of the cluster. Carbon
abundances presented here are based solely on the C {\sc ii} 4267 \AA~line and should only be used as a guide to the relative carbon abundance between the stars (see Sect.~\ref{individ_ab}).
Uncertainties on the abundances account for both random and systematic errors as discussed in
Sect.~\ref{aberror}. Abundances are presented as [X]=12 + $\log$([X/H]) in units of dex, \teff~in K,
\logg~in cms$^{-2}$ and \eps~in \kms. 
}  
\begin{tabular}{lllllcccccccc} \hline \hline
Star        & Sp.Typ     & \teff & \logg &  \eps$_{Ave}$ &  C {\sc ii} &N {\sc ii} & O {\sc ii} & Mg {\sc ii} & Si {\sc ii}  &Si {\sc iii}   &Si {\sc iv} &Fe {\sc iii}   \\   
\hline   \\
NGC3293-003 & B1   III   & 20500 & 2.75 & 13 & 7.84  & 7.52 $\pm$ 0.07 & 8.75 $\pm$ 0.28 & 7.29 $\pm$ 0.27 &		     & 7.48 $\pm$ 0.33 & 7.49 $\pm$ 0.64 & 7.27 $\pm$ 0.31 \\ 
NGC3293-004 & B1   III   & 22700 & 3.13 & 13 & 7.89  & 7.55 $\pm$ 0.09 & 8.74 $\pm$ 0.23 & 7.44 $\pm$ 0.29 &		     & 7.45 $\pm$ 0.22 & 7.45 $\pm$ 0.60 &		   \\
NGC3293-007 & B1   III   & 22600 & 3.10 & 11 & 7.86  & 7.50 $\pm$ 0.08 & 8.71 $\pm$ 0.22 & 7.26 $\pm$ 0.23 & 7.32 $\pm$ 0.39 & 7.44 $\pm$ 0.24 & 7.44 $\pm$ 0.59 & 7.24 $\pm$ 0.24 \\
NGC3293-010 & B1   III   & 21450 & 3.20 & 11 & 7.66  & 7.45 $\pm$ 0.11 & 8.82 $\pm$ 0.31 & 7.17 $\pm$ 0.24 & 6.89 $\pm$ 0.26 & 7.49 $\pm$ 0.33 & 7.53 $\pm$ 0.58 & 7.30 $\pm$ 0.24 \\
NGC3293-012 & B1   III   & 21500 & 3.30 & 11 & 7.67  & 7.45 $\pm$ 0.16 & 8.83 $\pm$ 0.33 & 7.29 $\pm$ 0.21 &		     & 7.50 $\pm$ 0.33 & 7.51 $\pm$ 0.56 &		   \\
NGC3293-018 & B1   V     & 23450 & 3.75 &  5 & 7.66  & 7.56 $\pm$ 0.11 & 8.71 $\pm$ 0.28 & 7.15 $\pm$ 0.18 & 6.88 $\pm$ 0.23 & 7.49 $\pm$ 0.36 & 7.49 $\pm$ 0.57 & 7.70 $\pm$ 0.26 \\
NGC3293-026 & B2   III   & 22100 & 3.65 &  2 & 7.79  & 7.66 $\pm$ 0.17 & 8.75 $\pm$ 0.31 & 7.22 $\pm$ 0.29 & 6.95 $\pm$ 0.25 & 7.47 $\pm$ 0.35 & 7.46 $\pm$ 0.64 & 7.66 $\pm$ 0.18 \\
NGC3293-043 & B3   V     & 19500 & 4.05 &$<$ 0& 7.91 & 7.55 $\pm$ 0.26 & 8.50 $\pm$ 0.34 & 7.14 $\pm$ 0.26 & 7.31 $\pm$ 0.37 & 7.31 $\pm$ 0.34 &		   & 7.51 $\pm$ 0.25 \\
\\								  		  			 								    
\hline  							  		  			 										     
\\								  		  			 										     
NGC4755-002 & B3   Ia    & 15950 & 2.20 & 18 & 7.87  & 8.18 $\pm$ 0.31 & 8.59 $\pm$ 0.43 & 7.30 $\pm$ 0.32 & 7.40 $\pm$ 0.30 & 7.40 $\pm$ 0.46 &		 & 7.54 $\pm$ 0.22 \\ 
NGC4755-003 & B2   III   & 17700 & 2.50 & 15 & 7.78  & 8.12 $\pm$ 0.27 & 8.52 $\pm$ 0.36 & 7.33 $\pm$ 0.26 & 7.40 $\pm$ 0.26 & 7.40 $\pm$ 0.43 &		 & 7.47 $\pm$ 0.18 \\
NGC4755-004 & B1.5 Ib    & 19550 & 2.60 & 17 & 7.76  & 8.07 $\pm$ 0.14 & 8.64 $\pm$ 0.28 & 7.29 $\pm$ 0.29 & 7.39 $\pm$ 0.35 & 7.38 $\pm$ 0.38 & 7.32 $\pm$ 0.67 & 7.38 $\pm$ 0.25 \\
NGC4755-006 & B1   III   & 19900 & 2.95 & 17 & 7.60  & 7.63 $\pm$ 0.19 & 8.90 $\pm$ 0.32 & 7.21 $\pm$ 0.21 &		     & 7.43 $\pm$ 0.37 & 7.47 $\pm$ 0.52 &		   \\
NGC4755-015 & B1   V     & 22400 & 3.70 &  5 & 7.64  & 7.81 $\pm$ 0.21 & 8.80 $\pm$ 0.33 & 7.09 $\pm$ 0.23 &		     & 7.47 $\pm$ 0.38 & 7.46 $\pm$ 0.64 & 7.73 $\pm$ 0.38 \\
NGC4755-017 & B1.5 V     & 20400 & 3.90 &  3 & 7.93  & 7.74 $\pm$ 0.37 & 8.71 $\pm$ 0.45 & 7.32 $\pm$ 0.34 & 7.44 $\pm$ 0.36 & 7.44 $\pm$ 0.44 &		 &		   \\
NGC4755-020 & B2   V     & 21700 & 3.95 &  1 & 8.01  & 7.68 $\pm$ 0.22 & 8.51 $\pm$ 0.31 & 7.44 $\pm$ 0.29 & 7.38 $\pm$ 0.29 & 7.39 $\pm$ 0.34 &		 & 7.84 $\pm$ 0.25 \\
NGC4755-033 & B3   V     & 17300 & 3.85 &  6 & 8.17  & 7.44 $\pm$ 0.36 & 8.72 $\pm$ 0.43 & 7.15 $\pm$ 0.35 & 7.38 $\pm$ 0.49 & 7.39 $\pm$ 0.39 &		 &		   \\
NGC4755-040 & B2.5 V     & 18900 & 4.10 &  2 & 8.31  & 7.69 $\pm$ 0.39 & 8.98 $\pm$ 0.40 & 7.19 $\pm$ 0.40 & 7.43 $\pm$ 0.44 & 7.44 $\pm$ 0.34 &		 &		   \\
NGC4755-048 & B3 V       & 17800 & 3.95 &  4 & 8.30  & 7.45 $\pm$ 0.34 & 9.00 $\pm$ 0.36 & 7.08 $\pm$ 0.30 & 7.36 $\pm$ 0.40 & 7.39 $\pm$ 0.34 &		 &		   \\
\\						        													  
\hline	        				        															   
\\						        															   
NGC2004-003 & B5   Ia	 & 14450 & 2.10 & 15 & 7.66  & 7.85 $\pm$ 0.40 & 8.45 $\pm$ 0.55 & 7.02 $\pm$ 0.26 & 7.22 $\pm$ 0.25 & 7.22 $\pm$ 0.50 &		 & 7.29 $\pm$ 0.23 \\
NGC2004-005 & B8   Ia	 & 12390 & 1.90 & 12 & 7.77  & 7.86 $\pm$ 0.37 & 8.84 $\pm$ 0.57 & 6.98 $\pm$ 0.46 & 7.22 $\pm$ 0.36 & 7.22 $\pm$ 0.44 &		 & 7.20 $\pm$ 0.23 \\
NGC2004-007 & B8   Ia	 & 12250 & 2.00 & 10 & 7.73  & 7.78 $\pm$ 0.28 & 8.84 $\pm$ 0.53 & 6.95 $\pm$ 0.43 & 7.21 $\pm$ 0.39 & 7.21 $\pm$ 0.44 &		 & 7.11 $\pm$ 0.26 \\
NGC2004-010 & B2.5 Iab   & 17160 & 2.40 & 14 & 7.53  & 8.16 $\pm$ 0.29 & 8.29 $\pm$ 0.37 & 7.11 $\pm$ 0.26 & 7.21 $\pm$ 0.25 & 7.20 $\pm$ 0.43 &		 & 7.26 $\pm$ 0.12 \\
NGC2004-011 & B1.5 Ia	 & 21250 & 2.75 & 13 & 7.45  & 7.73 $\pm$ 0.07 & 8.41 $\pm$ 0.23 & 7.15 $\pm$ 0.26 &		     & 7.19 $\pm$ 0.34 & 7.22 $\pm$ 0.66 & 7.14 $\pm$ 0.23 \\
NGC2004-012 & B1.5 Iab   & 21270 & 2.87 & 12 & 7.44  & 7.75 $\pm$ 0.09 & 8.43 $\pm$ 0.24 & 7.08 $\pm$ 0.23 &		     & 7.19 $\pm$ 0.35 & 7.21 $\pm$ 0.64 & 7.12 $\pm$ 0.19 \\
NGC2004-014 & B3   Ib	 & 17800 & 2.85 & 10 & 7.58  & 7.52 $\pm$ 0.26 & 8.24 $\pm$ 0.37 & 7.06 $\pm$ 0.22 & 7.22 $\pm$ 0.22 & 7.20 $\pm$ 0.40 &		 & 7.17 $\pm$ 0.18 \\
NGC2004-021 & B1.5 Ib	 & 21450 & 3.00 & 14 & 7.55  & 7.20 $\pm$ 0.10 & 8.47 $\pm$ 0.26 & 7.07 $\pm$ 0.22 &		     & 7.18 $\pm$ 0.33 & 7.23 $\pm$ 0.57 & 7.19 $\pm$ 0.17 \\
NGC2004-022 & B1.5 Ib	 & 21780 & 3.15 & 11 & 7.32  & 7.68 $\pm$ 0.11 & 8.55 $\pm$ 0.27 & 6.94 $\pm$ 0.19 &		     & 7.23 $\pm$ 0.34 & 7.22 $\pm$ 0.58 & 7.00 $\pm$ 0.22 \\
NGC2004-026 & B2   II	 & 22900 & 3.65 &  0 & 7.53  & 7.01 $\pm$ 0.12 & 8.18 $\pm$ 0.27 & 7.08 $\pm$ 0.17 & 7.24 $\pm$ 0.22 & 7.25 $\pm$ 0.36 &		 & 7.35 $\pm$ 0.17 \\
NGC2004-029 & B1.5 e	 & 23100 & 3.50 &  1 & 7.43  & 6.81 $\pm$ 0.05 & 8.29 $\pm$ 0.26 & 7.03 $\pm$ 0.19 &		     & 7.21 $\pm$ 0.35 & 7.20 $\pm$ 0.56 & 7.35 $\pm$ 0.17 \\
NGC2004-036 & B1.5 III   & 22870 & 3.35 &  7 & 7.51  & 7.38 $\pm$ 0.07 & 8.48 $\pm$ 0.26 & 6.93 $\pm$ 0.18 &		     & 7.21 $\pm$ 0.32 & 7.18 $\pm$ 0.56 & 7.29 $\pm$ 0.21 \\
NGC2004-042 & B2.5 III   & 20980 & 3.45 &  2 & 7.64  & 6.90 $\pm$ 0.19 & 8.16 $\pm$ 0.32 & 7.14 $\pm$ 0.22 & 7.18 $\pm$ 0.24 & 7.19 $\pm$ 0.36 &		 & 7.20 $\pm$ 0.21 \\
NGC2004-046 & B1.5 III   & 26090 & 3.85 &  2 & 7.52  & 7.60 $\pm$ 0.11 & 8.44 $\pm$ 0.13 & 7.02 $\pm$ 0.19 &		     & 7.22 $\pm$ 0.26 & 7.22 $\pm$ 0.50 &		   \\
\\
\hline
\end{tabular}
%\end{center}
\end{table}  

\end{landscape}
\begin{landscape}
\vspace{-5mm}
\begin{table}
\setcounter{table}{5}
%\begin{center}
\caption[Abundances of NGC3293, NGC4755, NGC2004 \& NGC330 stars. Fixing all atmospheric parameters]
{\label{fixalla}   Contd.}  
\begin{tabular}{lllllcccccccc} \hline \hline
Star        & Sp.Typ   &\teff & \logg &  \eps$_{\rm Ave}$&  C {\sc ii} &N {\sc ii} & O {\sc ii} & Mg {\sc ii} & Si {\sc ii}  &Si {\sc iii}	&Si {\sc iv} & Fe {\sc iii}  \\   
\hline   \\
NGC2004-053 & B0.2 Ve	  & 31500 & 4.15 &  6    &  7.83  & 7.64 $\pm$ 0.15 & 8.39 $\pm$ 0.17 & 7.24 $\pm$ 0.22 &	       & 7.18 $\pm$ 0.26 & 7.18 $\pm$ 0.53 &		     \\
NGC2004-061 & B2   III    & 20990 & 3.35 &  1    &  7.75  & 6.99 $\pm$ 0.24 & 8.30 $\pm$ 0.45 & 7.13 $\pm$ 0.28 & 7.16 $\pm$ 0.29 & 7.17 $\pm$ 0.47 &		   &		     \\
%NGC2004-062 & B0.2 V	  & 31670 & 4.38 &  1    &  8.07  & 6.70 $\pm$ 0.09 & 8.03 $\pm$ 0.13 & 6.83 $\pm$ 0.05 &	       & 7.21 $\pm$ 0.21 & 7.21 $\pm$ 0.36 &		     \\
NGC2004-064 & B0.7-B1 III & 25900 & 3.70 &  6    &  7.49  & 7.53 $\pm$ 0.09 & 8.37 $\pm$ 0.12 & 7.08 $\pm$ 0.25 &	       & 7.18 $\pm$ 0.25 & 7.18 $\pm$ 0.53 &		     \\
NGC2004-070 & B0.7-B1 III & 27400 & 3.90 &  4    &  7.58  & 7.43 $\pm$ 0.11 & 8.50 $\pm$ 0.20 & 7.15 $\pm$ 0.22 &	       & 7.19 $\pm$ 0.36 & 7.20 $\pm$ 0.56 &		     \\
NGC2004-084 & B1.5 III    & 27395 & 4.00 &  3    &  7.64  & 7.28 $\pm$ 0.15 & 8.42 $\pm$ 0.13 & 7.17 $\pm$ 0.20 &	       & 7.25 $\pm$ 0.24 & 7.25 $\pm$ 0.47 &		     \\
NGC2004-090 & O9.5 III    & 32500 & 4.10 & $<$ 0 &  7.63  & 7.65 $\pm$ 0.21 & 8.44 $\pm$ 0.24 & 6.97 $\pm$ 0.20 &	       & 7.15 $\pm$ 0.27 & 7.16 $\pm$ 0.41 &		     \\
NGC2004-091 & B1.5 III    & 26520 & 4.05 &  0    &  7.80  & 7.29 $\pm$ 0.09 & 8.38 $\pm$ 0.14 & 7.29 $\pm$ 0.25 &	       & 7.26 $\pm$ 0.27 & 7.26 $\pm$ 0.52 &		     \\
NGC2004-108 & B2.5 III    & 22600 & 4.00 & $<$ 0 &  7.29  & 6.89 $\pm$ 0.14 & 8.23 $\pm$ 0.30 & 6.99 $\pm$ 0.20 & 6.98 $\pm$ 0.23 & 6.98 $\pm$ 0.31 &		   &		     \\
NGC2004-119 & B2   III    & 23210 & 3.75 & $<$ 0 &  7.48  & 6.95 $\pm$ 0.09 & 8.35 $\pm$ 0.27 & 7.17 $\pm$ 0.25 & 7.14 $\pm$ 0.24 & 7.14 $\pm$ 0.34 &		   &		     \\
\\ 			      
\hline			      
\\			      
 NGC330-002 & B3   Ib     & 14590 & 2.15 &16  &  7.09  & 7.59 $\pm$ 0.33 & 7.98 $\pm$ 0.49 & 6.60 $\pm$ 0.23 & 6.81 $\pm$ 0.23 & 6.82 $\pm$ 0.46 &		   & 6.70 $\pm$ 0.24 \\
 NGC330-003 & B2   Ib     & 17210 & 2.25 &20  &  7.25  & 7.69 $\pm$ 0.19 & 8.07 $\pm$ 0.34 & 6.79 $\pm$ 0.27 & 6.82 $\pm$ 0.25 & 6.81 $\pm$ 0.39 &		   & 6.87 $\pm$ 0.22 \\
 NGC330-004 & B2.5 Ib     & 17000 & 2.30 &16  &  6.84  & 7.83 $\pm$ 0.14 & 7.86 $\pm$ 0.32 & 6.73 $\pm$ 0.21 & 6.76 $\pm$ 0.21 & 6.82 $\pm$ 0.39 &		   & 6.83 $\pm$ 0.14 \\
 NGC330-005 & B5   Ib     & 13700 & 2.25 & 8  &  6.77  & 7.49 $\pm$ 0.37 & 7.82 $\pm$ 0.45 & 6.58 $\pm$ 0.19 & 6.76 $\pm$ 0.23 & 6.81 $\pm$ 0.43 &		   &		     \\
 NGC330-009 & B5   Ib     & 13940 & 2.45 & 6  &  7.05  & 7.20 $\pm$ 0.41 & 8.30 $\pm$ 0.53 & 6.66 $\pm$ 0.31 & 6.80 $\pm$ 0.33 & 6.81 $\pm$ 0.49 &		   &		     \\
 NGC330-010 & B5   Ib     & 14820 & 2.60 & 4  &  7.17  & 7.14 $\pm$ 0.35 & 7.71 $\pm$ 0.43 & 6.63 $\pm$ 0.21 & 6.81 $\pm$ 0.30 & 6.84 $\pm$ 0.43 &		   &		     \\
 NGC330-014 & B1.5 Ib     & 20130 & 2.75 &18  &  6.94  & 7.53 $\pm$ 0.13 & 8.11 $\pm$ 0.24 & 6.72 $\pm$ 0.21 &  	       & 6.83 $\pm$ 0.29 & 6.83 $\pm$ 0.52 &		     \\
 NGC330-016 & B5:  II     & 14220 & 2.60 & 5  &  7.36  & 7.18 $\pm$ 0.37 & 8.07 $\pm$ 0.51 & 6.69 $\pm$ 0.21 & 6.80 $\pm$ 0.31 & 6.81 $\pm$ 0.47 &		   &		     \\
 NGC330-017 & B2   II     & 22000 & 3.35 &$<$0&  7.07  & 7.12 $\pm$ 0.12 & 7.64 $\pm$ 0.25 & 6.47 $\pm$ 0.17 &  	       & 6.74 $\pm$ 0.33 &		   & 6.90 $\pm$ 0.21 \\
 NGC330-018 & B3   II     & 18000 & 2.95 & 5  &  7.17  & 7.24 $\pm$ 0.26 & 8.05 $\pm$ 0.37 & 6.61 $\pm$ 0.18 & 6.74 $\pm$ 0.18 & 6.80 $\pm$ 0.36 &		   & 7.14 $\pm$ 0.21 \\
 NGC330-020 & B3   II     & 16720 & 2.85 & 5  &  7.05  & 7.01 $\pm$ 0.44 & 7.80 $\pm$ 0.59 & 6.64 $\pm$ 0.28 & 6.82 $\pm$ 0.31 & 6.83 $\pm$ 0.57 &		   &		     \\
 NGC330-022 & B3   II     & 18860 & 3.00 & 1  &  7.14  & 7.29 $\pm$ 0.25 & 7.76 $\pm$ 0.39 & 6.85 $\pm$ 0.17 & 6.82 $\pm$ 0.20 & 6.83 $\pm$ 0.39 &		   &		     \\
 NGC330-026 & B2.5 II     & 22500 & 3.40 &$<$0&  7.30  & 7.23 $\pm$ 0.19 & 7.74 $\pm$ 0.30 & 7.02 $\pm$ 0.23 &  	       & 6.64 $\pm$ 0.39 &		   &		     \\
 NGC330-027 & B1   V      & 22040 & 3.20 & 7  &  7.13  & 7.40 $\pm$ 0.21 & 8.27 $\pm$ 0.43 & 6.44 $\pm$ 0.21 &  	       & 6.81 $\pm$ 0.36 & 6.82 $\pm$ 0.78 &		     \\
 NGC330-032 & B0.5 V      & 29700 & 4.15 &$<$0&  7.16  & 7.37 $\pm$ 0.10 & 7.90 $\pm$ 0.07 & 6.80 $\pm$ 0.20 &  	       & 6.77 $\pm$ 0.19 & 6.77 $\pm$ 0.45 &		     \\
 NGC330-042 & B2   II     & 25450 & 3.75 & 1  &  7.04  & 7.20 $\pm$ 0.05 & 7.73 $\pm$ 0.15 & 6.88 $\pm$ 0.18 &  	       & 6.82 $\pm$ 0.34 & 6.83 $\pm$ 0.52 &		     \\
 NGC330-047 & B1   V      & 26700 & 4.05 & 0  &  7.20  & 6.76 $\pm$ 0.11 & 8.07 $\pm$ 0.13 & 6.57 $\pm$ 0.15 &  	       & 6.83 $\pm$ 0.23 & 6.84 $\pm$ 0.46 &		     \\
 NGC330-074 & B0   V      & 32020 & 4.20 & 4  &  7.29  & 7.45 $\pm$ 0.19 & 7.87 $\pm$ 0.17 & 6.81 $\pm$ 0.18 &  	       & 6.83 $\pm$ 0.25 & 6.83 $\pm$ 0.47 &		     \\
 NGC330-114 & B2   III    & 23800 & 3.90 & 4  &  7.23  & 7.32 $\pm$ 0.18 & 8.00 $\pm$ 0.27 & 6.97 $\pm$ 0.22 &  	       & 6.83 $\pm$ 0.32 &		   &		     \\
 NGC330-124 & B0.2 V      & 30980 & 4.25 & 2  &  7.55  & 7.36 $\pm$ 0.25 & 7.75 $\pm$ 0.17 & 6.75 $\pm$ 0.18 &  	       & 6.84 $\pm$ 0.23 & 6.84 $\pm$ 0.47 &		     \\ 
\\ \hline
\end{tabular}
%\end{center}
\end{table}  

\end{landscape}
\noindent Indeed from theoretical calculations involving 10,000 configurations
the ratio of these two lines is found to be 0.69-0.70 (private comm.: A.
Hibbert, Queen's University Belfast). However, from the observations of these
stars we would require a ratio of 0.89 for these gf-values, to obtain consistent
abundances from this multiplet.  It is beyond the scope of this paper but this
discrepancy clearly merits further investigation of the oscillator strengths in
the future. In estimating the effective temperatures of these stars we have
simply taken the mean of the two lines to represent the Si abundance as
determined from the Si \2 spectrum. 

{\bf Iron} lines from Fe \3 at 4419 and 4430~\AA\ have been detected in many of
our targets, particularly in the more metal rich environs of the LMC and Milky
Way Clusters. Unfortunately,  whilst the Fe model atom is included in TLUSTY,
apart from the ground level the higher energy states are bundled into
super-levels to simplify and speed up the calculations. As a result it is
difficult to isolate the levels associated with these transitions, and hence
their departure coefficients. It was therefore necessary to calculate the
abundances in LTE, as \citet{tho07} have shown that non-LTE effects appear to
be small.

In the Galactic clusters the Fe abundances from the higher gravity
objects (\logg~$>$ 3.3 dex), are higher in comparison to the lower
gravity targets. For example, in NGC3293 the mean of the Fe abundance
is 7.27 dex from the 3 stars with log~g~$<$~3.3 dex, whereas the mean
is 7.62 dex from the 3 stars with log~g~$>$~3.3 dex.  The weighted
mean of this cluster is 7.49 dex and is reasonable for the Galaxy, yet
clearly there is a large spread amongst the objects, which could be
due to the microturbulences adopted. In addition to the objects
analysed for this paper we have estimated the Fe abundances for a
number of the objects from Paper IV (see Table~\ref{youngfe} \&
\ref{meanab}).

%----------------------------------------------------------------------------------

\section{Discussion}
\label{discuss}

\subsection{Cluster Metallicities}

In the following discussion we will compare the abundances derived for the older
clusters analysed here with the younger clusters presented in Paper~IV. In
addition we will discuss all these results in relation to those available in
the  literature. The mean abundances for each cluster are presented in
Table~\ref{meanab}, and are based on the results presented in
Table~\ref{fixall} weighted by their degree of uncertainty. Carbon has been omitted from
Table~\ref{meanab} due to the uncertainties in the absolute abundances as
discussed in Sect.~\ref{individ_ab}. Objects for which the median silicon
abundance of the cluster could not be reproduced, were omitted from the
calculation of the mean cluster abundance (\ie those objects with $\xi_{ave} <
0$). The abundances of the young clusters (NGC6611, N11 \& NGC346) were taken
from Paper~IV except for the Fe abundances which were calculated in LTE using
the methods described in Sect.~\ref{individ_ab}. \begin{table}
\begin{center}
\caption[Comparison of NGC2004 analyses.]
{\label{ngc2004comp} Comparison of NGC2004 analyses in this Paper with that from 
\citet[K00]{korn00}. In K00, NGC2004-022 is labeled with the \citet{r74}
 identification - B30. K00 use LTE model atmospheres with NLTE line formations.
  Abundances are presented as [X]=12 + $\log$([X/H]) in units of dex. }  
\begin{tabular}{lcc} \hline \hline
        &\multicolumn{2}{c}{NGC2004-022}     \\
        & This Paper      &  K00            \\
\hline \\
\teff  (K)	      &    21780	&    23450	  \\
\logg  (cms$^{-2}$)   &     3.15	&     3.45	  \\
\eps   (\kms)	      &       11	&	14	  \\
     	&                 &                 \\
C    	&  7.32           & 7.60:           \\
N    	& 7.68 $\pm$ 0.11 & 7.50 $\pm$ 0.20 \\
O    	& 8.55 $\pm$ 0.27 & 8.35 $\pm$ 0.20 \\
Mg   	& 6.94 $\pm$ 0.19 & 7.00 $\pm$ 0.30 \\
Si   	& 7.23 $\pm$ 0.58 & 6.90 $\pm$ 0.20 \\  				     
Fe   	& 7.00 $\pm$ 0.22 & 7.08 $\pm$ 0.30 \\  				     
\\ \hline			  
\end{tabular}			  
\end{center}			  
\end{table}

\subsubsection{Galactic Clusters: NGC3293, NGC4755, NGC6611}
\label{galab}
\begin{table*}[bht]
\begin{center}
\caption[Comparison of NGC330 analyses.]
{\label{ngc330comp} Comparison of NGC330 analyses in this Paper with those from
\citet[L03]{l03}, and \citet[K00]{korn00}. In L03 and K00, NGC330-002,-004 and
-018 are labeled with \citet{r74} identifications, which are A02, B37 and B30. Fe
has been omitted from this comparison as the other analyses did not consider this
element.  Abundances are presented as [X]=12 + $\log$([X/H]) in units of dex.}  
\begin{tabular}{lcccccccc} \hline \hline
        &\multicolumn{2}{c}{NGC330-002}     & & \multicolumn{2}{c}{NGC330-004}    & &    \multicolumn{2}{c}{NGC330-018}  \\
        & This Paper      &  L03            & &  This Paper     &     L03	  & &   This Paper    &	K00              \\
 \hline \\
\teff  (K)	    &	 14590        &    16000	& &    17000	    &	 18000        & &    18000	  &    16950	     \\
\logg  (cms$^{-2}$) &	  2.15        &     2.30	& &	2.30	    &	  2.40        & &     2.95	  &	2.77	     \\
\eps   (\kms)	    &	    16        &        8	& &	  16	    &	    10        & &	 5	  &	   6	     \\
     	&                 &                 & &                 &                 & &                 & 	         \\
C    	& 7.09 		  & 6.91  	    & & 6.84 		& 6.71  	  & & 7.17            & 7.30:		 \\
N    	& 7.59 $\pm$ 0.33 & 7.71  	    & & 7.83 $\pm$ 0.14 & 7.89  	  & & 7.24 $\pm$ 0.26 & 		 \\
O    	& 7.98 $\pm$ 0.49 & 8.14  	    & & 7.86 $\pm$ 0.32 & 7.98  	  & & 8.05 $\pm$ 0.37 & 8.25 $\pm$ 0.30  \\
Mg   	& 6.60 $\pm$ 0.23 & 6.69  	    & & 6.73 $\pm$ 0.21 & 6.79  	  & & 6.61 $\pm$ 0.18 & 6.85 $\pm$ 0.40  \\
Si   	& 6.82 $\pm$ 0.46 & 6.82  	    & & 6.82 $\pm$ 0.39 & 7.12  	  & & 6.80 $\pm$ 0.36 & 6.82 $\pm$ 0.30  \\      				  
\\ \hline			  
\end{tabular}			  
\end{center}			  
\end{table*}  			  

\begin{table*}[bht]
\begin{center}
\caption[Average Abundances of Flames clusters]
{\label{meanab}   Weighted average abundances of the FLAMES clusters. Results for
the clusters NGC6611, N11 \& NGC346 are taken from Paper III, except for the Fe
abundances which were derived from LTE calculations following the method
described in Sect.~\ref{individ_ab}. Solar abundances are taken from
\citet{asp05}.  Abundances are presented as [X]=12 + $\log$([X/H]) in units of dex.}  
\begin{tabular}{lcccccccccc} \hline \hline
    &                 &\multicolumn{3}{c}{Milky Way}                       & &      \multicolumn{2}{c}{LMC}	& & 	 \multicolumn{2}{c}{SMC}    \\
    & Solar           &NGC6611         & NGC3293         &   NGC4755       & &       N11       &     NGC2004	& & 	NGC346      &	   NGC330     \\
 \hline \\
N   & 7.78 $\pm$ 0.06 & 7.59 $\pm$ 0.10 & 7.52 $\pm$ 0.07 & 7.85 $\pm$ 0.26 & & 7.54 $\pm$ 0.40 & 7.33 $\pm$ 0.36 & & 7.17 $\pm$ 0.29 & 7.27 $\pm$ 0.24 \\
O   & 8.66 $\pm$ 0.05 & 8.55 $\pm$ 0.04 & 8.76 $\pm$ 0.05 & 8.73 $\pm$ 0.18 & & 8.33 $\pm$ 0.08 & 8.40 $\pm$ 0.18 & & 8.06 $\pm$ 0.10 & 7.90 $\pm$ 0.19 \\
Mg  & 7.53 $\pm$ 0.09 & 7.32 $\pm$ 0.06 & 7.24 $\pm$ 0.10 & 7.23 $\pm$ 0.12 & & 7.06 $\pm$ 0.09 & 7.08 $\pm$ 0.10 & & 6.74 $\pm$ 0.07 & 6.72 $\pm$ 0.12 \\
Si  & 7.51 $\pm$ 0.04 & 7.41 $\pm$ 0.05 & 7.47 $\pm$ 0.06 & 7.41 $\pm$ 0.04 & & 7.19 $\pm$ 0.07 & 7.21 $\pm$ 0.03 & & 6.79 $\pm$ 0.05 & 6.81 $\pm$ 0.02 \\
Fe  & 7.45 $\pm$ 0.15 & 7.62 $\pm$ 0.15 & 7.49 $\pm$ 0.21 & 7.56 $\pm$ 0.19 & & 7.23 $\pm$ 0.10 & 7.24 $\pm$ 0.12 & & 6.98 $\pm$ 0.13 & 6.88 $\pm$ 0.16\\
\\ \hline
\end{tabular}
\end{center}
\end{table*}

The $\alpha$-processed elements, Mg \& Si, and the heavier Fe nuclei provide
good estimates of the metallicity of a cluster as they should be unaffected by
any nuclear processes in the cores of these relatively young stars. As discussed
above the Si results are sensitive to the atmospheric parameters, but
nevertheless the mean Si abundances of the three clusters are in very good
agreement as are those for Mg. In comparison to the solar composition from
\citet[]{asp05} these B-type star estimates are approximately 0.3 and 0.1 dex
lower for Mg and Si. Indeed solar abundances  suggest that the Mg to Si
ratio is almost 1:1 \citep[and references therein]{asp05}, but the B-type star
results suggest something closer to 0.69:1. As discussed in
Sect.~\ref{individ_ab}, there is a large spread in Fe abundances within a
cluster and between the Galactic clusters, although the mean Fe abundances from
each cluster agree within the uncertainties, and with the solar estimate. 

In Paper~IV, we found that the oxygen abundances of NGC6611 were in very good
agreement with those determined from H \2 regions. Yet both the stellar and H
\2  regions whilst agreeing within the errors, were approximately 0.1 dex lower
than the solar abundances from \citeauthor{asp05}. The oxygen
abundances from the two older clusters are 8.73 dex, approximately 0.2
dex higher than those in NGC6611. As these two clusters are situated
in the Carina arm of the Milky Way, it would be conceivable that the
NGC6611 cluster has a different composition to these clusters. However
the good agreement of the Si \& Mg abundances would not support
this. There are 3 objects with particularly high oxygen abundance in
NGC4755 (\#006, \#040, \& \#048). The latter two objects have a very
weak oxygen spectrum, whilst the oxygen abundance in \#006 is very
sensitive to the adopted microturbulence. Omitting these three objects
would lower the mean abundance of NGC4755 to 8.66 dex, in better
agreement with the solar abundance. It is therefore more likely that
the difference is due to an artifact of the small number of objects
per cluster, and the scatter from object to object, rather than a real
metallicity difference between these clusters and their environments.

An LTE analysis of B-type stars in these three clusters by \citet[]{rol00},
shows very good agreement in the oxygen abundances between these clusters.
Besides the work by \citeauthor[]{rol00}, a number of the objects in NGC3293 and
NGC4755 have been analysed by \citet[]{mat02}. However there are significant
differences in the techniques used to determine the parameters in this work as
they use Str\"omgren photometry to determine the atmospheric parameters.  Whilst
the microturbulences agree well in all the objects, for the hotter objects the
\teff~ and \logg~values differ significantly. This maybe a result of the
insensitivity of the Str\"omgren colour index at high temperatures. Due to the
large differences between these analyses we have not attempted any further
comparisons.

\subsubsection{LMC clusters: NGC2004, N11}
\label{lmcab}

From a literature search we found only one object in NGC2004 which had been
analysed previously, (\#022), by \citet[]{korn00} using LTE model atmospheres
and non-LTE line formations. The comparison of their results with ours is shown
in Table~\ref{ngc2004comp}.  Whilst there are significant differences in the
derived atmospheric  parameters, the abundances agree well within the errors,
particularly the Fe \& Mg abundances. A comparison of the silicon equivalent
widths (A. Korn, private communication) show very good agreement for the Si~\3
lines, although the Si \4 4116 \AA\ line is weaker in the UVES data of
\citeauthor[]{korn00}, than in our spectrum. This however would not explain the
effective temperature difference of $\sim$ 1700K between the two analyses and
would in fact require that the estimated temperatures of \citeauthor[]{korn00}
should be lower. We conclude that there is a significant difference in the
fitting to the Balmer lines to determine the surface gravity, being 0.2 dex
higher in the analysis by \citeauthor[]{korn00} and this would lead to the
higher \teff. Such a high gravity does not seem appropriate from the FLAMES
data. There is also a difference in the adopted microturbulence which may
explain some of the discrepancies in the abundances. Using the Si \3 equivalent
widths provided by \citeauthor[]{korn00} however we reproduce the value of 11
\kms~as obtained from the FLAMES data.

Apart from the nitrogen abundances, the mean abundances in the two LMC clusters
are in excellent agreement.
\begin{figure*}
\begin{center}
\epsfig{file=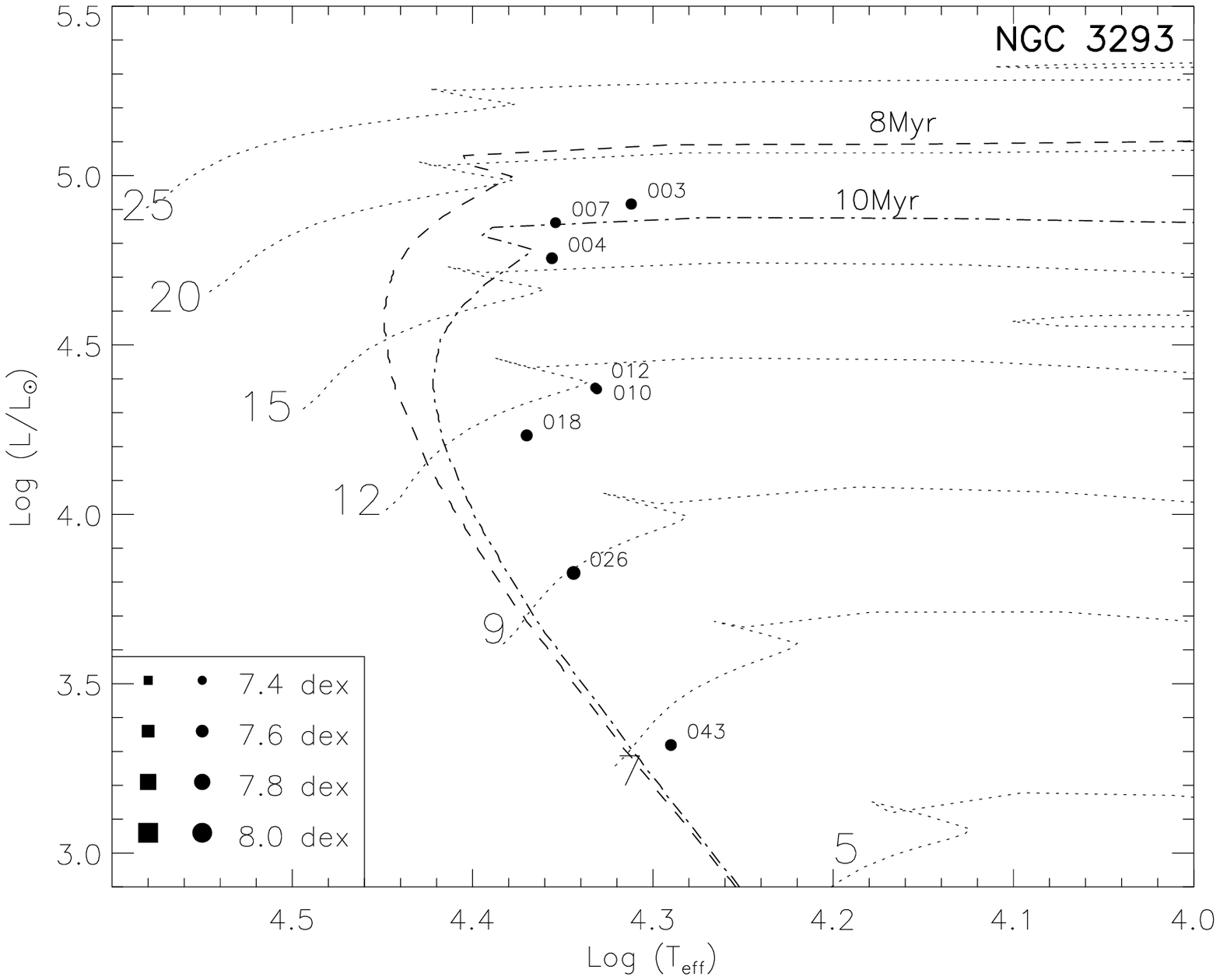, width=85mm, angle=0}
\epsfig{file=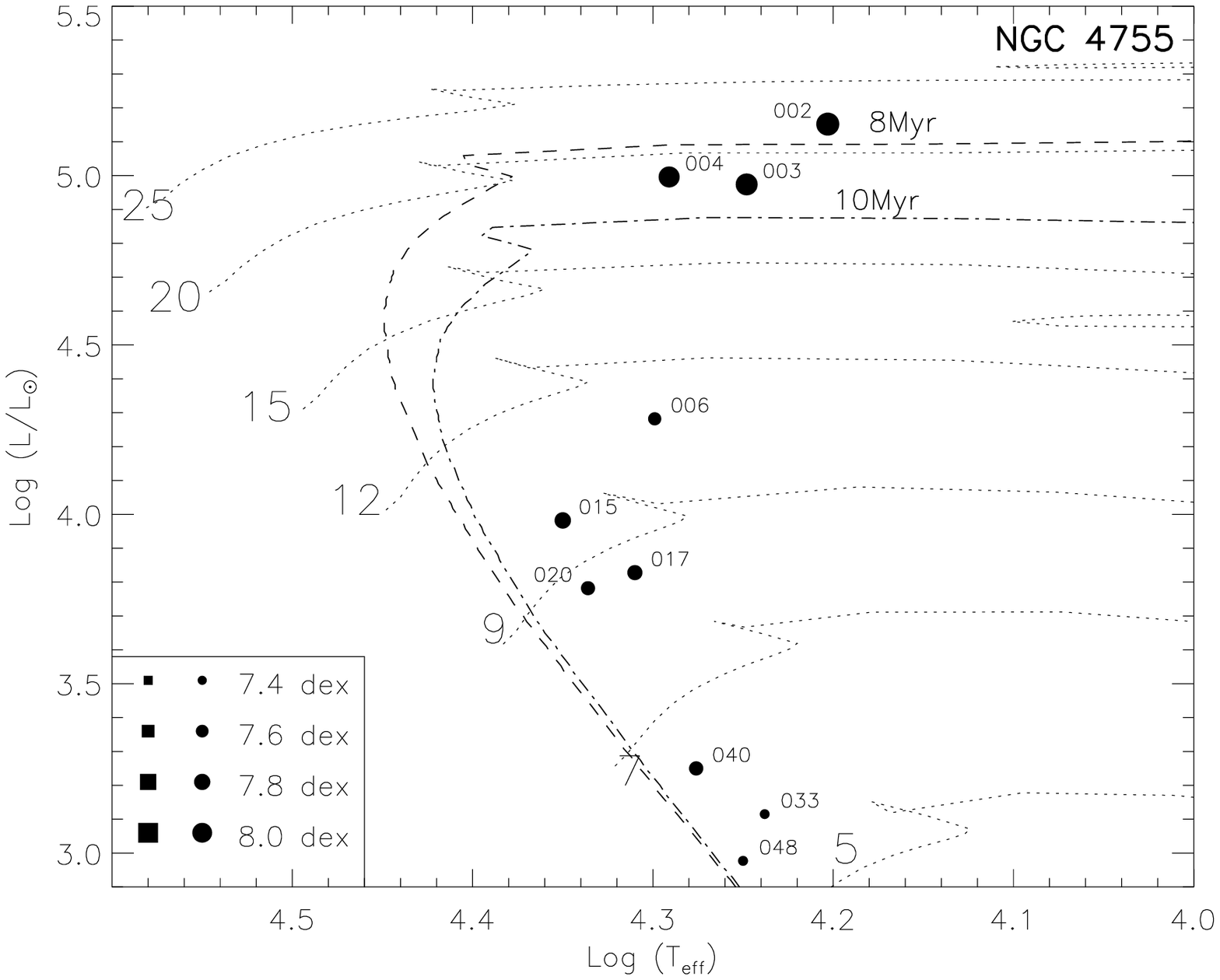, width=85mm, angle=0}
\epsfig{file=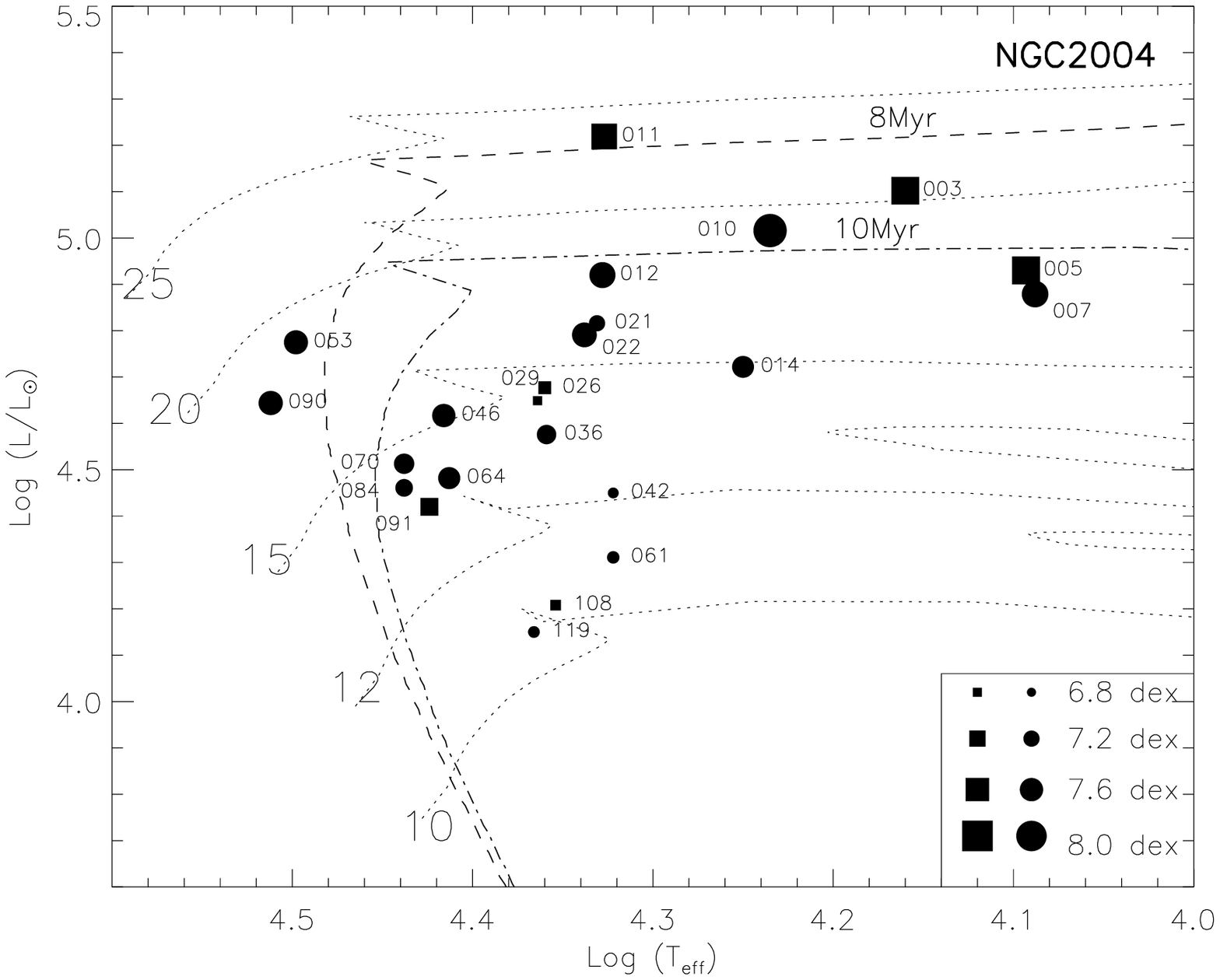, width=85mm, angle=0}
\epsfig{file=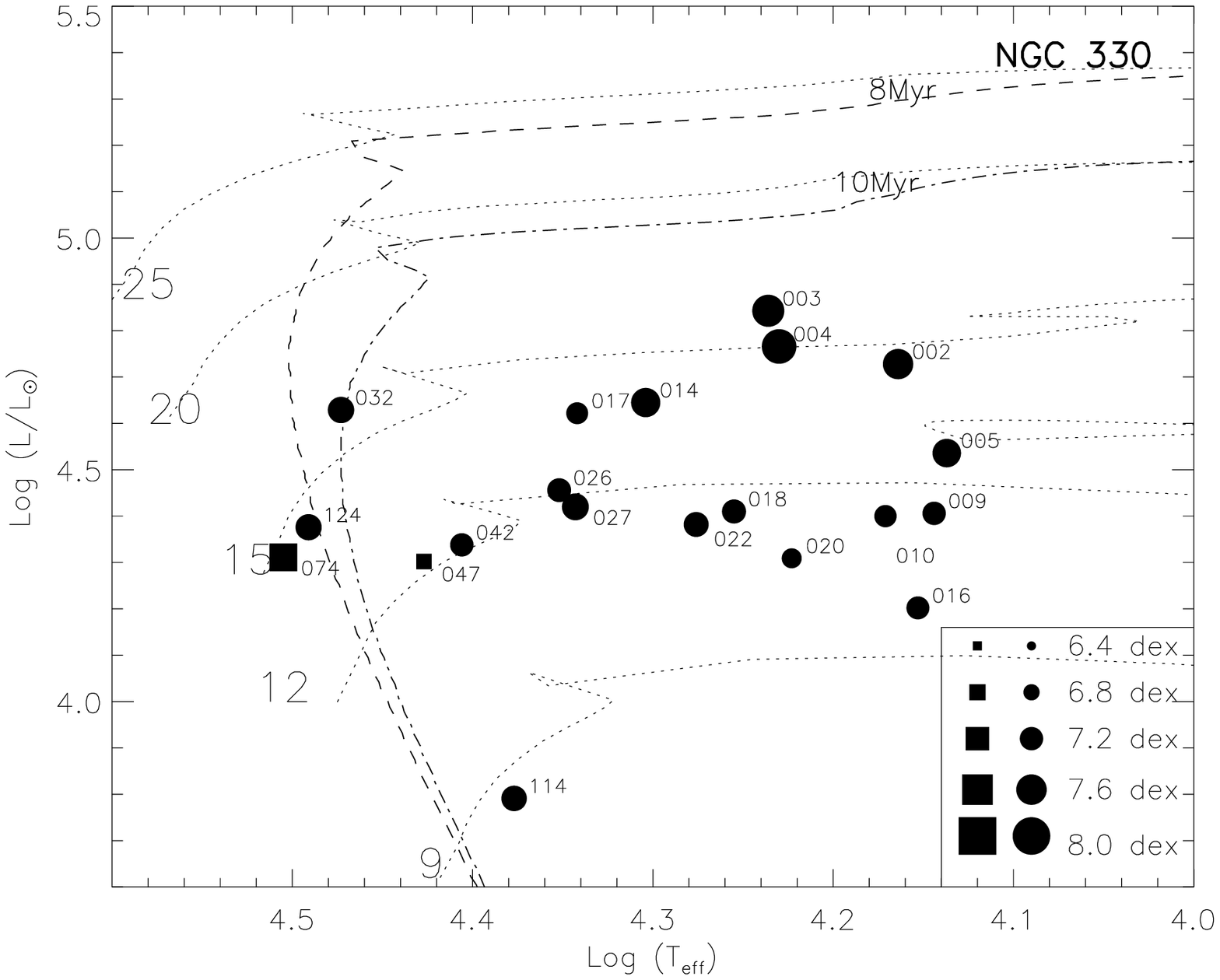 , width=85mm, angle=0}
\caption[]
{ Hertzsprung-Russell Diagrams of NGC3293, NGC4755, NGC2004 \& NGC330
objects.  Binaries are represented by squares, and single stars by
solid circles. The nitrogen abundances in the stars are represented by
the size of the symbols, as per the key in each diagram (\ie the
larger the symbol the more nitrogen at the surface of the star). The
evolutionary tracks are shown as dotted lines and represent the
non-rotating tracks from \citet[]{mey94} and \citet[]{sch92} for the
Milky Way clusters, and \citet[]{sch93} and \citet[]{char93} for
NGC2004 and NGC330, respectively. Isochrones are presented as dashed
lines, but in the case of the Magellanic cloud HRDs they should be
considered to be illustrative only (see Sect.~\ref{Nabund}).  }
\label{hrd}
\end{center}
\end{figure*}

\subsubsection{SMC Clusters: NGC330, NGC346}
\label{smcab}

The mean elemental abundances of NGC330 presented here are in very good
agreement, within the errors, with those derived for NGC346 in Paper~IV. The O
\& Fe abundances in NGC346 are 0.10 \& 0.16 dex higher than those in NGC330, but
this is within the uncertainties and the excellent agreement between the
magnesium and silicon abundances suggest it is insignificant. 

Three of our targets have been analysed previously, NGC330-002, NGC330-004 and
NGC330-018. The former two were analysed by \citet[]{l03} using LTE model
atmospheres and making some non-LTE corrections to the abundances. Their
estimated effective temperatures are hotter in both cases, whilst the
microturbulences are lower (see Table~\ref{ngc330comp}), these differences are
mainly due to the different analysis techniques. In particular, they use oxygen
lines to determine the microturbulence and flux distributions to determine the
effective temperature. Nevertheless, the abundances are in good agreement.
NGC330-018 was analysed by \citeauthor[]{korn00} and again, whilst quite
different techniques and model atmospheres were applied, there is good agreement
in the abundances.

%--------------------------------------------------------------------------------------

\subsection{Evolution of Nitrogen Abundances.}
\label{Nabund}

\begin{figure*}
\begin{center}
\epsfig{file=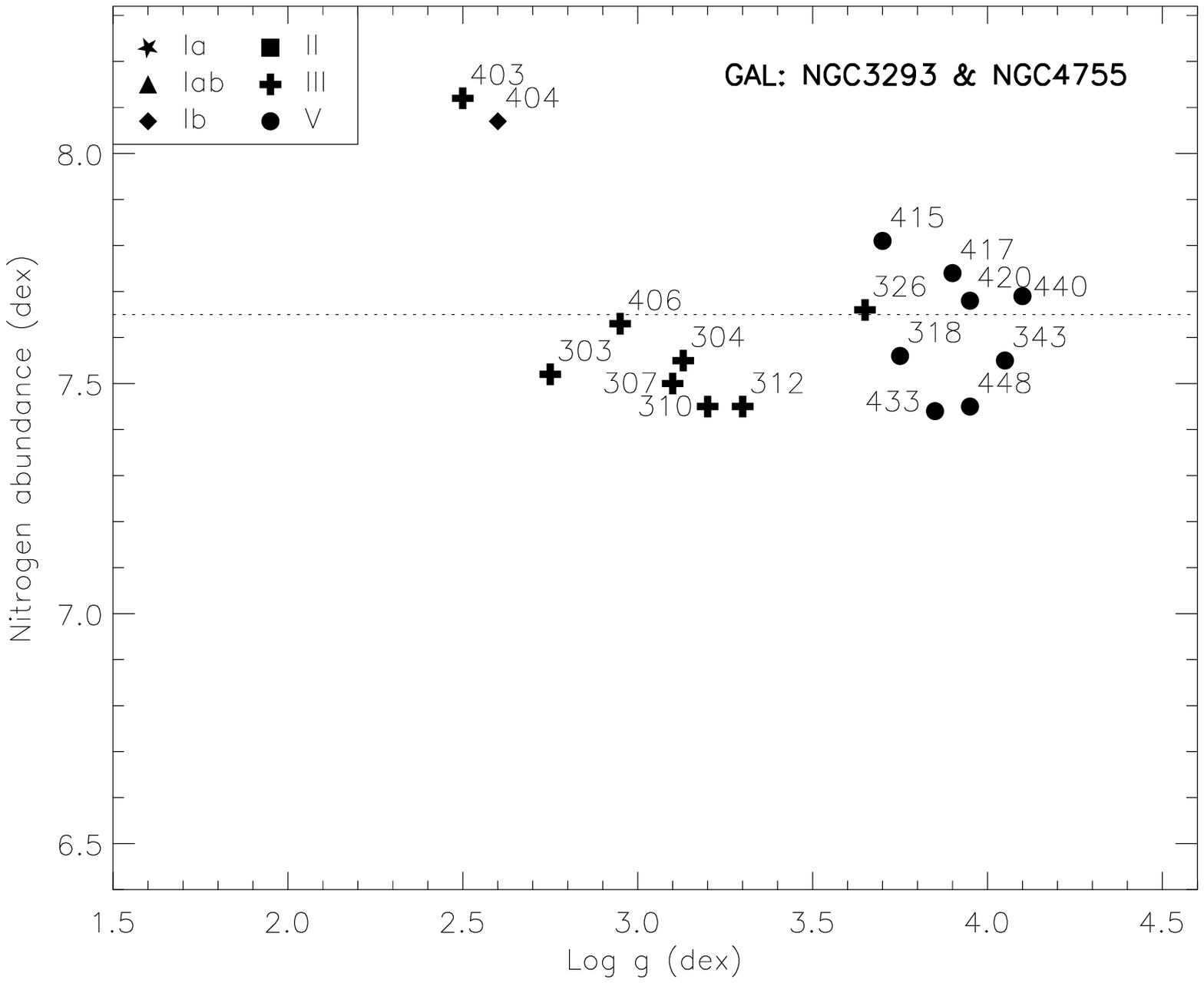, width=84mm, angle=0}
\epsfig{file=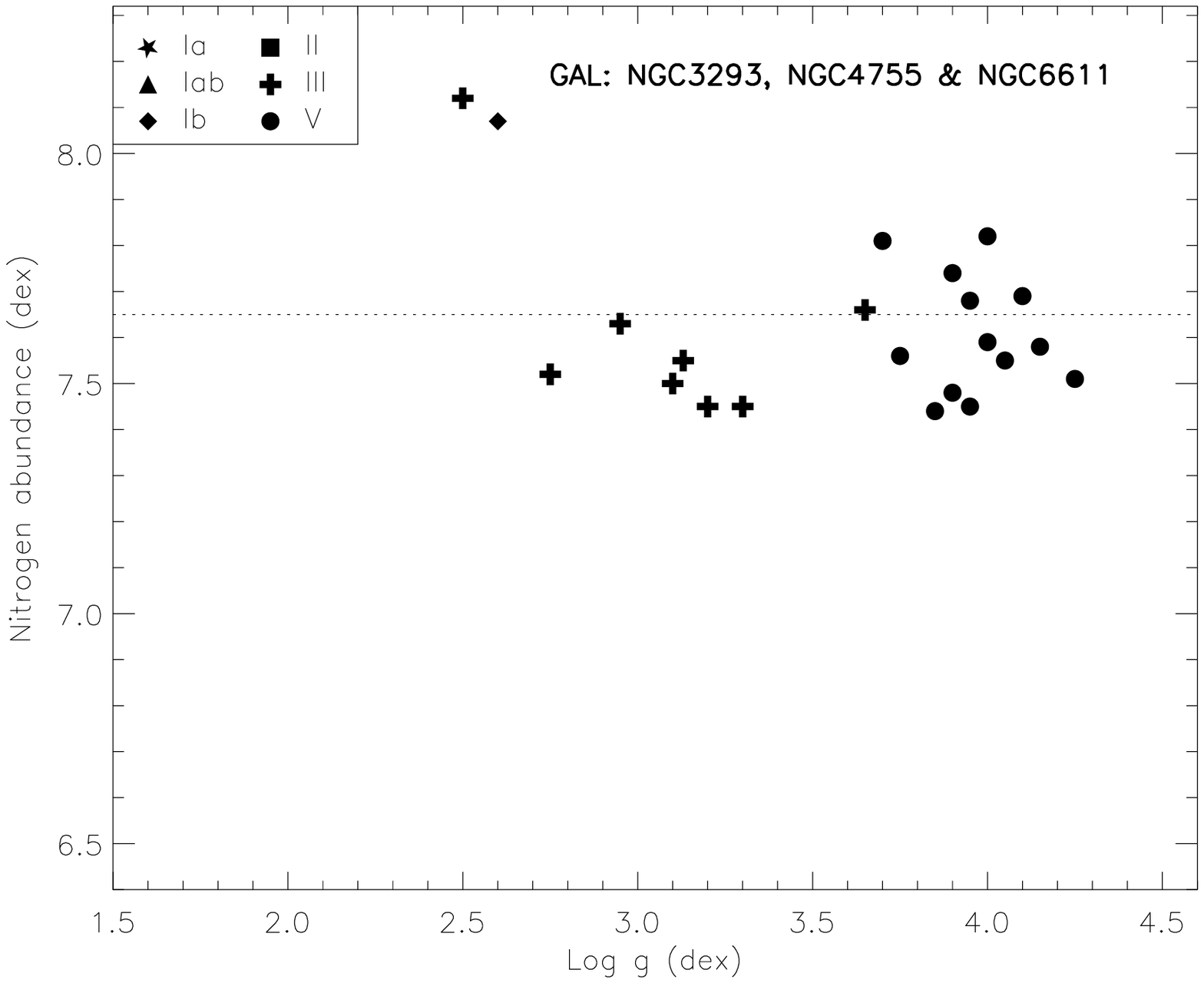, width=84mm, angle=0}
\epsfig{file=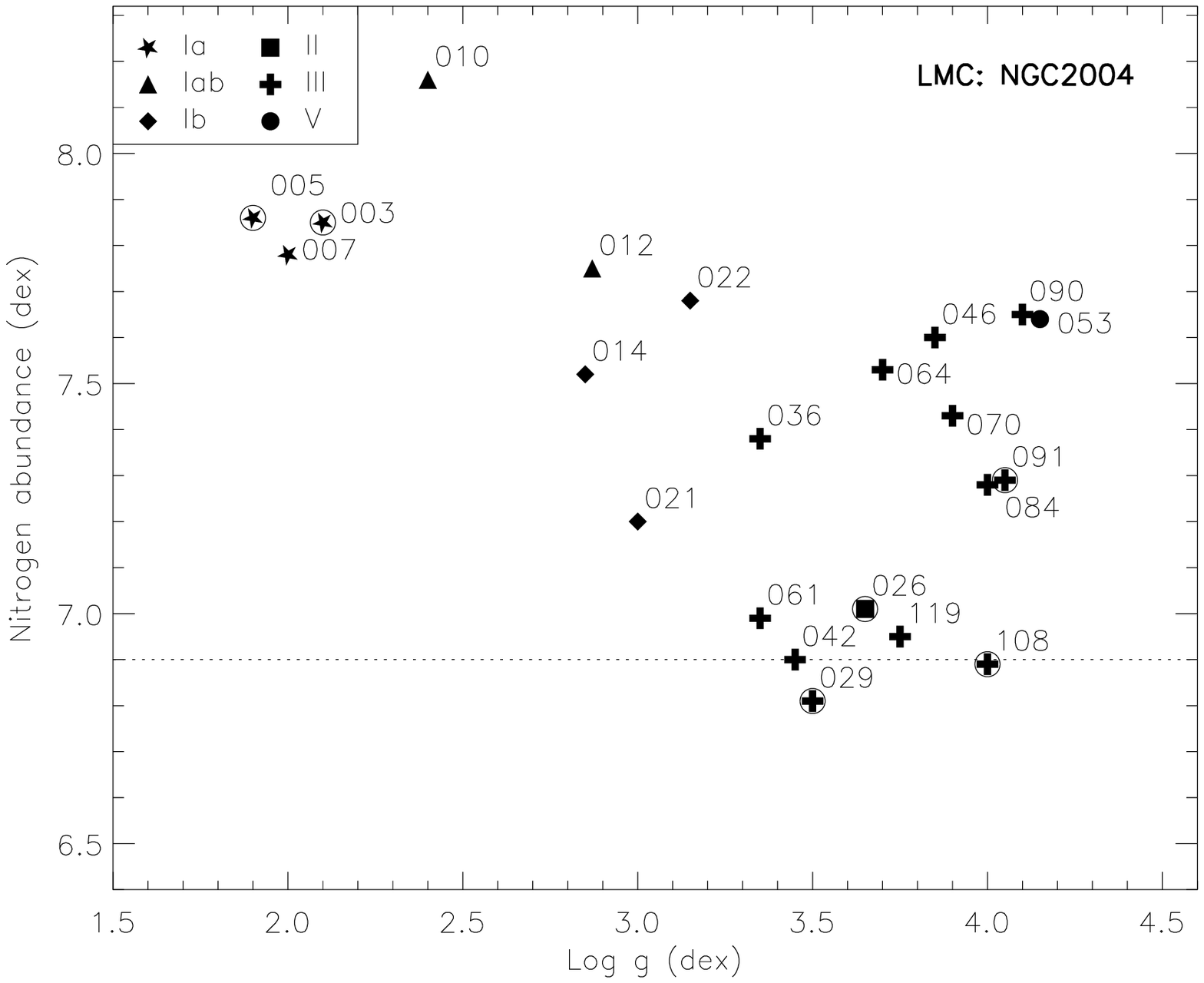, width=84mm, angle=0}
\epsfig{file=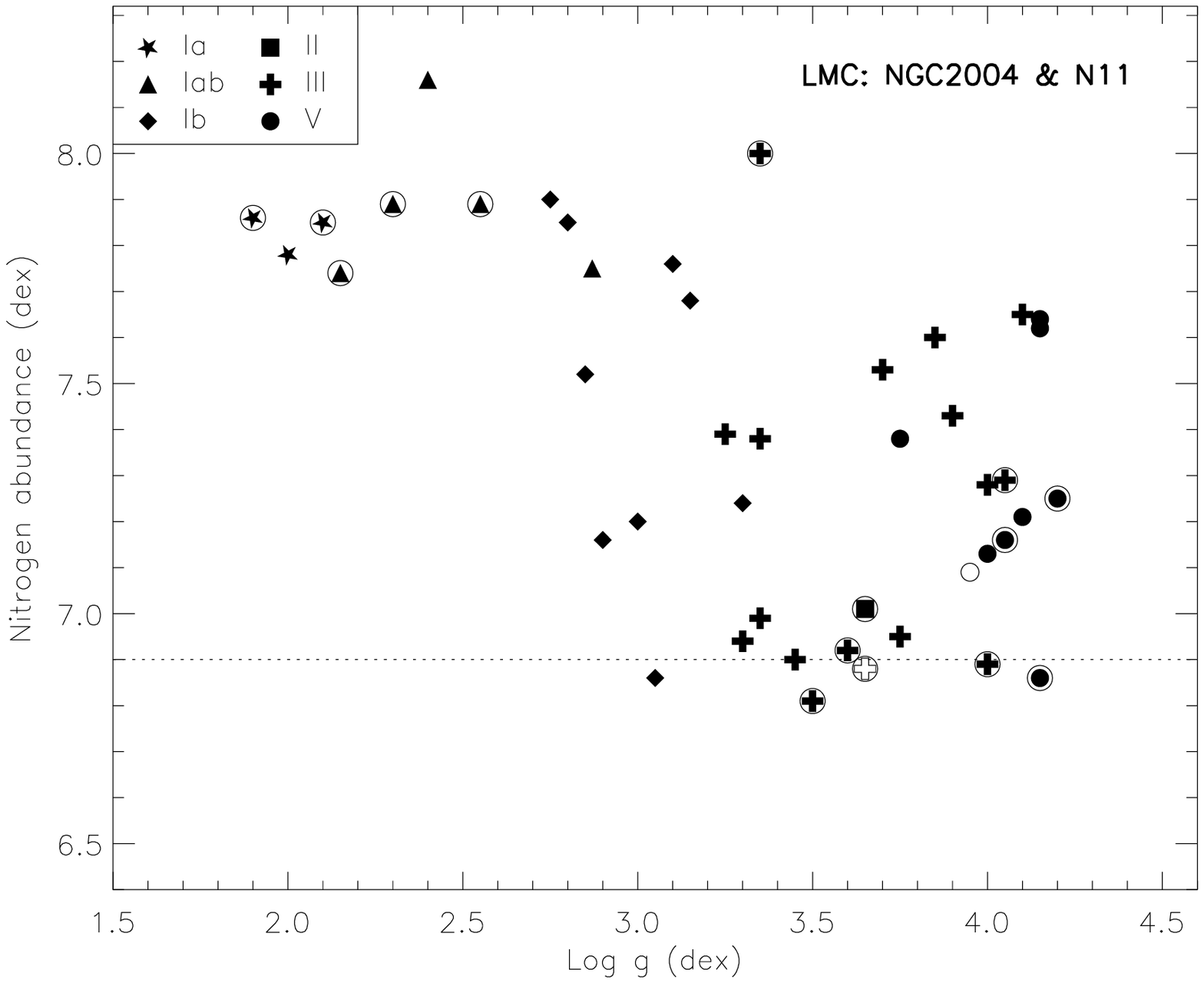, width=84mm, angle=0}
\epsfig{file=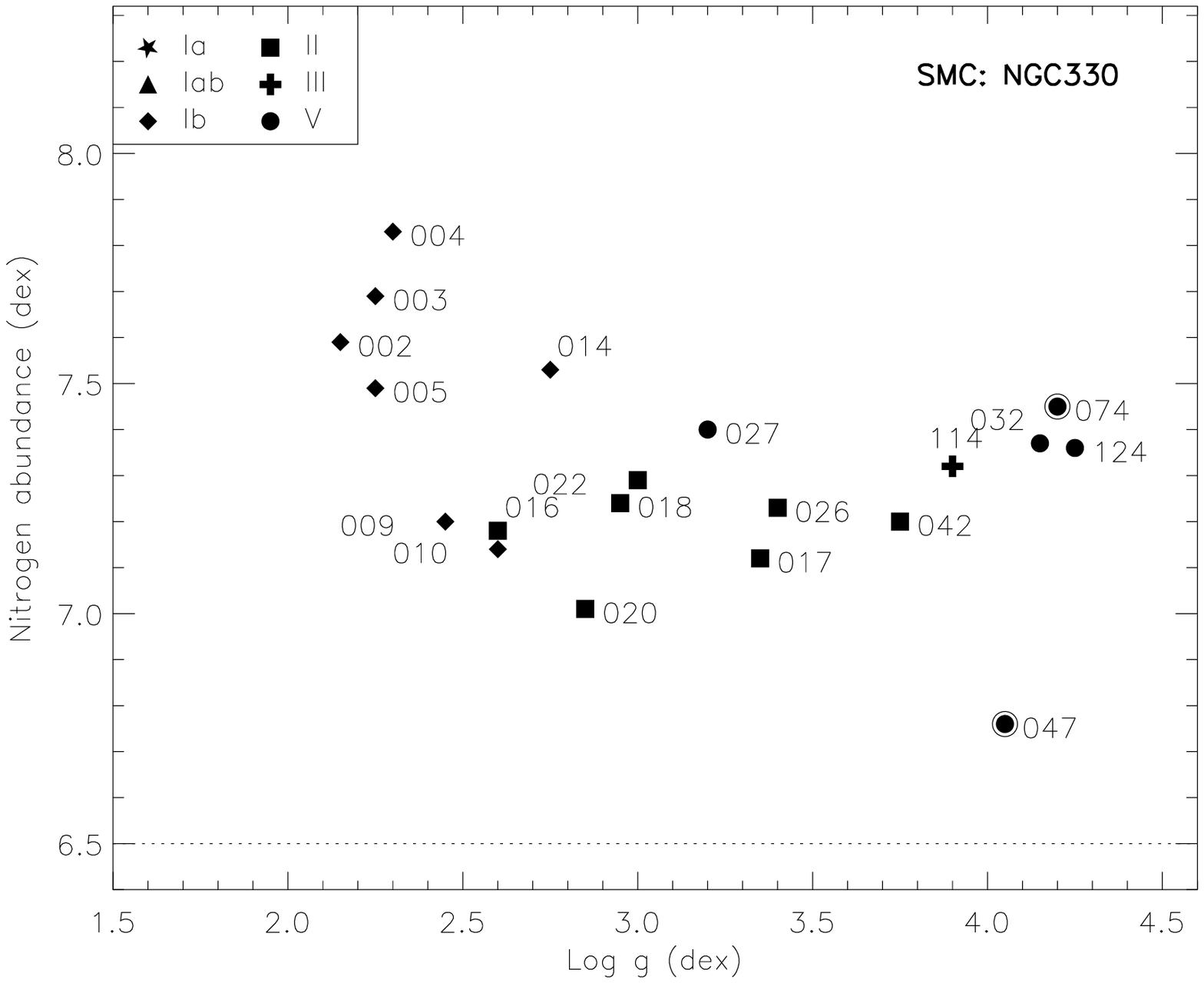, width=84mm, angle=0}
\epsfig{file=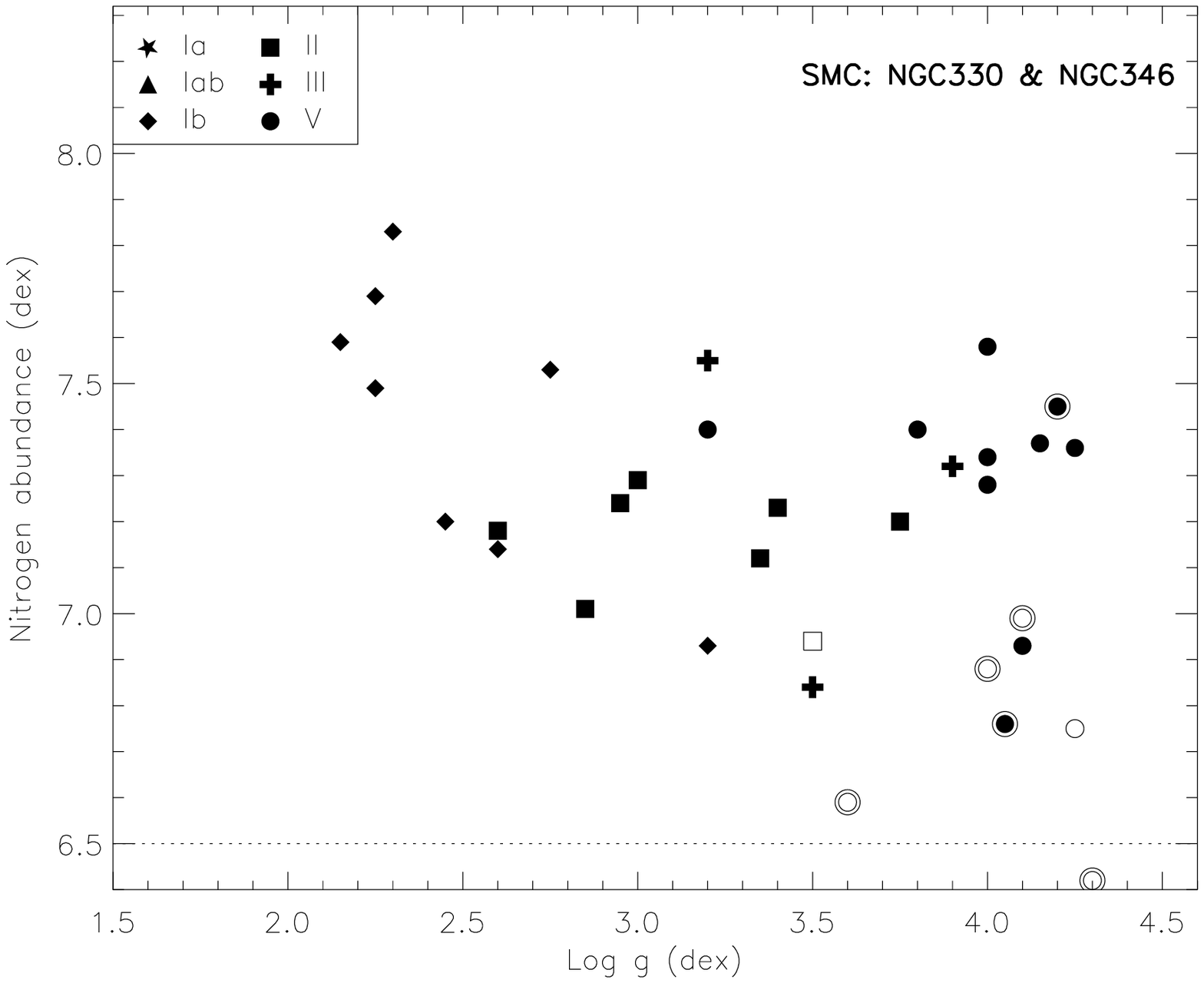, width=84mm, angle=0}
\caption[]
{
Nitrogen abundances as a function of surface gravity (\logg) for each galaxy.
Only those objects with masses less than 20 solar masses are included. The left
panel shows a plot for each of the galaxies with only those objects analysed in
this paper included, whereas the right panel also contains objects from Paper IV
with similar masses. Open circles represent upper limits to the nitrogen abundances,
circled symbols are possible composite objects. Dotted lines in each plot
represent the baseline abundance of the appropriate galaxy as estimated from H \2
regions.
}
\label{nvslogg}
\end{center}
\end{figure*}
\begin{figure*}
\begin{center}
\epsfig{file=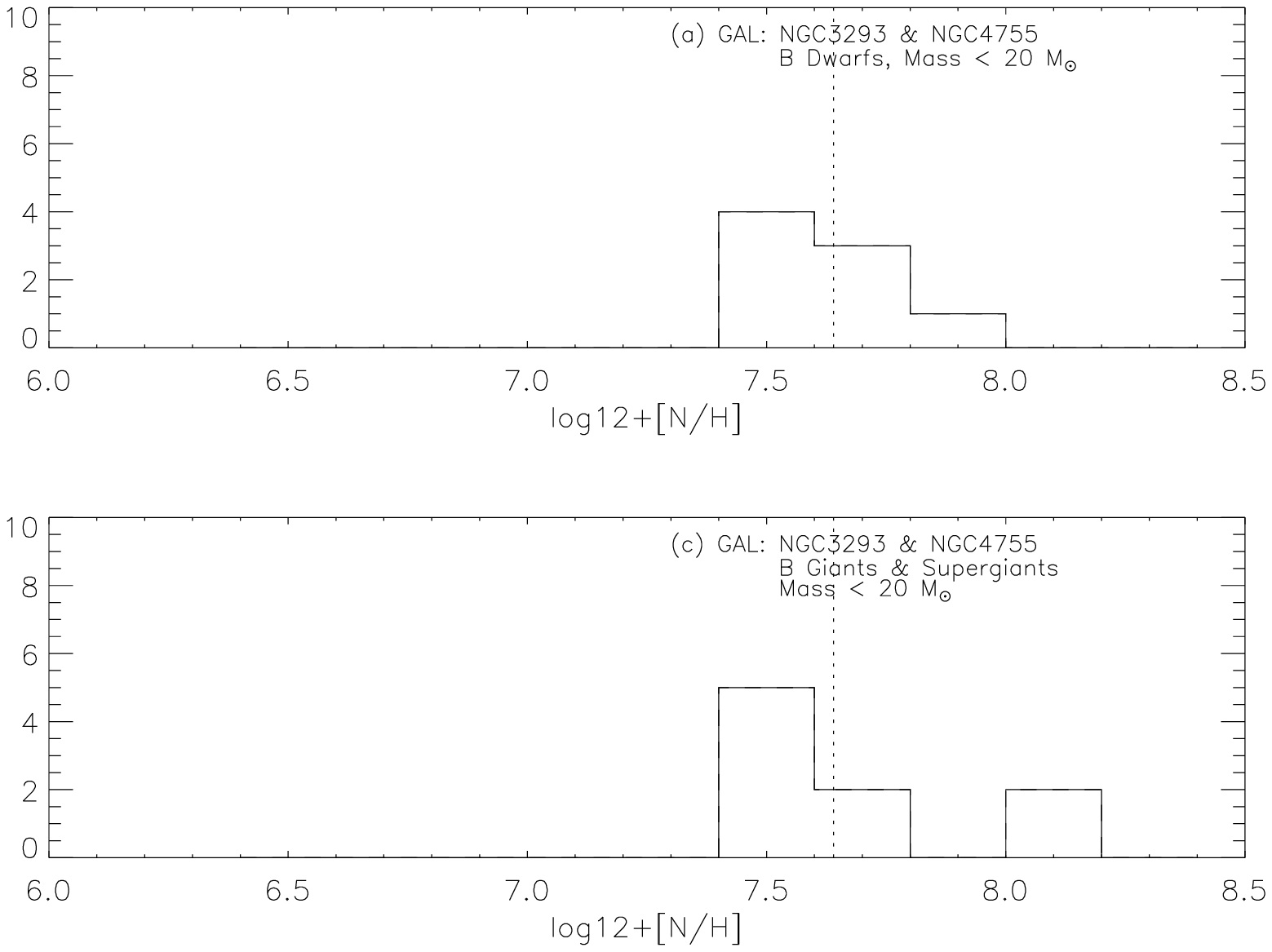, width=84mm, angle=0}
\epsfig{file=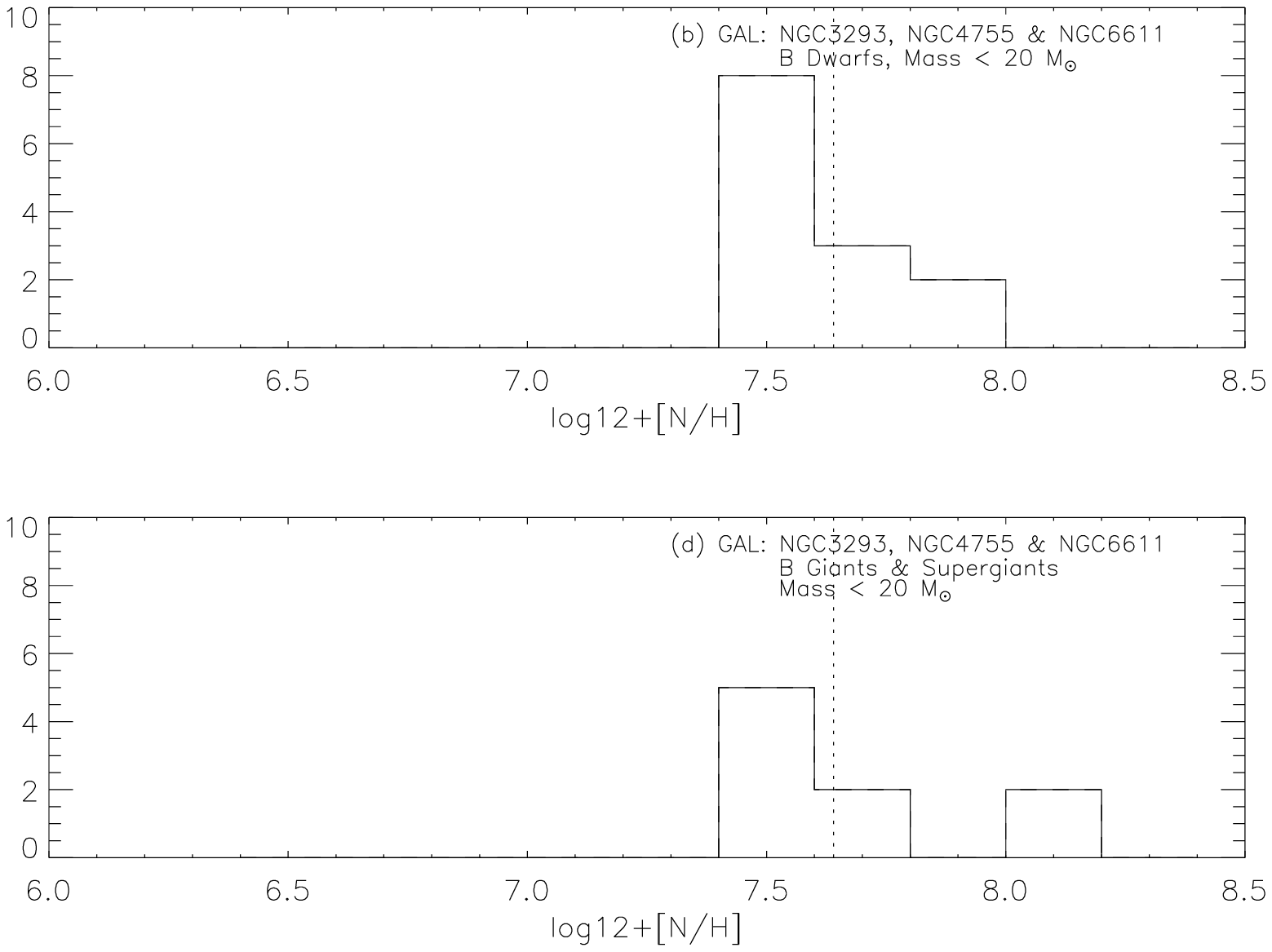, width=84mm, angle=0}
\epsfig{file=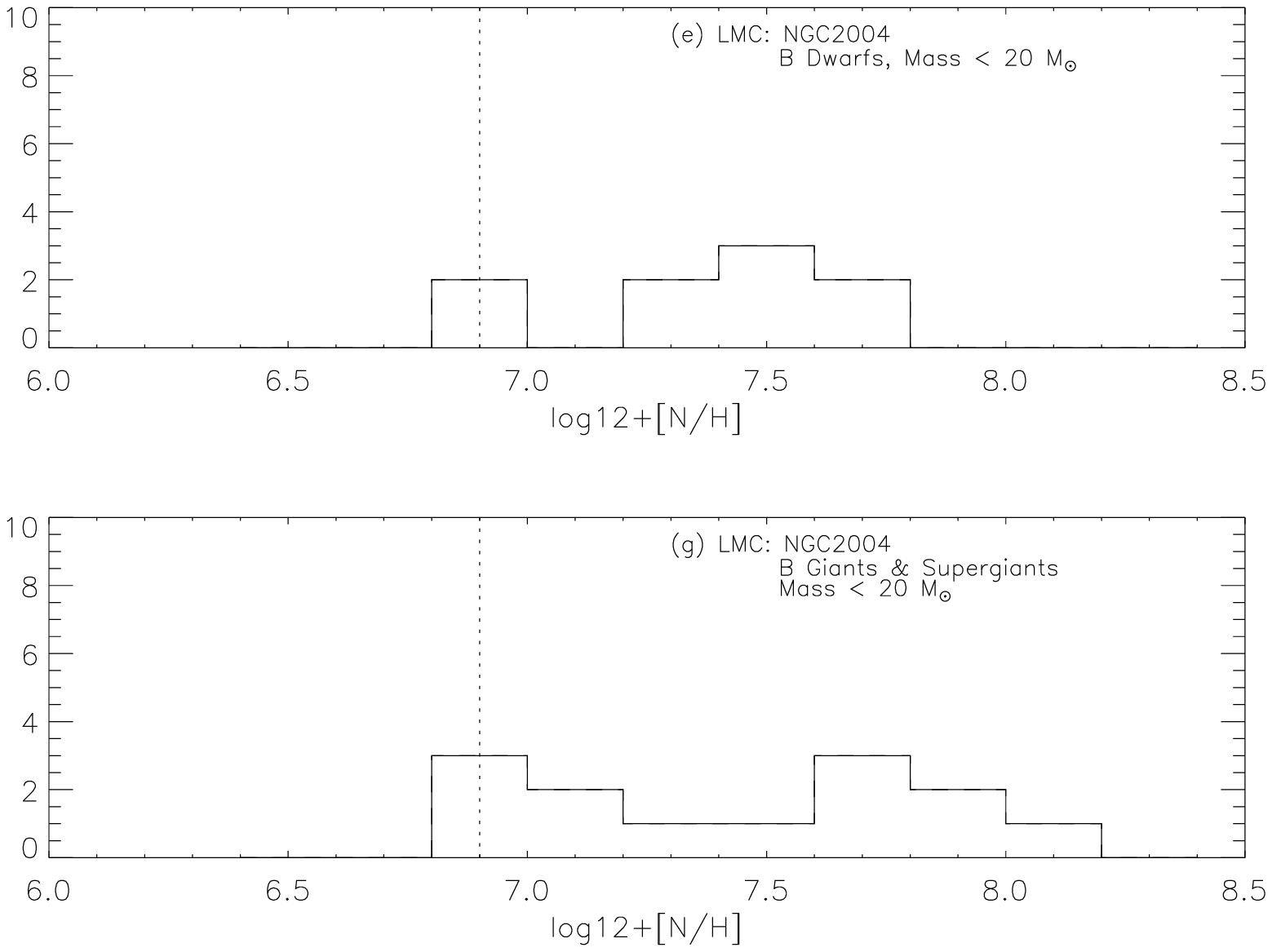,width=84mm, angle=0}
\epsfig{file=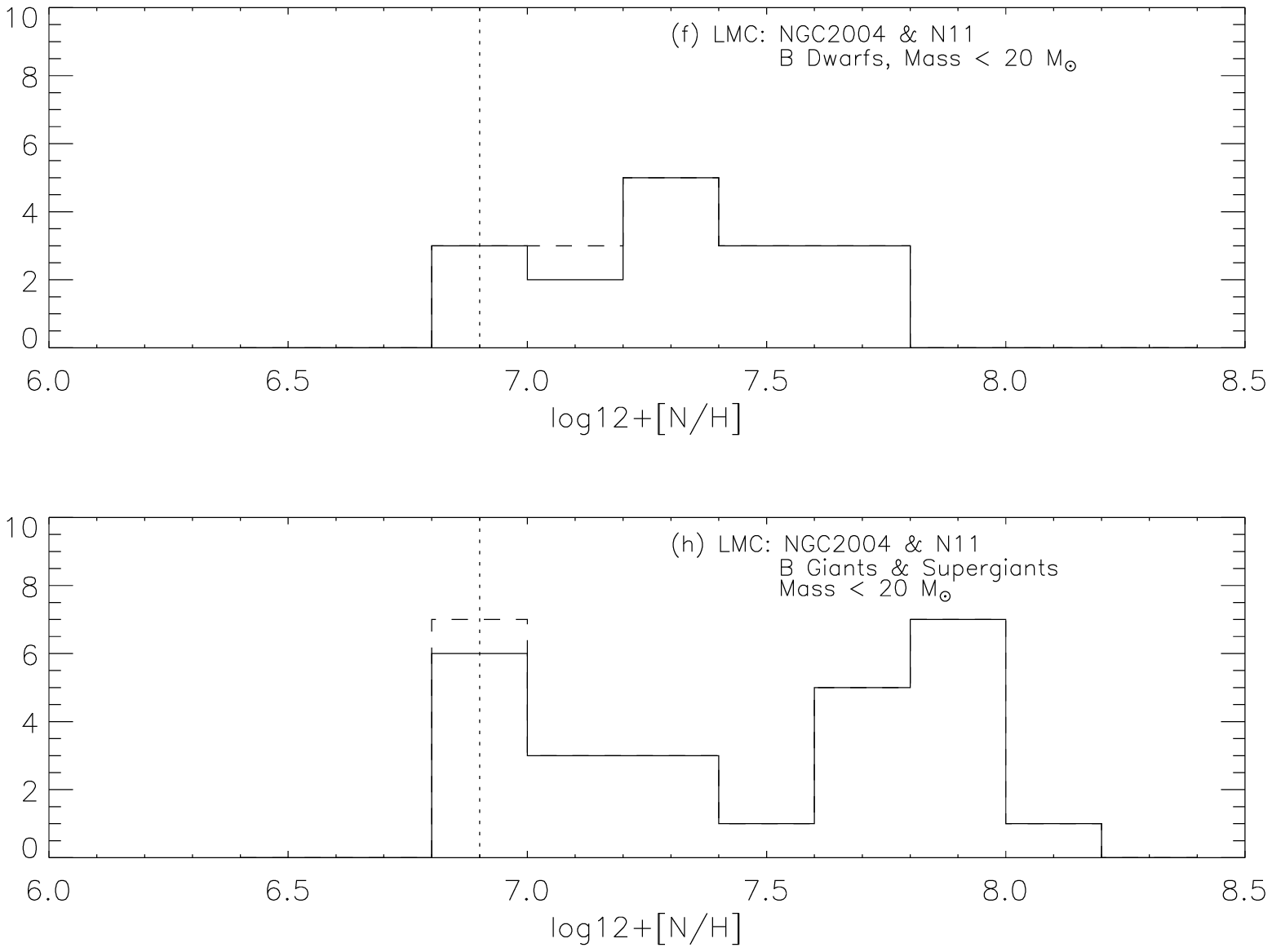, width=84mm, angle=0}
\epsfig{file=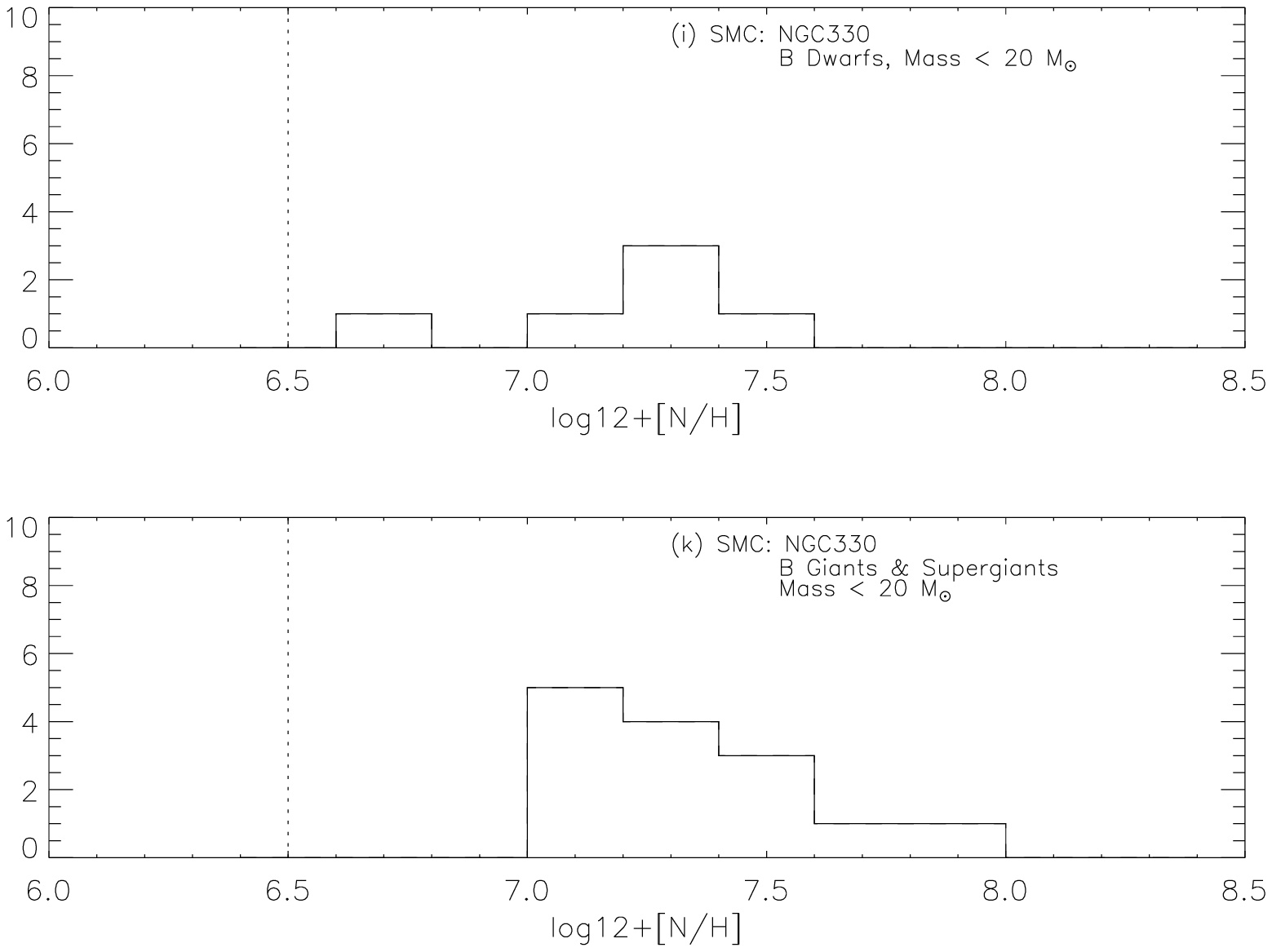, width=84mm, angle=0}
\epsfig{file=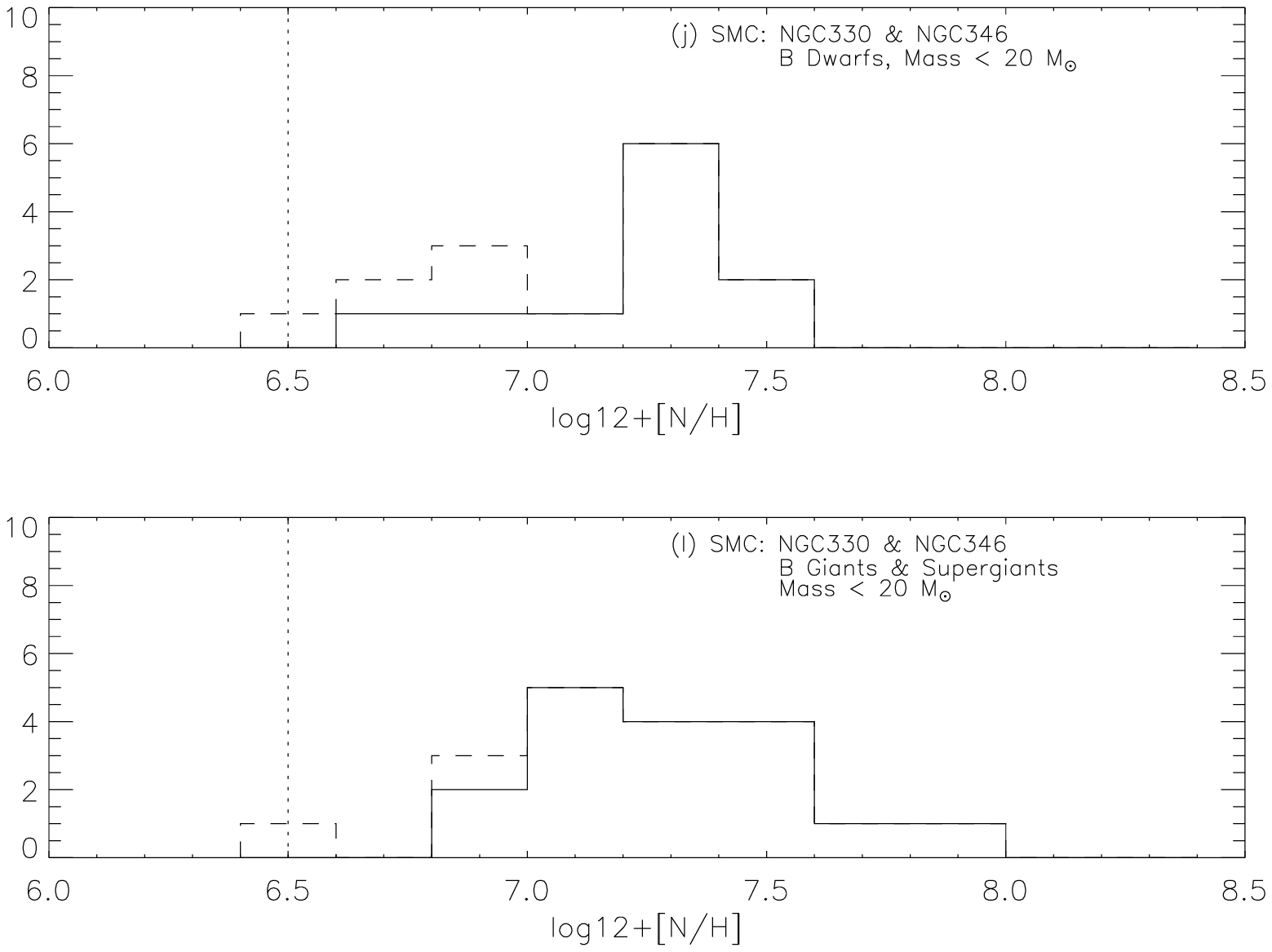, width=84mm, angle=0}
\caption[]
{ Histograms of the 12+log[N/H] distributions in the Milky Way objects
{\bf (a-d)}, the LMC objects {\bf (e-h)} and the SMC objects {\bf
(i-l)}. The left panels show plots for each of the galaxies including
the objects analysed in this paper, whereas the right panel also
contains objects with similar masses from Paper IV. The dotted line
represents the baseline nitrogen abundance of the appropriate galaxy,
and the dashed lines represent upper limits.  }
\label{hist}
\end{center}
\end{figure*}

Photospheric nitrogen abundances of B-type stars are known to be good tracers of
their evolution due to the wide range of abundances that have been estimated from 
their spectra, as evident from Table~\ref{fixall}. Whilst it is clear that
nitrogen can be produced in the core of these stars via the CNO cycle, the
mechanism for transferring this nitrogen enriched material to the surface is still
under debate. In this section we will discuss the nitrogen abundances of these
stars in terms of their evolution in an attempt to decipher the correlation of
nitrogen with surface gravity and mass.

The targets from each field are shown on separate Hertzsprung-Russell
diagrams (HRD) in Fig.~\ref{hrd}, with the relative sizes of the
symbols representing the degree of nitrogen in their atmospheres. The
evolutionary tracks included are non-rotating tracks from
\citet[]{mey94} and \citet[]{sch92} for the Milky Way clusters, and
\citet[]{sch93} and \citet[]{char93} for NGC2004 and NGC330,
respectively. Isochrones are also presented, but in the case of the
Magellanic cloud HRD's they should be used for information only. The
majority of the stars in these two sets of Magellanic Cloud FLAMES
observations are expected to be field stars (the compactness of the
clusters caused difficulties in putting fibres on objects close to
their centre, see discussion in Paper II). On the contrary, amongst
the Galactic objects only one star is thought to be a non-cluster
member, this is NGC4755-033 (see Paper~III).

It is important to note that these are intermediate age clusters or in the case
of the Magellanic clouds, fields around such clusters and  hence these objects
are relatively low mass and can only be compared to the similarly low mass stars
analysed in Paper~IV. This can be seen in Fig.~\ref{hrd} where all but two
objects, NGC4755-002 and NGC2004-011, lie below or on the 20 solar mass track.
\citet[]{mey00} and \citet[]{mae01} have shown that mass-loss becomes an
important factor in the evolution of stars with initial solar masses greater
than 20 to 25 M$_{\odot}$, particularly for the evolution of the surface
rotational velocities and angular momentum. We therefore exclude NGC4755-002 and
NGC2004-011 from most of the following discussions and assume that mass-loss
does not play an important role in the evolution of the surface nitrogen
abundances of these stars.

From Fig.~\ref{hrd}, it is immediately clear that the more massive and evolved
objects have the larger absolute nitrogen abundances. This has been discussed
by \citet[]{mey00}, who showed that the surface nitrogen enrichments are more
pronounced as the stellar mass is increased in their models. The HRDs of the
Galactic clusters show that there is little variation in the nitrogen
abundances of these objects, except for the three most massive and evolved
objects in NGC4755 (\#002, \#003, \& \#004).  

In Fig.~\ref{nvslogg} the absolute nitrogen abundances are shown as a function
of the stellar surface gravity for each cluster, hence the main-sequence objects
are to the right of the plot at high surface gravities and tend towards the more
evolved objects at low values of \logg. The baseline nitrogen abundances of the
Galaxy, LMC and SMC are marked on these plots as adopted from H \2 region
analyses (see Paper~IV and references therein). The Galactic nitrogen abundance
as estimated from H \2 regions is higher than the mean nitrogen abundances in
these clusters. As can be seen from Fig.~\ref{nvslogg}, the majority of Galactic
objects have absolute nitrogen abundances below this baseline value. Omitting
the three low gravity objects with $\log$ [N/H] $>$ 8.0 dex
(NGC4755-002,-003,-004) the mean nitrogen abundance from the three clusters is
7.58 dex, whereas that from H \2 regions is approx 7.64 dex, which is consistent
within the errors of the analysis. 

The NGC2004 \& NGC330 objects span a very large range of nitrogen abundances,
even if we exclude the supergiant objects. In NGC2004 the luminosity class V-II
objects span a range of nitrogen abundances from the baseline LMC abundance to
approximately 7.65 dex, whilst similar objects in NGC330 span a range from 6.75
dex (0.15 dex above baseline SMC abundances) to approx. 7.45 dex. We do not find
many ``normal" N abundance objects in NGC330 and indeed there is only one object (\#47)
with a nitrogen abundance in the range of 6.5 to 7 dex. However there are very
few unevolved objects in our NGC330 sample, as can be seen from Fig.~\ref{hrd}. 

In the right panel of Fig.~\ref{nvslogg} we compare our results  with those from
Paper~IV for the objects with masses less than 20 M$_{\odot}$. The LMC and
Galactic objects from the young clusters analysed in Paper~IV and the relatively
older clusters studied here have the same range of abundances.  In the SMC, the
young cluster NGC346 has more objects than NGC330 with abundances close to the
SMC baseline N abundance due to the larger sample of unevolved objects. The
range of abundances for the relatively unevolved objects in the Magellanic
clouds could be explained by the stellar sample having a range of initial
rotational velocities. Hence if rotation is the dominant factor in producing the
surface nitrogen enrichment one would expect a range in the nitrogen abundances.
However, as stated in Paper~IV and reconfirmed here, at least for the Magellanic
Clouds, large nitrogen enrichment occurs for the stars close to the
zero-age main-sequence. Additionally the nitrogen excesses for the luminosity
class V-II objects in the Magellanic Clouds would results in an increase of a
factor of two or less for the baseline Galactic nitrogen abundance, which is
consistent with our results but indistinguishable from our uncertainties.

The majority of our supergiant or luminosity class I objects have higher surface
nitrogen enrichments than seen in the dwarf or giant objects. In Fig.~\ref{hist}
we present histograms of the nitrogen abundances, separating the targets from
each galaxy into a group of dwarfs and another of giants and supergiants. In the
left panels of Fig.~\ref{hist} are histograms which only include the objects
analysed in this paper, it is clear from these that the more evolved objects
have higher nitrogen abundances. As the histograms for these intermediate aged
stars agree well with those presented in Paper~IV for the younger objects in the
survey, we present histograms of the complete sample of low \vsini~objects in
the right hand-side panel of Fig.~\ref{hist}. Whilst the giant/supergiant
samples have objects with a similar range of surface nitrogen abundances as the
dwarf objects, they also have a number of objects with higher nitrogen
abundances. These latter objects are normally the more luminous supergiants,
however not all the supergiants have highly nitrogen enriched atmospheres as a
number of the Ib supergiants have abundances similar to the giant and dwarf
objects (see Fig.~\ref{nvslogg}). These low nitrogen abundances can again be
accounted for by the objects evolving from a population of stars with a range of
initial rotational velocities. 

Our observations indicate that there are quite significant nitrogen enhancements
during the supergiant phase in all three galaxies.  The SMC evolutionary models
from the Geneva group \citep[]{mey00, mae01} show that between the end of the
main sequence and the red supergiant phase at $\log$(\teff) = 3.9, there is a
low level of nitrogen enrichment for 15-25 $M_\odot$ stars of approx 0.1 dex in
the SMC. Lower mass objects that go through even a partial blue-loop show more
significant nitrogen enrichments (see Fig.16 of Maeder \& Meynet, 2001). However
these N excesses are not produced in the Galactic models for stars crossing the
HR diagram. Hence it is unclear if the unevolved Galactic objects exhibit
surface nitrogen enrichments or the scatter in their abundances is simply a
result of the uncertainties in the analysis.

In paper IV two distinct populations of radial velocity variable objects were
found, (1) a group of unevolved objects with normal nitrogen abundances and (2)
a group of evolved objects with large surface nitrogen enrichments. In this work
we have discovered a number of objects which also fall into these two
categories, as can be seen in Fig.~\ref{nvslogg}. The first group are
conceivably binary objects with low initial rotational velocities and hence have
small or negligible surface nitrogen enrichments. The small rotational
velocities may arise from tidal locking with the orbital velocity of the system
within the main-sequence lifetime of the primary, thus slowing down these
objects \citep{zah92, hua06}. There are two unevolved objects that don't fall
into these categories, these are N11-075 and NGC330-074 both of which have
reasonably high nitrogen abundances. The second group of objects are all
luminous objects, most of which have very small radial velocity variations of $<$
5 \kms. However further observations are required to determine if these objects
are true binary systems. If these objects are in binary systems, the nitrogen
excesses we are observing could suggest that they have gone through a
mass-transfer process. That said, the nitrogen abundances are similar to other
luminous objects for which we have not detected any radial velocity variations
and may not necessarily be connected to binarity.

%--------------------------------------------------------------------------------------
\begin{figure}
\begin{center}
\epsfig{file=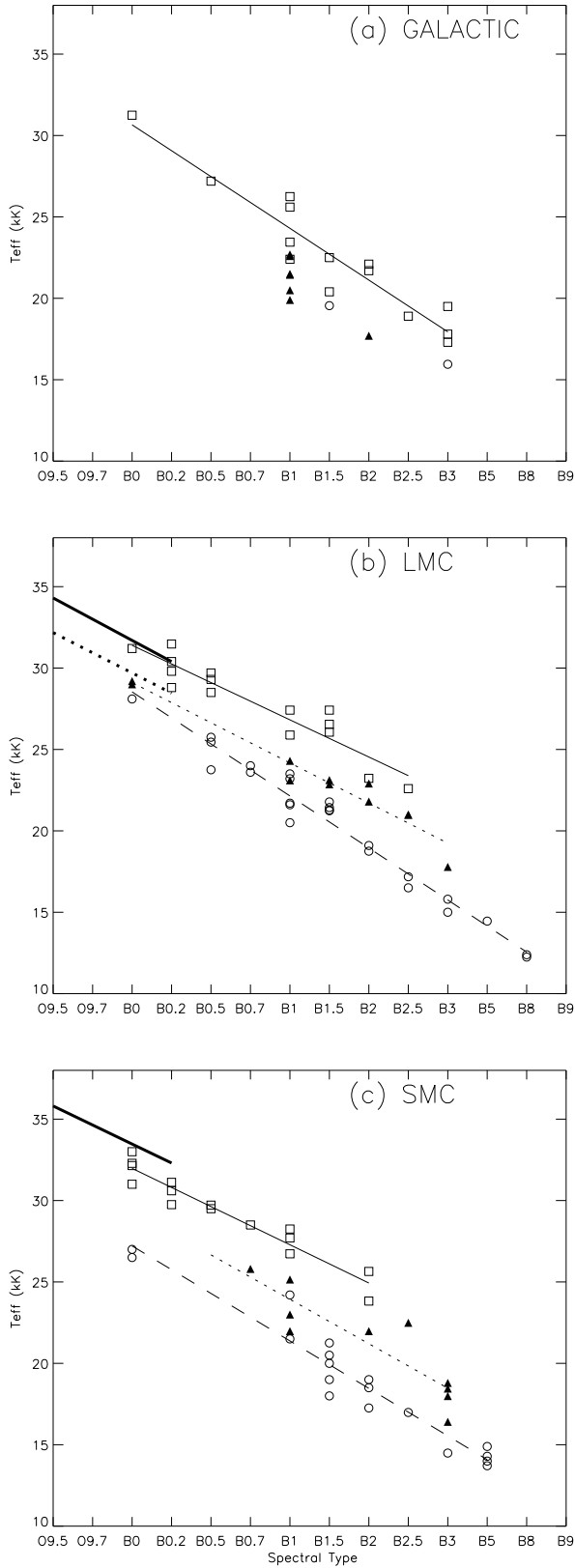}
\caption[]
{
Temperature Scales based on 107 B-type stars from the Flames Survey of 7 clusters; NGC6611,
NGC3293, NGC4755, N11, NGC2004, NGC346 and NGC330. Also included are 9 SMC stars from
\citet[]{duf05}. Comparison of the temperature scales as a function of {\bf
luminosity class} in the three galaxies (a) Milky Way, (b) LMC, (c) SMC. The thin lines
are fits to the mean at each spectral type for the B-type stars and the thick lines are fits
to the O star data presented in \citet[]{rmsmc,rmlmc}.  The different classes are
identified by: V - open squares \& solid line; III - filled triangles \& dotted line; I
- open circles \& dashed line.}
\label{teffplot1} 
\end{center} 
\end{figure}

\subsection{Effective Temperatures Scales.}
\label{teffscale}

Effective temperature scales of OB-type stars as a function of spectral type are
important for determining their properties, such as luminosities and the
number of ionising photons. In turn, these stellar properties are crucial for many
topics in astrophysics; for comparisons with stellar evolution models, determining
cluster properties such as age and distance and for understanding the properties
of ionising stars. Due to most of their flux being emitted in the far-UV, the
effective temperatures of OB-type stars can only be determined using indirect
techniques involving the complex modelling of their atmospheres. These
sophisticated model atmospheres need to take into account non-LTE effects, line
blanketing and in some cases stellar wind effects.
\begin{figure}
\begin{center}
\epsfig{file=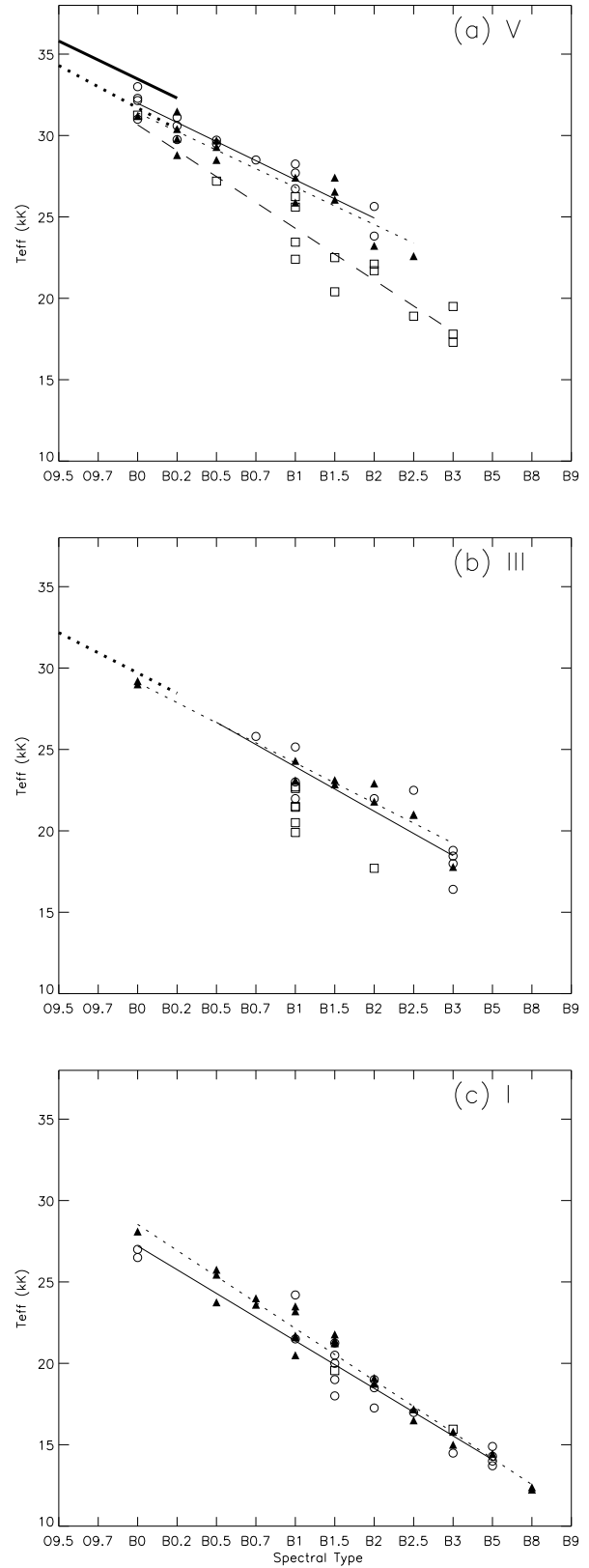}
\caption[]
{
Temperature Scales based on 107 B-type stars from the Flames Survey of 7 clusters; NGC6611,
NGC3293, NGC4755, N11, NGC2004, NGC346 and NGC330. Also included are 9 SMC stars from
\citet[]{duf05}.  Comparison of the temperature scales as a function of {\bf
metallicity} for the three luminosity classes (a) V , (b) III  and (c) I objects. The
thin lines are fits to the mean at each spectral type for the B-type stars and the thick line
are fits to the O star data presented in \citet[]{rmsmc, rmlmc}. The open circles \&
solid lines, filled triangles \& dotted lines and open squares \& dashed lines represent
the SMC, LMC and Milky Way, respectively. } 

\label{teffplot2}

\end{center}
\end{figure}
In recent years several revisions have been made to O and B-type star temperature
scales in particular due to the inclusion of line blanketing and non-LTE effects
in the model atmospheres \citep[]{mar02, crow02, crow06, mas04, mas05}. These have revealed
clear differences in the temperature scales depending on luminosity class and
metallicity. In Paper III, the temperatures of 49 B-type stars in the spectral
range of O9.5 to B2 were presented. The best sampled luminosity group in that
work was the main sequence objects for which a clear difference in the temperature
scales with metallicity was shown for the SMC and Galaxy, where the latter was 
cooler. Here we present the effective temperature scales based on the consistent
analysis of 107 B-type stars studied in the Galaxy, LMC and SMC from the FLAMES
dataset (60 from this paper and 47 from Paper~IV, thus excluding O9 stars). In
addition we have included nine  supergiants from the work of \citet[]{duf05},
bringing the total to 116. \citeauthor[]{duf05} applied the same analysis
techniques and so their results should be consistent with those presented here.

Firstly, the objects need to be separated into different luminosity classes.
This is, in theory, easily done by adopting the luminosity classes assigned to the
objects in Paper~I \& II based on the widths of the high order hydrogen Balmer
lines. A small number of objects have atmospheric parameters that are
inconsistent with their assigned classifications. This reflects the
uncertainties of precise classification (e.g. with respect to rotation) and also
the continuous nature of the physical gravities when compared to the discrete
luminosity classes. For the purposes of investigating physical trends we have
therefore reassigned some of the objects. Any luminosity class III objects with
high gravities ($>$ 3.7) were grouped with the class V objects, whilst any
luminosity class V stars with low surface gravities ($<$ 3.7) were grouped with
the luminosity class III objects. The luminosity class II objects, as assigned
in Paper~I \& II, were assigned to the luminosity class III or I scales
depending on their surface gravity. Apart from the luminosity class II objects,
this affected only 4 objects in the SMC: NGC346-039, NGC330-027, -042, -114, the
former two having very high gravities for giants, and the latter two having too
low gravities for dwarfs. In the LMC, 5 objects were affected; NGC2004-014, -046,
-064, -119 having higher gravities than those expected for a giant, whereas
N11-037 has a lower gravity and was included with the supergiants.

In Fig.~\ref{teffplot1} the effective temperatures of our sample are plotted as
a function of spectral type for the Galaxy, LMC and SMC, where the luminosity
classes are identified with different symbols. For the LMC and SMC, there is a
clear separation of temperature scales depending on the luminosity class of the
objects; the more luminous the object the cooler it is.  From B0 to B2.5 the
effective temperatures of the dwarfs are higher than those of the giants by
$\sim$ 2300-3000 K in the LMC and $\sim$ 3100-3700 K in the SMC, whereas the
supergiants are cooler than the giants by $\sim$ 550-3000 K in the LMC and
$\sim$ 1600-2650 K in the SMC. For the Galactic sample there is insufficient
sampling of luminosity classes III or I to characterize this separation,
although the luminosity class V objects are clearly the hottest objects. The
fact that the more evolved giants and supergiants are progressively cooler than
the dwarfs is as expected since the former two classes are more evolved objects
with lower gravities and hence require lower temperatures to fit the silicon
ionisation equilibrium. \begin{table*}
\begin{center}
\caption[Effective Temperature Scale.]
{\label{tefftab}   Effective temperature scales of B-type stars in the Milky Way, LMC and SMC for
different luminosity classes. These scales are based on parameters presented in this
Paper and Paper IV, as well as a number of SMC stars from \citet{duf05}. The
values in parenthesis are those for which we have no objects in the spectral type bin
but for which we could interpolate.  The \teff~values are presented in units of K.}  
\begin{tabular}{lccccccc} \hline \hline
        &\multicolumn{1}{c}{Milky Way} &        \multicolumn{3}{c}{LMC}    &  	 \multicolumn{3}{c}{SMC}      \\
Sp. Typ.&\multicolumn{1}{c}{V}&\multicolumn{1}{c}{I}&\multicolumn{1}{c}{III}&\multicolumn{1}{c}{V}&\multicolumn{1}{c}{I}&\multicolumn{1}{c}{III}&\multicolumn{1}{c}{V}\\
 \hline \\
B0      &    30650    &  28550	  &  29100  &  31400 &	27200 &       &  32000 \\
B0.2    &   [29050]   & [26950]   & [27850] &  30250 & [25750]&       &  30800 \\
B0.5    &    27500    &  25350	  & [26650] &  29100 & [24300]&       &  29650 \\
B0.7    &   [25900]   &  23750	  & [25400] & [27950]& [22850]& 25300 &  28450 \\
B1      &    24300    &  22150	  &  24150  &  26800 &  22350 & 23950 &  27300 \\
B1.5    &    22700    &  20550	  &  22950  &  25700 &  20650 &[22550]& [26100]\\
B2      &    22100    &  18950	  &  21700  &  24550 &  18950 & 21200 &  24950 \\
B2.5    &    19550    &  17350	  &  20450  &  23400 &  17200 & 19850 &        \\
B3      &    17950    &  15750	  &  19250  &	     &  15500 & 18450 &        \\
B5      &             &  14150	  &  	    & 	     &  13800 &       &	       \\
B8      &             &  12550	  &  	    & 	     &        &	      &	       \\
\\ \hline
\end{tabular}
\end{center}
\end{table*}

In the SMC and LMC plots we also show fits to the O-type star data, presented
in \citet[]{rmsmc, rmlmc}, for the different luminosity classes and which are
based on all the data in the O3-B0.2 range. The sample size only allows a good
fit to the dwarf and giant objects in the LMC and the dwarf objects in the SMC.
There appears to be good agreement (within 500-700 K) in the crossover region
from O to B-type stars (i.e O9-B0.2) for the dwarf and giant \teff~scales in
the LMC. However, in the SMC, the O star dwarf scale is approximately 1500 K
hotter than for the B-type star dwarf scale at B0 where the two scales cross.
\citet[]{rmsmc} have analysed two B0 stars with FASTWIND, the \teff~of one
agreeing well with the mean of the four B0 objects analysed with TLUSTY, the
other however has a \teff~of 34400, 2300 K higher than derived with TLUSTY.
However, the error on the \teff~estimation of this object is 2200 K. In
addition the slope of the fit to the mean of the O-type stars is affected by
the low \teff~of two O4 stars, omitting these objects would make the agreement
at the crossover much better, and well within the uncertainties in the
\teff~estimations. \citet[]{mar05} have produced observational effective
temperature scales for Galactic O-type stars and their O9.5 dwarf point on the
observational scale appears to be in good agreement, as it is 1200 K hotter
than the B0 point on our scale. 

In Fig.~\ref{teffplot2} the \teff~scale from each luminosity class is considered
as a function of metallicity.  In the case of the dwarf objects the SMC stars are
substantially hotter than the Galactic stars ($\sim$ 1300 at B0 to 4000 K at B2),
as expected due to the increased line blanketing in the more metal rich Galactic
objects. Surprisingly, there is no apparent metallicity dependence on the 
\teff~between the SMC and LMC objects, with all luminosity classes showing a close
agreement between the LMC and SMC objects. The largest difference is in the dwarf
objects where the SMC stars are hotter by $\sim$ 800 K at B0 to 100 K at B2.  

% If anything
%the LMC B supergiants are hotter than the SMC B supergiants, this is also the case for
%the O supergiants, though the numbers of O supergiants are few.
%
The analysis in this paper and Paper~IV has provided an unprecedented, large
and consistent set of atmospheric parameters based on non-LTE and line-blanketed
model atmospheres. As such it is important to provide an up-to-date calibration
of effective temperatures with spectral type. Using the linear fits to the mean
at each spectral type we present the results in Table~\ref{tefftab}. This table
is based on 116 objects analysed between B0-B8 in this paper,
Paper~IV and \citet[]{duf05}. Whilst we have chosen to present the LMC and SMC
scales separately the reader should note the close agreement between these,
particularly in the case of the luminosity class III and I objects.

%----------------------------------------------------------------------------------------
\section{Conclusions}
\label{end}

In this paper we have completed a spectral analysis of 61 B-type stars, located
in four fields; two centered on the Galactic clusters, NGC3293 \& NGC4755, in
the Carina arm, one centered on NGC2004 in the Large Magellanic cloud and the
fourth centered on NGC330 in the Small Magellanic cloud.  We have presented both their
atmospheric parameters and abundances (C, N, O, Mg, Si) estimated using non-LTE
line-blanketed TLUSTY model atmospheres. In addition we have calculated the LTE
iron abundances for these objects plus 17 objects from Paper IV located in
three fields toward the younger clusters observed within the FLAMES project for
massive stars; NGC6611, N11, NGC346. In paper IV we presented the present day
chemical composition of the LMC and SMC, these results are verified  here and
can now be extended to include the Fe abundances for these galaxies. In the LMC
we derive an Fe abundance of 7.23 dex, and in the SMC 6.93 dex, this can be
compared with that derived from the Galactic B-type stars, 7.53 dex, suggesting
that the LMC is a factor of 2 less abundant than the Galaxy and the SMC is a
factor of factor of 4 less abundant. 

We have studied the nitrogen abundances of the 107 low \vsini~B-type stars
within the FLAMES survey of massive stars, and discussed them in terms of their
evolution. We have observed a large range of nitrogen abundances in the LMC and
SMC main-sequence objects which is likely to reflect the range of rotational
velocities, and hence rotational-mixing, initially present in these stars. Whilst
mass-loss may play some role in the evolution of these stars, the stars
considered in this work have masses less than 20 $M_\odot$ masses and it is
unlikely that it has a dominant role in altering their surface abundances. This
is not the case for higher mass objects considered in Paper IV. The large
spread of nitrogen abundances seen in the LMC and SMC main-sequence objects
does not seem to be present in the Galactic objects but if we assume that these
Galactic stars should undergo the same degree of enrichment as the LMC and SMC,
a simple calculation shows that this would amount to only a factor of two or 0.30
dex enhancement of nitrogen in the Galaxy. This is similar to our degree of
uncertainty so it is unclear if we are observing the scatter due to our
uncertainties or the range of rotational velocities of this population of
stars. The supergiants in the sample are significantly more enriched than their
main-sequence counterparts, and hence there must be some method of enhancing the
photospheric nitrogen abundance as the star evolves off the main-sequence and
across the HRD. Models of single stars in the SMC \citep{mey00,mae01}, do not
predict such large degrees of nitrogen enrichments and would require more
efficient mixing mechanisms to match these observations.

Two populations of binary objects were identified in Paper IV and in this paper we
have detected additional objects belonging to these groups; (1) unevolved, `normal'
nitrogen stars (2) evolved, nitrogen-rich stars. The former group have significant
radial velocity variations, whilst the evolved objects have very small shifts which
require verification by observations over multiple epochs. Binary evolution can be
invoked to explain these two groups of objects using tidal locking to the orbital
velocity to suppress nitrogen excesses in the first group, and mass-transfer to
enhance the nitrogen in the more evolved objects, although the abundances observed in
these objects are similar to those in the supposedly single stars. Clearly further
observations of these objects are required to disentangle the binary objects from
the single stars and to study the evolution of nitrogen abundances in these
populations separately.

The results presented here and in Paper IV provide a consistent analysis of a
large sample of Galactic objects, as well as the largest sample of Magellanic
Cloud B-type stars studied to date. As a result, we have constructed effective
temperature scales as a function of spectral type for B-type stars from B0 down
to B8,  for each of the galaxies and have shown that they are
consistent with effective temperatures from O-type stars. In each of the three
metallicity environments we have detected a dependency on luminosity, where the
most luminous objects are the coolest. Due to the lower metallicity of the
Magellanic Clouds and hence the expected reduction in line-blanketing, one would
expect to detect higher temperatures with decreasing metallicity. Whilst we do
see that the Galactic dwarf objects are cooler than the Magellanic cloud
objects, we do not detect a significant difference in the
temperature scales of the SMC and LMC supergiants and giants and the less luminous
dwarf objects in the two clouds differ at the most by 600 K. We therefore
present the effective temperature calibration of dwarf B-type stars in the
Galaxy, LMC and SMC and giant and supergiant calibrations for LMC and SMC
objects with the caveat that the latter are very similar. 

This work has focused on the low \vsini~population of B-type stars in 7 fields in 3
distinct metallicity environments. However we require knowledge of their
high \vsini~counterparts to get a complete picture of the evolution of their
surface abundances in terms of their rotational velocities. We also require
additional knowledge on whether or not these objects are binary or single stars,
to unravel the evolution of these two distinct groups of stars. Future work will
focus on these points, to place better constraints on the evolution of massive
stars.

%--------------------------------------------------------------------------------------
\section*{Acknowledgments}   

CT would like to thank A. Hibbert, N. Pryzbilla, M.F. Nieva and J. Eldridge for
their insightful discussions on this work.  We also acknowledge financial support
from the UK Particle Physics and Astronomy Research Council (PPARC). This work,
conducted as part of the award "Understanding the lives of massive stars from
birth to supernovae" (S.J. Smartt) made under the European Heads of Research
Councils and European Science Foundation EURYI (European Young Investigator)
Awards scheme, was supported by funds from the Participating Organisations of
EURYI and the EC Sixth Framework Programme. SJS also thanks the Leverhulme Trust
for funding. We are also grateful to the referee, Dr. P. Bonifacio, for his
constructive comments on this work.

\begin{appendix}
\section{Notes on Individual Objects:}
\label{appA}
\noindent{\bf NGC4755-005:} Revisiting this spectrum we noticed weak absorption
features from  Si \4 4116/4089 \AA\ and He \2 4686 \AA, that are inconsistent
with the B2 \2 classification from Paper~I. These lines are blue-shifted from
their rest wavelengths by 5 \kms, whereas the majority of the absorption lines
in the spectrum are red-shifted by 40 \kms. In addition, the metal lines are quite
asymmetric and broad. Contrary to the statement in Paper~I, this star appears to
be a double-lined spectroscopic binary. 
\\\\
\noindent{\bf NGC4755-011:} This object is also a possible double-lined
spectroscopic binary. There are clear signs of double features in the Si \3
multiplet at 4560 \AA\ and the Mg \2 line at 4481 \AA\ is highly asymmetric.

\end{appendix}

%---------------------------------------------------------------
%\begin{appendix}
%\section{To Do Notes:}
%
%Check the assignment of Paper~I-V etc for the FLAMES references\\
%
%figure out C abundances- waiting on N.Przybilla for ew and abundances.\\
%
%waiting for Si \2 oscillator strengths - Hibbert\\
%
%
%left ngc2004-011 out of Nvslogg plots as M $>$ 20\\
%
%
%a couple of ngc3293 objects couldn't be analysed with tlusty as beyond edge of grid -
%\#001, \#002. These were analysed by crowther with cmfgen in galactic B supergiant paper.
%HD91969 and HD91943.\\
%
%\end{appendix}
%---------------------------------------------------------------

\bibliographystyle{aa}
\bibliography{flames_oldclusters_draft2}

\newpage

\Online
\begin{appendix}%mean abundance tables
\begin{landscape}
\vspace{-20mm}
\begin{table}
\setcounter{table}{0}
%\begin{center}
\caption[Abundances of NGC3293, NGC4755, NGC2004 \& NGC330 stars.]
{\label{ablin}   Absolute Abundances of NGC3293, NGC4755, NGC2004 \& NGC330
stars. The abundances presented below are the mean absolute abundances of each species
studied, using the initial atmospheric parameters from Tables~\ref{galpar} \& ~\ref{mcpar}. Numbers in the parentheses represent the number of lines of each species
observed in the star, only 1 line was considered for C {\sc ii}, Mg {\sc ii} \& Si {\sc iv}, 2 lines for Si {\sc ii} and 3 for Si {\sc iii}.Uncertainties on the
abundances account for both random and systematic errors as discussed in Sect.~\ref{aberror}. Abundances
are presented as [X]=12 + $\log$([X/H]) in units of dex.  }  
\begin{tabular}{llcclclcccccl} \hline \hline
Star        & Sp.Typ     & C {\sc ii}  &N {\sc ii} && O {\sc ii} && Mg {\sc ii} & Si {\sc ii}  &Si {\sc iii}   &Si {\sc iv}&Fe {\sc iii}&   \\   
\hline   \\
NGC3293-003 & B1   III    & 7.83  & 7.51 $\pm$ 0.05 & (5) & 8.66 $\pm$ 0.24 & (27) & 7.30 $\pm$ 0.26 &  	       & 7.31 $\pm$ 0.34 & 7.30 $\pm$ 0.65& 7.27 $\pm$ 0.27 & (1)  \\ 
NGC3293-004 & B1   III    & 7.89  & 7.55 $\pm$ 0.09 & (5) & 8.74 $\pm$ 0.23 & (21) & 7.44 $\pm$ 0.29 &  	       & 7.45 $\pm$ 0.31 & 7.45 $\pm$ 0.62&		    &	   \\
NGC3293-007 & B1   III    & 7.86  & 7.50 $\pm$ 0.08 & (4) & 8.67 $\pm$ 0.20 & (28) & 7.25 $\pm$ 0.21 & 7.35 $\pm$ 0.41 & 7.36 $\pm$ 0.30 & 7.34 $\pm$ 0.60& 7.26 $\pm$ 0.22 & (2)  \\
NGC3293-010 & B1   III    & 7.66  & 7.46 $\pm$ 0.12 & (6) & 8.88 $\pm$ 0.33 & (28) & 7.17 $\pm$ 0.25 & 6.86 $\pm$ 0.26 & 7.60 $\pm$ 0.43 & 7.63 $\pm$ 0.62& 7.31 $\pm$ 0.26 & (1)  \\
NGC3293-012 & B1   III    & 7.67  & 7.50 $\pm$ 0.17 & (4) & 8.96 $\pm$ 0.36 & (25) & 7.28 $\pm$ 0.23 &  	       & 7.66 $\pm$ 0.42 & 7.69 $\pm$ 0.61&		    &	   \\
NGC3293-018 & B1   V      & 7.67  & 7.61 $\pm$ 0.13 & (5) & 8.84 $\pm$ 0.30 & (27) & 7.20 $\pm$ 0.20 & 6.87 $\pm$ 0.22 & 7.71 $\pm$ 0.37 & 7.70 $\pm$ 0.55& 7.68 $\pm$ 0.19 & (2)  \\
NGC3293-026 & B2   III    & 7.81  & 7.78 $\pm$ 0.21 & (4) & 8.96 $\pm$ 0.33 & (30) & 7.29 $\pm$ 0.31 & 6.95 $\pm$ 0.25 & 7.76 $\pm$ 0.38 & 7.80 $\pm$ 0.56& 7.66 $\pm$ 0.19 & (2)  \\
NGC3293-043 & B3   V      & 7.91  & 7.55 $\pm$ 0.26 & (7) & 8.50 $\pm$ 0.34 & (19) & 7.14 $\pm$ 0.26 & 7.31 $\pm$ 0.37 & 7.31 $\pm$ 0.34 &		  & 7.51 $\pm$ 0.25 & (1)  \\
% NGC3293-062 & B3   V      & 8.22 $\pm$ 0.29 & (4) & 7.66 $\pm$ 0.30 & (2) &	            &      & 6.90 $\pm$ 0.16 & 7.03 $\pm$ 0.24 & 7.07 $\pm$ 0.34 &           	  & 		    & 	   \\
\\																					    		    
\hline  																				    		    
\\																					    		      
NGC4755-002 & B3   Ia     & 7.84  & 8.16 $\pm$ 0.30 & (7) & 8.57 $\pm$ 0.43 & (26) & 7.28 $\pm$ 0.32 & 7.38 $\pm$ 0.28 & 7.37 $\pm$ 0.45 &		  & 7.53 $\pm$ 0.22 & (2)  \\ 
NGC4755-003 & B2   III    & 7.74  & 8.09 $\pm$ 0.26 & (7) & 8.50 $\pm$ 0.37 & (27) & 7.27 $\pm$ 0.26 & 7.34 $\pm$ 0.25 & 7.34 $\pm$ 0.41 &		  & 7.46 $\pm$ 0.18 & (2)  \\
NGC4755-004 & B1.5 Ib     & 7.73  & 8.06 $\pm$ 0.14 & (7) & 8.65 $\pm$ 0.28 & (31) & 7.25 $\pm$ 0.28 & 7.34 $\pm$ 0.34 & 7.37 $\pm$ 0.37 & 7.35 $\pm$ 0.69& 7.37 $\pm$ 0.24 & (1)  \\
NGC4755-006 & B1   III    & 7.64  & 7.76 $\pm$ 0.23 & (4) & 9.16 $\pm$ 0.38 & (24) & 7.18 $\pm$ 0.23 &  	       & 7.76 $\pm$ 0.43 & 7.78 $\pm$ 0.54&		    &	   \\
NGC4755-015 & B1   V      & 7.66  & 7.98 $\pm$ 0.25 & (5) & 9.07 $\pm$ 0.33 & (30) & 7.16 $\pm$ 0.26 &  	       & 7.83 $\pm$ 0.35 & 7.84 $\pm$ 0.52& 7.72 $\pm$ 0.23 & (1)  \\
NGC4755-017 & B1.5 V      & 7.83  & 7.63 $\pm$ 0.35 & (4) & 8.57 $\pm$ 0.45 & (14) & 7.19 $\pm$ 0.31 & 7.22 $\pm$ 0.29 & 7.20 $\pm$ 0.42 &		  &		    &	   \\
NGC4755-020 & B2   V      & 7.95  & 7.61 $\pm$ 0.20 & (6) & 8.42 $\pm$ 0.30 & (29) & 7.34 $\pm$ 0.27 & 7.24 $\pm$ 0.25 & 7.20 $\pm$ 0.32 &		  & 7.83 $\pm$ 0.29 & (1)  \\
NGC4755-033 & B3   V      & 7.92  & 7.28 $\pm$ 0.27 & (2) & 8.41 $\pm$ 0.36 & ( 3) & 7.02 $\pm$ 0.25 & 7.06 $\pm$ 0.22 & 7.05 $\pm$ 0.31 &		  &		    &	   \\
NGC4755-040 & B2.5 V      & 8.45  & 7.87 $\pm$ 0.39 & (2) & 9.19 $\pm$ 0.32 & ( 3) & 7.28 $\pm$ 0.42 & 7.71 $\pm$ 0.58 & 7.70 $\pm$ 0.35 &		  &		    &	   \\
NGC4755-048 & B3 V        & 8.15  & 7.32 $\pm$ 0.31 & (2) & 8.81 $\pm$ 0.38 & ( 4) & 7.03 $\pm$ 0.27 & 7.19 $\pm$ 0.30 & 7.18 $\pm$ 0.33 &		  &		    &	   \\
\\																					    		           
\hline																					    		    
\\	        																					      
NGC2004-003 & B5   Ia     & 7.68  & 7.86 $\pm$ 0.40 & (7) & 8.46 $\pm$ 0.55 & (20) & 7.04 $\pm$ 0.26 & 7.24 $\pm$ 0.26 & 7.25 $\pm$ 0.50 & 		  & 7.30 $\pm$ 0.23 & (2)  \\
NGC2004-005 & B8   Ia	  & 7.50  & 7.67 $\pm$ 0.37 & (5) & 8.56 $\pm$ 0.57 & ( 4) & 6.78 $\pm$ 0.46 & 6.95 $\pm$ 0.36 & 6.94 $\pm$ 0.44 & 		  & 7.13 $\pm$ 0.23 & (2)  \\
NGC2004-007 & B8   Ia	  & 7.36  & 7.47 $\pm$ 0.28 & (2) & 8.46 $\pm$ 0.53 & ( 3) & 6.67 $\pm$ 0.43 & 6.82 $\pm$ 0.39 & 6.82 $\pm$ 0.44 &		  & 7.02 $\pm$ 0.26 & (2)  \\
NGC2004-010 & B2.5 Iab    & 7.50  & 8.13 $\pm$ 0.28 & (7) & 8.29 $\pm$ 0.36 & (27) & 7.06 $\pm$ 0.26 & 7.15 $\pm$ 0.24 & 7.16 $\pm$ 0.41 & 		  & 7.25 $\pm$ 0.12 & (2)  \\
NGC2004-011 & B1.5 Ia     & 7.45  & 7.72 $\pm$ 0.07 & (7) & 8.39 $\pm$ 0.21 & (30) & 7.15 $\pm$ 0.26 &                 & 7.13 $\pm$ 0.32 & 7.16 $\pm$ 0.64& 7.14 $\pm$ 0.22 & (2)  \\
NGC2004-012 & B1.5 Iab 	  & 7.44  & 7.75 $\pm$ 0.09 & (7) & 8.43 $\pm$ 0.24 & (31) & 7.08 $\pm$ 0.23 &                 & 7.19 $\pm$ 0.35 & 7.21 $\pm$ 0.64& 7.12 $\pm$ 0.19 & (2)  \\
NGC2004-014 & B3   Ib	  & 7.52  & 7.51 $\pm$ 0.25 & (7) & 8.22 $\pm$ 0.37 & (23) & 6.96 $\pm$ 0.20 & 7.09 $\pm$ 0.19 & 7.08 $\pm$ 0.38 &                & 7.16 $\pm$ 0.17 & (2)  \\
NGC2004-021 & B1.5 Ib     & 7.56  & 7.20 $\pm$ 0.10 & (6) & 8.53 $\pm$ 0.28 & (29) & 7.09 $\pm$ 0.23 &                 & 7.31 $\pm$ 0.37 & 7.32 $\pm$ 0.60& 7.19 $\pm$ 0.20 & (2)  \\
NGC2004-022 & B1.5 Ib	  & 7.31  & 7.70 $\pm$ 0.13 & (7) & 8.59 $\pm$ 0.28 & (32) & 6.94 $\pm$ 0.19 &                 & 7.31 $\pm$ 0.37 & 7.31 $\pm$ 0.59& 7.00 $\pm$ 0.25 & (1)  \\
NGC2004-026 & B2   II	  & 7.51  & 7.02 $\pm$ 0.11 & (4) & 8.19 $\pm$ 0.28 & (30) & 7.03 $\pm$ 0.18 & 7.16 $\pm$ 0.21 & 7.17 $\pm$ 0.34 &                & 7.31 $\pm$ 0.15 & (2)  \\
NGC2004-029 & B1.5 e      & 7.43  & 6.81 $\pm$ 0.05 & (4) & 8.29 $\pm$ 0.26 & (30) & 7.03 $\pm$ 0.19 &                 & 7.21 $\pm$ 0.35 & 7.20 $\pm$ 0.56& 7.35 $\pm$ 0.17 & (2)  \\
NGC2004-036 & B1.5 III    & 7.46  & 7.43 $\pm$ 0.11 & (6) & 8.72 $\pm$ 0.32 & (28) & 6.88 $\pm$ 0.20 &                 & 7.57 $\pm$ 0.41 & 7.59 $\pm$ 0.58& 7.32 $\pm$ 0.28 & (1)  \\
NGC2004-042 & B2.5 III    & 7.62  & 6.89 $\pm$ 0.18 & (2) & 8.14 $\pm$ 0.31 & (16) & 7.09 $\pm$ 0.21 & 7.11 $\pm$ 0.22 & 7.11 $\pm$ 0.36 &                & 7.17 $\pm$ 0.20 & (1)  \\
NGC2004-046 & B1.5 III    & 7.54  & 7.62 $\pm$ 0.11 & (6) & 8.53 $\pm$ 0.17 & (30) & 7.05 $\pm$ 0.22 &                 & 7.43 $\pm$ 0.33 & 7.43 $\pm$ 0.53& 		    &      \\
\\ \hline
\end{tabular}
%z\end{center}
\end{table}

\end{landscape}

\begin{landscape}
\vspace{-5mm}
\begin{table}
\setcounter{table}{0}
%\begin{center}
\caption[Abundances of NGC3293, NGC4755, NGC2004 \& NGC330 stars.]
{\label{ablina} Abundances of NGC3293, NGC4755, NGC2004 \& NGC330 stars. Contd.}  
\begin{tabular}{llcclclcccccl} \hline \hline
Star        & Sp.Typ     & C {\sc ii}   &N {\sc ii} && O {\sc ii} && Mg {\sc ii} & Si {\sc ii}  &Si {\sc iii}   &Si {\sc iv}  &Fe {\sc iii}&\\
\hline   \\
NGC2004-053 & B0.2 Ve 	  & 7.93  & 7.73 $\pm$ 0.16 & (3) & 8.51 $\pm$ 0.19 & (28) & 7.31 $\pm$ 0.24 &  	       & 7.41 $\pm$ 0.30 & 7.41 $\pm$ 0.60 &		     &      \\
NGC2004-061 & B2   III	  & 7.78  & 7.00 $\pm$ 0.24 & (3) & 8.32 $\pm$ 0.46 & (16) & 7.20 $\pm$ 0.30 & 7.26 $\pm$ 0.31 & 7.27 $\pm$ 0.48 &		   & 7.53 $\pm$ 0.34 & (1)  \\
%NGC2004-062 & B0.2 V     & 8.08  & 6.71 $\pm$ 0.09 & (1) & 8.06 $\pm$ 0.13 & ( 2) & 6.84 $\pm$ 0.05 & 0.00 $\pm$ 0.00 & 7.28 $\pm$ 0.23 & 7.29 $\pm$ 0.37 &		     &      \\
NGC2004-064 & B2 III      & 7.50  & 7.56 $\pm$ 0.09 & (5) & 8.50 $\pm$ 0.16 & (29) & 7.12 $\pm$ 0.29 & 0.00 $\pm$ 0.00 & 7.47 $\pm$ 0.31 & 7.49 $\pm$ 0.55 &		     &      \\
NGC2004-070 & B0.7-B1 III & 7.59  & 7.48 $\pm$ 0.12 & (4) & 8.73 $\pm$ 0.19 & (21) & 7.23 $\pm$ 0.29 & 0.00 $\pm$ 0.00 & 7.60 $\pm$ 0.33 & 7.61 $\pm$ 0.51 &		     &      \\
NGC2004-084 & B0.7-B1 III & 7.64  & 7.28 $\pm$ 0.14 & (4) & 8.57 $\pm$ 0.16 & (26) & 7.22 $\pm$ 0.23 & 0.00 $\pm$ 0.00 & 7.56 $\pm$ 0.32 & 7.57 $\pm$ 0.48 &		     &      \\
NGC2004-090 & O9.5 III    & 7.53  & 7.54 $\pm$ 0.19 & (3) & 8.32 $\pm$ 0.23 & (24) & 6.91 $\pm$ 0.19 & 0.00 $\pm$ 0.00 & 6.94 $\pm$ 0.29 & 6.93 $\pm$ 0.39 &		     &      \\
NGC2004-091 & B1.5 III    & 7.80  & 7.28 $\pm$ 0.08 & (5) & 8.34 $\pm$ 0.13 & (27) & 7.26 $\pm$ 0.24 & 0.00 $\pm$ 0.00 & 7.16 $\pm$ 0.24 & 7.16 $\pm$ 0.51 &		     &      \\
NGC2004-108 & B2.5 III    & 7.29  & 6.89 $\pm$ 0.14 & (3) & 8.23 $\pm$ 0.30 & (28) & 6.99 $\pm$ 0.20 & 6.98 $\pm$ 0.23 & 6.98 $\pm$ 0.31 & 0.00 $\pm$ 0.00 & 7.35 $\pm$ 0.21 & (1)  \\
NGC2004-119 & B2   III	  & 7.48  & 6.95 $\pm$ 0.09 & (3) & 8.35 $\pm$ 0.27 & (29) & 7.17 $\pm$ 0.25 & 7.14 $\pm$ 0.24 & 7.14 $\pm$ 0.34 & 0.00 $\pm$ 0.00 & 7.34 $\pm$ 0.25 & (1)  \\
\\
\hline
\\
 NGC330-002 & B3   Ib	  & 7.08  & 7.58 $\pm$ 0.33 & (8) & 7.99 $\pm$ 0.52 & (14) & 6.55 $\pm$ 0.22 & 6.74 $\pm$ 0.22 & 6.78 $\pm$ 0.46 &		   & 6.70 $\pm$ 0.25 & (1)  \\
 NGC330-003 & B2   Ib	  & 7.28  & 7.74 $\pm$ 0.19 & (8) & 8.11 $\pm$ 0.36 & (24) & 6.82 $\pm$ 0.27 & 6.89 $\pm$ 0.27 & 6.95 $\pm$ 0.43 &		   & 6.90 $\pm$ 0.25 & (1)  \\
 NGC330-004 & B2.5 Ib	  & 6.84  & 7.83 $\pm$ 0.14 & (8) & 7.86 $\pm$ 0.32 & (15) & 6.73 $\pm$ 0.21 & 6.76 $\pm$ 0.21 & 6.82 $\pm$ 0.39 &		   & 6.83 $\pm$ 0.14 & (1)  \\
 NGC330-005 & B5   Ib	  & 6.77  & 7.49 $\pm$ 0.37 & (4) & 7.82 $\pm$ 0.45 & ( 2) & 6.58 $\pm$ 0.19 & 6.76 $\pm$ 0.23 & 6.81 $\pm$ 0.43 &		   & 		     &      \\
 NGC330-009 & B5   Ib	  & 6.99  & 7.14 $\pm$ 0.37 & (2) & 8.19 $\pm$ 0.52 & ( 3) & 6.57 $\pm$ 0.26 & 6.61 $\pm$ 0.22 & 6.65 $\pm$ 0.45 &		   & 		     &      \\
 NGC330-010 & B5   Ib	  & 7.07  & 7.05 $\pm$ 0.31 & (2) & 7.62 $\pm$ 0.40 & ( 3) & 6.52 $\pm$ 0.17 & 6.57 $\pm$ 0.18 & 6.65 $\pm$ 0.38 &		   & 		     &      \\
 NGC330-014 & B1.5 Ib	  & 6.93  & 7.56 $\pm$ 0.16 & (7) & 8.18 $\pm$ 0.27 & (20) & 6.72 $\pm$ 0.21 &  	       & 6.96 $\pm$ 0.33 & 6.96 $\pm$ 0.54 & 		     &      \\
 NGC330-016 & B5:  II	  & 7.24  & 7.08 $\pm$ 0.35 & (1) & 7.97 $\pm$ 0.50 & ( 2) & 6.57 $\pm$ 0.17 & 6.55 $\pm$ 0.16 & 6.61 $\pm$ 0.44 &		   & 		     &      \\
 NGC330-017 & B2   II	  & 7.07  & 7.12 $\pm$ 0.12 & (8) & 7.64 $\pm$ 0.25 & (24) & 6.47 $\pm$ 0.17 &  	       & 6.74 $\pm$ 0.33 &		   & 6.90 $\pm$ 0.21 & (1)  \\
 NGC330-018 & B3   II	  & 7.17  & 7.24 $\pm$ 0.26 & (3) & 8.05 $\pm$ 0.37 & ( 7) & 6.61 $\pm$ 0.18 & 6.74 $\pm$ 0.18 & 6.80 $\pm$ 0.36 &		   & 7.14 $\pm$ 0.21 & (1)  \\
 NGC330-020 & B3   II	  & 7.14  & 7.15 $\pm$ 0.49 & (2) & 7.99 $\pm$ 0.64 & ( 4) & 6.73 $\pm$ 0.34 & 7.05 $\pm$ 0.45 & 7.11 $\pm$ 0.60 &		   & 		     &      \\
 NGC330-022 & B3   II	  & 7.10  & 7.27 $\pm$ 0.24 & (8) & 7.76 $\pm$ 0.35 & ( 8) & 6.64 $\pm$ 0.14 & 6.53 $\pm$ 0.14 & 6.58 $\pm$ 0.34 &		   & 		     &      \\
 NGC330-026 & B2.5 II	  & 7.30  & 7.23 $\pm$ 0.19 & (3) & 7.74 $\pm$ 0.30 & ( 6) & 7.02 $\pm$ 0.23 &  	       & 6.64 $\pm$ 0.39 &		   & 		     &      \\
 NGC330-027 & B1   V	  & 7.13  & 7.45 $\pm$ 0.23 & (2) & 8.33 $\pm$ 0.45 & ( 8) & 6.44 $\pm$ 0.22 &  	       & 6.90 $\pm$ 0.39 & 6.88 $\pm$ 0.79 & 		     &      \\
 NGC330-032 & B0.5 V	  & 7.16  & 7.37 $\pm$ 0.10 & (3) & 7.90 $\pm$ 0.07 & (22) & 6.80 $\pm$ 0.20 &  	       & 6.77 $\pm$ 0.19 & 6.77 $\pm$ 0.45 & 		     &      \\
 NGC330-042 & B2   II	  & 7.06  & 7.20 $\pm$ 0.06 & (7) & 7.69 $\pm$ 0.12 & (19) & 6.86 $\pm$ 0.17 &  	       & 6.67 $\pm$ 0.28 & 6.66 $\pm$ 0.49 & 		     &      \\
 NGC330-047 & B1   V	  & 7.20  & 6.76 $\pm$ 0.11 & (2) & 8.07 $\pm$ 0.13 & (24) & 6.57 $\pm$ 0.15 &  	       & 6.83 $\pm$ 0.23 & 6.84 $\pm$ 0.46 & 		     &      \\
 NGC330-074 & B0   V	  & 7.29  & 7.52 $\pm$ 0.21 & (2) & 7.93 $\pm$ 0.17 & ( 7) & 6.83 $\pm$ 0.18 &  	       & 6.94 $\pm$ 0.24 & 6.95 $\pm$ 0.50 & 		     &      \\
 NGC330-114 & B2   III    & 7.24  & 7.34 $\pm$ 0.18 & (3) & 8.02 $\pm$ 0.28 & (18) & 6.99 $\pm$ 0.22 &  	       & 6.90 $\pm$ 0.34 &		   & 		     &      \\
 NGC330-124 & B0.2 V	  & 7.57  & 7.41 $\pm$ 0.26 & (1) & 7.81 $\pm$ 0.18 & ( 3) & 6.78 $\pm$ 0.18 &  	       & 6.98 $\pm$ 0.26 & 6.98 $\pm$ 0.48 & 		     &      \\       
\\ \hline																				     		    
\end{tabular}																				     		    
%\end{center}
\end{table}  

\end{landscape}

\begin{landscape}
\vspace{-5mm}
\begin{table}
\setcounter{table}{1}
%\begin{center}
\caption[Abundances of NGC3293, NGC4755, NGC2004 \& NGC330 stars. Fixing vt]
{\label{fixvtab}   Absolute Abundances of the NGC3293, NGC4755, NGC2004 \& NGC330 stars. The abundances presented below are the mean absolute abundances of each species
studied, using the initial atmospheric parameters from Tables~\ref{galpar} \&
~\ref{mcpar} but with the microturbulence corrected to obtain the mean Cluster silicon
abundance in each star Uncertainties on the abundances account for both random and systematic errors as
discussed in Sect.~\ref{aberror}. Abundances are presented as [X]=12 + $\log$([X/H]) in units of dex,
and \eps~in \kms. }  
\begin{tabular}{llllccccccccc} \hline \hline
Star        & Sp.Typ      & \eps$_{\rm Ave}$& \eps$_{\rm Si}$& C {\sc ii}   &N {\sc ii} & O {\sc ii} & Mg {\sc ii} & Si {\sc ii}  &Si {\sc iii}   &Si {\sc iv} &Fe {\sc iii}\\	
\hline   \\
NGC3293-003 & B1   III    & 13 & 15   & 7.87  & 7.53 $\pm$ 0.05 & 8.72 $\pm$ 0.26 &  7.33 $\pm$ 0.28 &  	       & 7.45 $\pm$ 0.38 & 7.39 $\pm$ 0.67 & 7.29 $\pm$ 0.31 \\ 
NGC3293-004 & B1   III    & 13 & 13   & 7.89  & 7.55 $\pm$ 0.09 & 8.74 $\pm$ 0.23 &  7.44 $\pm$ 0.29 &  	       & 7.45 $\pm$ 0.31 & 7.45 $\pm$ 0.62 &		     \\
NGC3293-007 & B1   III    & 11 & 12   & 7.88  & 7.51 $\pm$ 0.08 & 8.70 $\pm$ 0.21 &  7.27 $\pm$ 0.22 & 7.36 $\pm$ 0.40 & 7.43 $\pm$ 0.33 & 7.40 $\pm$ 0.61 & 7.26 $\pm$ 0.24 \\
NGC3293-010 & B1   III    & 11 & 10   & 7.65  & 7.45 $\pm$ 0.12 & 8.84 $\pm$ 0.32 &  7.15 $\pm$ 0.24 & 6.86 $\pm$ 0.26 & 7.51 $\pm$ 0.40 & 7.58 $\pm$ 0.61 & 7.30 $\pm$ 0.24 \\
NGC3293-012 & B1   III    & 11 & 10   & 7.65  & 7.49 $\pm$ 0.17 & 8.92 $\pm$ 0.35 &  7.25 $\pm$ 0.22 &  	       & 7.58 $\pm$ 0.41 & 7.64 $\pm$ 0.61 &		     \\
NGC3293-018 & B1   V      & 5  &  3   & 7.64  & 7.56 $\pm$ 0.12 & 8.74 $\pm$ 0.30 &  7.14 $\pm$ 0.19 & 6.83 $\pm$ 0.21 & 7.51 $\pm$ 0.37 & 7.58 $\pm$ 0.56 & 7.68 $\pm$ 0.27 \\
NGC3293-026 & B2   III    & 2  & $<0$ & 7.75  & 7.70 $\pm$ 0.20 & 8.85 $\pm$ 0.32 &  7.18 $\pm$ 0.28 & 6.88 $\pm$ 0.24 & 7.54 $\pm$ 0.36 & 7.67 $\pm$ 0.59 & 7.66 $\pm$ 0.19 \\
NGC3293-043 & B3   V      & $<$0&$<$ 0& 7.91  & 7.55 $\pm$ 0.26 & 8.50 $\pm$ 0.34 &  7.14 $\pm$ 0.26 & 7.31 $\pm$ 0.37 & 7.31 $\pm$ 0.34 &		   & 7.51 $\pm$ 0.25 \\
					        															     
\\					        													    
\hline  				        															     
\\					        															     
NGC4755-002 & B3   Ia     & 18 & 19   & 7.87  & 8.18 $\pm$ 0.31 & 8.59 $\pm$ 0.43 &  7.30 $\pm$ 0.32 & 7.40 $\pm$ 0.30 & 7.40 $\pm$ 0.46 &		   & 7.54 $\pm$ 0.22 \\ 
NGC4755-003 & B2   III    & 15 & 17   & 7.79  & 8.14 $\pm$ 0.28 & 8.54 $\pm$ 0.38 &  7.31 $\pm$ 0.26 & 7.39 $\pm$ 0.25 & 7.43 $\pm$ 0.43 &		   & 7.48 $\pm$ 0.18 \\
NGC4755-004 & B1.5 Ib     & 17 & 18   & 7.75  & 8.08 $\pm$ 0.15 & 8.67 $\pm$ 0.29 &  7.26 $\pm$ 0.28 & 7.35 $\pm$ 0.33 & 7.42 $\pm$ 0.39 & 7.38 $\pm$ 0.70 & 7.38 $\pm$ 0.26 \\
NGC4755-006 & B1   III    & 21 & 14   & 7.54  & 7.67 $\pm$ 0.19 & 8.95 $\pm$ 0.36 &  7.07 $\pm$ 0.20 &  	       & 7.41 $\pm$ 0.37 & 7.55 $\pm$ 0.56 &		     \\
NGC4755-015 & B1   V      &  6 &  2   & 7.58  & 7.80 $\pm$ 0.22 & 8.85 $\pm$ 0.33 &  7.02 $\pm$ 0.23 &  	       & 7.43 $\pm$ 0.36 & 7.60 $\pm$ 0.59 & 7.56 $\pm$ 0.31 \\
NGC4755-017 & B1.5 V      &  3 &  6   & 7.92  & 7.72 $\pm$ 0.37 & 8.68 $\pm$ 0.46 &  7.33 $\pm$ 0.33 & 7.45 $\pm$ 0.36 & 7.41 $\pm$ 0.44 &		   &		     \\
NGC4755-020 & B2   V      &  1 &  3   & 8.02  & 7.67 $\pm$ 0.21 & 8.49 $\pm$ 0.31 &  7.45 $\pm$ 0.29 & 7.39 $\pm$ 0.29 & 7.37 $\pm$ 0.34 &		   & 7.83 $\pm$ 0.25 \\
NGC4755-033 & B3   V      &  2 & 10   & 8.19  & 7.39 $\pm$ 0.33 & 8.63 $\pm$ 0.40 &  7.44 $\pm$ 0.43 & 8.00 $\pm$ 0.80 & 7.39 $\pm$ 0.39 &		   &		     \\
NGC4755-040 & B2.5 V      &  5 & $<0$ & 8.28  & 7.72 $\pm$ 0.36 & 9.02 $\pm$ 0.38 &  6.98 $\pm$ 0.35 & 7.03 $\pm$ 0.32 & 7.41 $\pm$ 0.33 &		   &		     \\
NGC4755-048 & B3 V        &  2 &  6   & 8.29  & 7.39 $\pm$ 0.33 & 8.93 $\pm$ 0.37 &  7.25 $\pm$ 0.34 & 7.74 $\pm$ 0.51 & 7.38 $\pm$ 0.34 &		   &		     \\
\\					        													    
\hline	        			        															     
\\					        															     
NGC2004-003 & B5   Ia     & 15 & 14   & 7.66  & 7.85 $\pm$ 0.40 & 8.45 $\pm$ 0.55 &  7.02 $\pm$ 0.26 & 7.22 $\pm$ 0.25 & 7.22 $\pm$ 0.50 &		   & 7.29 $\pm$ 0.23 \\
NGC2004-005 & B8   Ia	  &  9 & 24   & 7.79  & 7.82 $\pm$ 0.37 & 8.78 $\pm$ 0.57 &  7.17 $\pm$ 0.46 & 7.49 $\pm$ 0.36 & 7.20 $\pm$ 0.44 &		   & 7.19 $\pm$ 0.23 \\
NGC2004-007 & B8   Ia	  &  6 & 29   & 7.74  & 7.75 $\pm$ 0.28 & 8.74 $\pm$ 0.53 &  7.37 $\pm$ 0.43 & 8.02 $\pm$ 0.39 & 7.20 $\pm$ 0.44 &		   & 7.10 $\pm$ 0.26 \\
NGC2004-010 & B2.5 Iab    & 14 & 16   & 7.53  & 8.18 $\pm$ 0.29 & 8.32 $\pm$ 0.37 &  7.10 $\pm$ 0.27 & 7.19 $\pm$ 0.25 & 7.24 $\pm$ 0.43 &		   & 7.26 $\pm$ 0.12 \\
NGC2004-011 & B1.5 Ia     & 13 & 14   & 7.46  & 7.73 $\pm$ 0.07 & 8.41 $\pm$ 0.22 &  7.16 $\pm$ 0.26 &  	       & 7.18 $\pm$ 0.33 & 7.20 $\pm$ 0.64 & 7.14 $\pm$ 0.23 \\
NGC2004-012 & B1.5 Iab 	  & 12 & 12   & 7.44  & 7.75 $\pm$ 0.09 & 8.43 $\pm$ 0.24 &  7.08 $\pm$ 0.23 &  	       & 7.19 $\pm$ 0.35 & 7.21 $\pm$ 0.64 & 7.12 $\pm$ 0.19 \\
NGC2004-014 & B3   Ib	  & 11 & 14   & 7.58  & 7.54 $\pm$ 0.26 & 8.26 $\pm$ 0.39 &  7.02 $\pm$ 0.22 & 7.16 $\pm$ 0.21 & 7.19 $\pm$ 0.40 &		   & 7.18 $\pm$ 0.17 \\
NGC2004-021 & B1.5 Ib     & 13 & 12   & 7.55  & 7.20 $\pm$ 0.10 & 8.50 $\pm$ 0.27 &  7.08 $\pm$ 0.23 &  	       & 7.25 $\pm$ 0.35 & 7.28 $\pm$ 0.59 & 7.19 $\pm$ 0.18 \\
NGC2004-022 & B1.5 Ib	  & 11 & 10   & 7.31  & 7.68 $\pm$ 0.12 & 8.56 $\pm$ 0.28 &  6.94 $\pm$ 0.19 &  	       & 7.24 $\pm$ 0.34 & 7.26 $\pm$ 0.58 & 7.00 $\pm$ 0.22 \\
NGC2004-026 & B2   II	  &  0 &  1   & 7.52  & 7.02 $\pm$ 0.12 & 8.22 $\pm$ 0.28 &  7.07 $\pm$ 0.19 & 7.21 $\pm$ 0.22 & 7.29 $\pm$ 0.37 &		   & 7.35 $\pm$ 0.17 \\
NGC2004-029 & B1.5 e      &  1 &  1   & 7.43  & 6.81 $\pm$ 0.05 & 8.29 $\pm$ 0.26 &  7.03 $\pm$ 0.19 &  	       & 7.21 $\pm$ 0.35 & 7.20 $\pm$ 0.56 & 7.35 $\pm$ 0.17 \\
NGC2004-036 & B1.5 III    &  8 &  5   & 7.42  & 7.38 $\pm$ 0.09 & 8.57 $\pm$ 0.27 &  6.84 $\pm$ 0.16 &  	       & 7.22 $\pm$ 0.31 & 7.38 $\pm$ 0.58 & 7.26 $\pm$ 0.20 \\
NGC2004-042 & B2.5 III    &  2 &  3   & 7.64  & 6.91 $\pm$ 0.19 & 8.17 $\pm$ 0.32 &  7.13 $\pm$ 0.22 & 7.17 $\pm$ 0.23 & 7.20 $\pm$ 0.36 &		   & 7.21 $\pm$ 0.21 \\
NGC2004-046 & B1.5 III    &  2 & $<0$ & 7.50  & 7.57 $\pm$ 0.10 & 8.43 $\pm$ 0.14 &  7.01 $\pm$ 0.20 &  	       & 7.20 $\pm$ 0.26 & 7.26 $\pm$ 0.51 &		     \\
\\ \hline		    	 																		        
\end{tabular}		    	 
%\end{center}		    
\end{table}

\end{landscape}
\begin{landscape}
\vspace{-5mm}
\begin{table}
\setcounter{table}{1}
%\begin{center}
\caption[Abundances of NGC3293, NGC4755, NGC2004 \& NGC330 stars. Fixing vt]
{\label{fixvt_a}   Contd.}  
\begin{tabular}{llllcccccccc} \hline \hline
Star        & Sp.Typ     & \eps$_{\rm Ave}$& \eps$_{\rm Si}$ & C {\sc ii} &N {\sc ii} & O {\sc ii} & Mg {\sc ii} & Si {\sc ii}  &Si {\sc iii}   &Si {\sc iv}  &Fe {\sc iii} \\
\hline   \\
NGC2004-053 & B0.2 Ve	  & 7  & 3     & 7.89  & 7.70 $\pm$ 0.16 & 8.44 $\pm$ 0.18 & 7.25 $\pm$ 0.22 &  		    & 7.23 $\pm$ 0.26 & 6.98 $\pm$ 0.46 &		  \\
NGC2004-061 & B2   III    & 1  & $<$0  & 7.75  & 6.98 $\pm$ 0.23 & 8.28 $\pm$ 0.45 & 7.14 $\pm$ 0.28 & 7.18 $\pm$ 0.29 & 7.15 $\pm$ 0.47 &		& 7.44 $\pm$ 0.31 \\
%NGC2004-062 & B0.2 V	  & 1  & 0  & 8.07 $\pm$ 0.13 & 6.71 $\pm$ 0.09 & 8.05 $\pm$ 0.13 & 6.84 $\pm$ 0.06 &		      & 7.22 $\pm$ 0.21 & 7.19 $\pm$ 0.36 &		    \\
NGC2004-064 & B2 III	  & 6  & 3     & 7.48  & 7.51 $\pm$ 0.08 & 8.37 $\pm$ 0.13 & 7.06 $\pm$ 0.25 &		    & 7.17 $\pm$ 0.25 & 7.24 $\pm$ 0.54 &		  \\
NGC2004-070 & B0.7-B1 III & 4  & $<$0  & 7.56  & 7.42 $\pm$ 0.11 & 8.51 $\pm$ 0.20 & 7.14 $\pm$ 0.22 &  		    & 7.19 $\pm$ 0.36 & 7.26 $\pm$ 0.56 &		  \\
NGC2004-084 & B0.7-B1 III & 3  & $<$0  & 7.61  & 7.26 $\pm$ 0.14 & 8.43 $\pm$ 0.13 & 7.15 $\pm$ 0.19 &  		    & 7.24 $\pm$ 0.24 & 7.31 $\pm$ 0.47 &		  \\
NGC2004-090 & O9.5 III    & $<$0 & 3   & 7.54  & 7.57 $\pm$ 0.19 & 8.36 $\pm$ 0.23 & 6.94 $\pm$ 0.19 &  		    & 7.05 $\pm$ 0.29 & 7.24 $\pm$ 0.42 &		  \\
NGC2004-091 & B1.5 III    & 0  & 1     & 7.81  & 7.29 $\pm$ 0.09 & 8.38 $\pm$ 0.14 & 7.29 $\pm$ 0.25 &  		    & 7.26 $\pm$ 0.27 & 7.24 $\pm$ 0.52 &		  \\
NGC2004-108 & B2.5 III    & $<$0& $<$0 & 7.29  & 6.89 $\pm$ 0.14 & 8.23 $\pm$ 0.30 & 6.99 $\pm$ 0.20 & 6.98 $\pm$ 0.23 & 6.98 $\pm$ 0.31 &		& 7.35 $\pm$ 0.21 \\
NGC2004-119 & B2   III    & $<$0& $<$0 & 7.48  & 6.95 $\pm$ 0.09 & 8.35 $\pm$ 0.27 & 7.17 $\pm$ 0.25 & 7.14 $\pm$ 0.24 & 7.14 $\pm$ 0.34 &		& 7.34 $\pm$ 0.25 \\
\\			 	
\hline			 	
\\			 	
 NGC330-002 & B3   Ib	 & 18 & 20  & 7.09  & 7.60 $\pm$ 0.32 & 8.00 $\pm$ 0.52 & 6.57 $\pm$ 0.23 & 6.77 $\pm$ 0.23 & 6.81 $\pm$ 0.47 & 		& 6.70 $\pm$ 0.24 \\
 NGC330-003 & B2   Ib	 & 19 & 15  & 7.25  & 7.69 $\pm$ 0.19 & 8.07 $\pm$ 0.34 & 6.80 $\pm$ 0.27 & 6.85 $\pm$ 0.26 & 6.82 $\pm$ 0.39 & 		& 6.88 $\pm$ 0.22 \\
 NGC330-004 & B2.5 Ib	 & 16 & 16  & 6.84  & 7.83 $\pm$ 0.14 & 7.86 $\pm$ 0.32 & 6.73 $\pm$ 0.21 & 6.76 $\pm$ 0.21 & 6.82 $\pm$ 0.39 & 		& 6.83 $\pm$ 0.14 \\
 NGC330-005 & B5   Ib	 &  8 &  8  & 6.77  & 7.49 $\pm$ 0.37 & 7.82 $\pm$ 0.45 & 6.58 $\pm$ 0.19 & 6.76 $\pm$ 0.23 & 6.81 $\pm$ 0.43 & 		&		  \\
 NGC330-009 & B5   Ib	 &  6 & 10  & 7.04  & 7.18 $\pm$ 0.41 & 8.27 $\pm$ 0.53 & 6.67 $\pm$ 0.31 & 6.81 $\pm$ 0.32 & 6.78 $\pm$ 0.49 & 		&		  \\
 NGC330-010 & B5   Ib	 &  4 &  9  & 7.15  & 7.11 $\pm$ 0.33 & 7.69 $\pm$ 0.41 & 6.64 $\pm$ 0.20 & 6.82 $\pm$ 0.31 & 6.82 $\pm$ 0.42 & 		&		  \\
 NGC330-014 & B1.5 Ib	 & 19 & 15  & 6.93  & 7.53 $\pm$ 0.15 & 8.12 $\pm$ 0.24 & 6.70 $\pm$ 0.21 &		    & 6.82 $\pm$ 0.29 & 6.89 $\pm$ 0.51 &		  \\
 NGC330-016 & B5:  II	 &  4 & 10  & 7.36  & 7.17 $\pm$ 0.38 & 8.06 $\pm$ 0.52 & 6.77 $\pm$ 0.25 & 6.91 $\pm$ 0.38 & 6.82 $\pm$ 0.48 & 		&		  \\
 NGC330-017 & B2   II	 &$<$0 &$<$0& 7.07  & 7.12 $\pm$ 0.12 & 7.64 $\pm$ 0.25 & 6.47 $\pm$ 0.17 &		    & 6.74 $\pm$ 0.33 & 		& 6.90 $\pm$ 0.21 \\
 NGC330-018 & B3   II	 &  5 &  5  & 7.17  & 7.24 $\pm$ 0.26 & 8.05 $\pm$ 0.37 & 6.61 $\pm$ 0.18 & 6.74 $\pm$ 0.18 & 6.80 $\pm$ 0.36 & 		& 7.14 $\pm$ 0.19 \\
 NGC330-020 & B3   II	 &  8 &  2  & 7.06  & 7.06 $\pm$ 0.44 & 7.87 $\pm$ 0.61 & 6.55 $\pm$ 0.26 & 6.66 $\pm$ 0.27 & 6.81 $\pm$ 0.55 & 		&		  \\
 NGC330-022 & B3   II	 &  3 &  7  & 7.13  & 7.33 $\pm$ 0.26 & 7.84 $\pm$ 0.37 & 6.73 $\pm$ 0.16 & 6.66 $\pm$ 0.17 & 6.80 $\pm$ 0.37 & 		&		  \\
 NGC330-026 & B2.5 II	 &$<$0&$<$0 & 7.30  & 7.23 $\pm$ 0.19 & 7.74 $\pm$ 0.30 & 7.02 $\pm$ 0.23 &		    & 6.64 $\pm$ 0.39 & 		&		  \\
 NGC330-027 & B1   V	 &  7 &  6  & 7.13  & 7.40 $\pm$ 0.22 & 8.28 $\pm$ 0.44 & 6.44 $\pm$ 0.22 &		    & 6.82 $\pm$ 0.36 & 6.84 $\pm$ 0.78 &		  \\
 NGC330-032 & B0.5 V	 &$<$0&$<$ 0& 7.16  & 7.37 $\pm$ 0.10 & 7.90 $\pm$ 0.07 & 6.80 $\pm$ 0.20 &		    & 6.77 $\pm$ 0.19 & 6.77 $\pm$ 0.50 &		  \\
 NGC330-042 & B2   II	 &  1 &  3  & 7.06  & 7.22 $\pm$ 0.06 & 7.73 $\pm$ 0.13 & 6.89 $\pm$ 0.18 &		    & 6.83 $\pm$ 0.34 & 6.76 $\pm$ 0.52 &		  \\
 NGC330-047 & B1   V	 &  0 &  0  & 7.20  & 6.76 $\pm$ 0.11 & 8.07 $\pm$ 0.13 & 6.57 $\pm$ 0.15 &		    & 6.83 $\pm$ 0.23 & 6.84 $\pm$ 0.46 &		  \\
 NGC330-074 & B0   V	 &  6 &  2  & 7.31  & 7.49 $\pm$ 0.21 & 7.88 $\pm$ 0.17 & 6.81 $\pm$ 0.19 &		    & 6.81 $\pm$ 0.25 & 6.62 $\pm$ 0.39 &		  \\
 NGC330-114 & B2   III   &  4 &  3  & 7.23  & 7.32 $\pm$ 0.18 & 8.00 $\pm$ 0.27 & 6.97 $\pm$ 0.22 &		    & 6.83 $\pm$ 0.32 & 		&		  \\
 NGC330-124 & B0.2 V	 &  3 &$<$0 & 7.57  & 7.37 $\pm$ 0.25 & 7.76 $\pm$ 0.18 & 6.75 $\pm$ 0.18 &		    & 6.82 $\pm$ 0.23 & 6.73 $\pm$ 0.45 &		  \\
\\ \hline
\end{tabular}
%\end{center}
\end{table}  

\end{landscape}
\end{appendix}

\begin{appendix}%fe from young cluster objects
\begin{table*}
\begin{center}
\caption[Fe abundances of young cluster objects]
{\label{youngfe} Iron equivalent widths and absolute abundances for the objects analysed in Paper IV.
The last column contains the weighted mean abundance and uncertainties of the iron abundance in each
star.  }  
\begin{tabular}{lccccccc} \hline \hline
ID           & \multicolumn{2}{c}{Fe \3 4419 \AA} & \multicolumn{2}{c}{Fe \3 4431 \AA} &          \\
             &  EW(m\AA)          & Abund.                &  EW(m\AA)     & Abund      &   [Fe/H]	 \\
 \hline \\
NGC6611-030  &   24               &	     7.55	  &	19 	  &	7.69   & 7.62 $\pm$ 0.15	 \\
N11-001      &	92		  &		7.59	  &	34  	  &	7.30   & 7.45 $\pm$ 0.17     \\
N11-002      &  55		  &		7.27	  &	41	  &	7.41   & 7.34 $\pm$ 0.15     \\
N11-009      &  38		  &		7.20	  &	26	  &	7.29   & 7.24 $\pm$ 0.13     \\
N11-012      &  42		  &		7.20	  &	14	  &	7.00   & 7.10 $\pm$ 0.27     \\
N11-014      &  52		  &		7.25	  &	28	  &	7.22   & 7.23 $\pm$ 0.17    \\
N11-015      &  23		  &		7.21	  &		  &	       & 7.21 $\pm$ 0.26    \\
N11-016      &  44		  &		7.26	  &		  &	       & 7.26 $\pm$ 0.28    \\
N11-017      &  47		  &		7.21	  &	26	  &	7.20   & 7.20 $\pm$ 0.15    \\
N11-024      &  36		  &		7.15	  &	16	  &	7.07   & 7.11 $\pm$ 0.16    \\
N11-036      &  23		  &		7.17	  &		  &	       & 7.17 $\pm$ 0.23    \\
N11-110      &  27		  &		7.17	  &	12	  &	7.08   & 7.12 $\pm$ 0.16   \\
NGC346-021   &  17		  &		7.14      &		  &	       & 7.14 $\pm$ 0.18    \\
NGC346-037   &  11		  &		6.83	  &		  &	       & 6.83 $\pm$ 0.21    \\
NGC346-039   &  10		  &		6.94	  &		  &	       & 6.94 $\pm$ 0.17    \\
NGC346-044   &  18		  &		6.99	  &		  &	       & 6.99 $\pm$ 0.18    \\

\\ \hline
\end{tabular}
\end{center}
\end{table*}  

\end{appendix}

\begin{appendix}%Abundance and equivalent width tables
\begin{table*}
\setcounter{table}{0}
\begin{center}
\caption[EW's \& Abundances of NGC3293. ]
{\label{ngc3293a}   Equivalent widths and line by line absolute abundances of the NGC3293 stars:\#3, \#4, \#7. The first and second columns
represent the ion and wavelength of the line. There are four columns per star the first represents the equivalent width of the line,
the following three are abundance  estimates from the lines and they show the results from the 3 steps in determining 
the abundances as described in Sect.~\ref{par}. The energy levels and oscillator strenghts relating to the
transitions for the metal lines are available online at
http://star.pst.qub.ac.uk/~pld/line\_identifications.html.}  
\begin{tabular}{llcccccccccccc} \hline \hline\\
          &          & \multicolumn{4}{c}{NGC3293-003} & \multicolumn{4}{c}{NGC3293-004} & \multicolumn{4}{c}{NGC3293-007}  \\
ION       & $\lambda$& EW (m\AA) & (1) & (2) & (3) & EW (m\AA) &  (1) & (2) & (3) & EW (m\AA) & (1) & (2) & (3) \\
 \hline
\\
C  II     & 3920.68 & 146 & 8.03 & 8.07 & 8.03 &     &      &      &      & 110 & 8.03 & 8.04 & 8.03 \\
C  II(B)  & 4267.00 & 237 & 7.83 & 7.87 & 7.84 & 214 & 7.89 & 7.89 & 7.89 & 200 & 7.86 & 7.88 & 7.86 \\
C  II     & 6578.05 &     &      &      &      &     &      &      &      &	&      &      &      \\
C  II     & 6582.88 &     &      &      &      &     &      &      &      &	&      &      &      \\
\hline
\end{tabular}
\end{center}
\end{table*}  

\begin{table*}
\setcounter{table}{0}
\begin{center}
\caption[EW's \& Abundances of NGC3293. ]
{\label{ngc3293_b}   Contd. STARS:\#10, \#12, \#18}  
\begin{tabular}{llcccccccccccc} \hline \hline\\
          &          & \multicolumn{4}{c}{NGC3293-010} & \multicolumn{4}{c}{NGC3293-012} & \multicolumn{4}{c}{NGC3293-018}  \\
ION       & $\lambda$& EW (m\AA) & (1) & (2) & (3) & EW (m\AA) &  (1) & (2) & (3) & EW (m\AA) & (1) & (2) & (3) \\
\\
\hline
\\
C  II     & 3920.68 &  97 & 7.71 & 7.70 & 7.71 &     &      &	   &	  &  72 & 7.80 & 7.74 & 7.77 \\
C  II(B)  & 4267.00 & 194 & 7.66 & 7.65 & 7.66 & 201 & 7.67 & 7.65 & 7.67 & 155 & 7.67 & 7.64 & 7.66 \\
C  II     & 6578.05 &     &	 &	&      &     &      &	   &	  &	&      &      &      \\
\hline
\end{tabular}
\end{center}
\end{table*}  

\begin{table*}
\setcounter{table}{0}
\begin{center}
\caption[EW's \& Abundances of NGC3293.]
{\label{ngc3293c}   Contd. STARS:\#26, \#43}  
\begin{tabular}{llcccccccc} \hline \hline\\
          &          & \multicolumn{4}{c}{NGC3293-026} & \multicolumn{4}{c}{NGC3293-043}	  \\ 
ION       & $\lambda$& EW (m\AA) & (1) & (2) & (3) & EW (m\AA) &  (1) & (2) & (3)  \\
\\
\hline
\\
C  II     & 3920.68 &  90 & 8.06 & 7.94 & 7.97 &   74 & 8.02 & 8.02  & 8.02 \\
C  II(B)  & 4267.00 & 181 & 7.81 & 7.75 & 7.79 &  169 & 7.91 & 7.91  & 7.91 \\
C  II     & 6578.05 & 131 & 8.01 & 7.90 & 7.93 &  104 & 8.14 & 8.14  & 8.14 \\
\hline
\end{tabular}
\end{center}
\end{table*}  

\begin{table*}
\setcounter{table}{1}
\begin{center}
\caption[EW's \& Abundances of NGC4755. ]
{\label{ngc4755a}   Equivalent widths and line by line absolute abundances of NGC4755 for stars:\#2, \#3, \#4. 
Columns are as for Table~\ref{ngc3293a}.}  
\begin{tabular}{llcccccccccccc} \hline \hline\\
          &          & \multicolumn{4}{c}{NGC4755-002} & \multicolumn{4}{c}{NGC4755-003} & \multicolumn{4}{c}{NGC4755-004}  \\
ION       & $\lambda$& EW (m\AA) & (1) & (2) & (3) & EW (m\AA) &  (1) & (2) & (3) & EW (m\AA) & (1) & (2) & (3) \\
\\
\hline
\\
C  II     & 3920.68 & 164 & 7.86 & 7.88 & 7.88 & 172 & 7.88 & 7.93 & 7.93 & 115 & 7.71 & 7.72 & 7.74 \\ 
C  II(B)  & 4267.00 & 289 & 7.84 & 7.87 & 7.87 & 274 & 7.74 & 7.79 & 7.78 & 240 & 7.73 & 7.75 & 7.76 \\ 
C  II     & 6578.05 &     &      &      &      &     &      &      &      &     &      &      &      \\
\hline      
\end{tabular}
\end{center}
\end{table*} 

\begin{table*}
\setcounter{table}{1}
\begin{center}
\caption[EW's \& Abundances of NGC4755. ]
{\label{ngc4755b}   Contd. STARS:\#6, \#15, \#17}  
\begin{tabular}{llcccccccccccc} \hline \hline\\
          &          & \multicolumn{4}{c}{NGC4755-006} & \multicolumn{4}{c}{NGC4755-015} & \multicolumn{4}{c}{NGC4755-017}  \\
ION       & $\lambda$& EW (m\AA) & (1) & (2) & (3) & EW (m\AA) &  (1) & (2) & (3) & EW (m\AA) & (1) & (2) & (3) \\
\\
\hline
\\
C  II     & 3920.68 & 286 & 8.59 & 8.27 & 8.41 &  63 & 7.59 & 7.50 & 7.57 &  95 & 7.86 & 8.01 & 8.02 \\ 
C  II(B)  & 4267.00 & 223 & 7.64 & 7.54 & 7.60 & 165 & 7.66 & 7.58 & 7.64 & 203 & 7.83 & 7.92 & 7.93 \\ 
C  II     & 6578.05 &	  &	 &	&      &     &      &	   &	  &	&      &      &      \\
\hline    
\end{tabular}
\end{center}													
\end{table*} 

\begin{table*}
\setcounter{table}{1}
\begin{center}
\caption[EW's \& Abundances of NGC4755. ]
{\label{ngc4755c}   Contd. STARS:\#20, \#33, \#40}  
\begin{tabular}{llcccccccccccc} \hline \hline\\
          &          & \multicolumn{4}{c}{NGC4755-020} & \multicolumn{4}{c}{NGC4755-033} & \multicolumn{4}{c}{NGC4755-040}  \\
ION       & $\lambda$& EW (m\AA) & (1) & (2) & (3) & EW (m\AA) &  (1) & (2) & (3) & EW (m\AA) & (1) & (2) & (3) \\
\hline
\\
C  II     & 3920.68 & 96 & 7.99 & 8.11 & 8.11 &  76 & 7.75 & 8.10 & 8.01 &  99 & 8.58 & 8.26 & 8.38 \\ 
C  II(B)  & 4267.00 &213 & 7.95 & 8.02 & 8.01 & 190 & 7.92 & 8.19 & 8.17 & 219 & 8.45 & 8.28 & 8.31 \\ 
C  II     & 6578.05 &155 & 8.14 & 8.28 & 8.28 & 111 & 8.02 & 8.29 & 8.26 & 144 & 8.80 & 8.45 & 8.57 \\ 
C  II     & 6582.88 &137 & 8.24 & 8.39 & 8.39 &  98 & 8.10 & 8.40 & 8.35 & 138 & 9.03 & 8.64 & 8.79 \\
\hline      
\end{tabular}
\end{center}
\end{table*} 

\begin{table*}
\setcounter{table}{1}
\begin{center}
\caption[EW's \& Abundances of NGC4755. ]
{\label{ngc4755d}   Contd. STARS:\#48}  
\begin{tabular}{llcccc} \hline \hline\\
          &          & \multicolumn{4}{c}{NGC4755-048}   \\
ION       & $\lambda$& EW (m\AA) & (1) & (2) & (3)  \\
\\
\hline
\\
C  II     & 3920.68   &  90 & 8.07 & 8.29 & 8.25\\ 
C  II(B)  & 4267.00   & 207 & 8.15 & 8.29 & 8.30\\ 
C  II     & 6578.05   & 109 & 8.12 & 8.27 & 8.27\\
\hline      
\end{tabular}
\end{center}
\end{table*} 

\begin{table*}
\setcounter{table}{2}
\begin{center}
\caption[EW's \& Abundances of NGC2004. ]
{\label{ngc2004a}   Equivalent widths and line by line absolute abundances of NGC2004 for stars \#3, \#5, \& \#7.
Columns are as for Table~\ref{ngc3293a}.}  
\begin{tabular}{llcccccccccccc} \hline \hline\\
          &          & \multicolumn{4}{c}{NGC2004-003} & \multicolumn{4}{c}{NGC2004-005} &	  \multicolumn{4}{c}{NGC2004-007}  \\
ION       & $\lambda$& EW (m\AA) & \multicolumn{3}{c}{Abundance}& EW (m\AA) & \multicolumn{3}{c}{Abundance}& EW (m\AA) & \multicolumn{3}{c}{Abundance}\\
\\
\hline
\\
C  II     & 3920.68 & 90 & 7.68 & 7.66 & 7.66 &  49 & 7.48 & 7.70 & 7.68 &  42 & 7.40 & 7.74 & 7.69\\
C  II(B)  & 4267.00 &173 & 7.68 & 7.66 & 7.66 & 105 & 7.50 & 7.79 & 7.77 &  86 & 7.36 & 7.74 & 7.73\\
C  II     & 6578.05 &    &      &      &      &     &      &      &      &     &      &      &     \\
\end{tabular}
\end{center}
\end{table*}  

\begin{table*}
\setcounter{table}{2}
\begin{center}
\caption[EW's \& Abundances of NGC2004. ]
{\label{ngc2004b}   Contd. STARS:\#10, \#11, \#12}  
\begin{tabular}{llcccccccccccc} \hline \hline\\
          &          & \multicolumn{4}{c}{NGC2004-010} & \multicolumn{4}{c}{NGC2004-011} & \multicolumn{4}{c}{NGC2004-012}  \\
ION       & $\lambda$& EW (m\AA) & (1) & (2) & (3) & EW (m\AA) &  (1) & (2) & (3) & EW (m\AA) & (1) & (2) & (3) \\
\\
\hline
\\
C  II     & 3920.68 & 91 & 7.46 & 7.48 & 7.48 &  51 & 7.46 & 7.46 & 7.45 &    &      &      &       \\
C  II(B)  & 4267.00 &194 & 7.50 & 7.53 & 7.53 & 130 & 7.45 & 7.46 & 7.45 & 134& 7.44 & 7.44 & 7.44  \\
C  II     & 6578.05 &    &      &      &      &     &      &      &      &  75& 7.46 & 7.46 & 7.46  \\
\hline									
\end{tabular}
\end{center}
\end{table*}  

\begin{table*}
\setcounter{table}{2}
\begin{center}
\caption[EW's \& Abundances of NGC2004. ]
{\label{ngc2004c}   Contd. STARS:\#14, \#21, \#22}  
\begin{tabular}{llcccccccccccc} \hline \hline\\
          &          & \multicolumn{4}{c}{NGC2004-014} & \multicolumn{4}{c}{NGC2004-021} & \multicolumn{4}{c}{NGC2004-022}  \\
ION       & $\lambda$& EW (m\AA) & (1) & (2) & (3) & EW (m\AA) &  (1) & (2) & (3) & EW (m\AA) & (1) & (2) & (3) \\
\\
\hline
\\
C  II     & 3920.68 &  98 & 7.55 & 7.60 & 7.61 &  83 & 7.63 & 7.62 & 7.62 &  46 & 7.29 & 7.29 & 7.30  \\
C  II(B)  & 4267.00 & 191 & 7.52 & 7.58 & 7.58 & 167 & 7.56 & 7.55 & 7.55 & 119 & 7.31 & 7.31 & 7.32  \\
C  II     & 6578.05 & 273 & 8.06 & 8.16 & 8.17 &     &      &      &      &  61 & 7.31 & 7.31 & 7.31  \\
\hline											
\end{tabular}
\end{center}
\end{table*}  

\begin{table*}
\setcounter{table}{2}
\begin{center}
\caption[EW's \& Abundances of NGC2004. ]
{\label{ngc2004d}   Contd. STARS:\#26, \#29, \#36}  
\begin{tabular}{llcccccccccccc} \hline \hline\\
          &          & \multicolumn{4}{c}{NGC2004-026} & \multicolumn{4}{c}{NGC2004-029} &	  \multicolumn{4}{c}{NGC2004-036}  \\
ION       & $\lambda$& EW (m\AA) & (1) & (2) & (3) & EW (m\AA) &  (1) & (2) & (3) & EW (m\AA) & (1) & (2) & (3) \\
\\
\hline
\\
C  II     & 3920.68 &  64 & 7.70 & 7.74 & 7.75&  56 & 7.64 & 7.64 & 7.64 &  56 & 7.49 & 7.45 & 7.54  \\
C  II(B)  & 4267.00 & 131 & 7.51 & 7.52 & 7.53& 112 & 7.43 & 7.43 & 7.43 & 131 & 7.46 & 7.42 & 7.51  \\
C  II     & 6578.05 & 124 & 7.91 & 7.97 & 7.98&  85 & 7.59 & 7.59 & 7.59 &  95 & 7.54 & 7.49 & 7.56  \\
\end{tabular}
\end{center}
\end{table*}  

\clearpage
\begin{table*}
\setcounter{table}{2}
\begin{center}
\caption[EW's \& Abundances of NGC2004. ]
{\label{ngc2004e}   Contd. STARS:\#42, \#46, \#53}  
\begin{tabular}{llcccccccccccc} \hline \hline\\
          &          & \multicolumn{4}{c}{NGC2004-042} & \multicolumn{4}{c}{NGC2004-046} &	  \multicolumn{4}{c}{NGC2004-053}  \\
ION       & $\lambda$& EW (m\AA) & (1) & (2) & (3) & EW (m\AA) &  (1) & (2) & (3) & EW (m\AA) & (1) & (2) & (3) \\
\\
\hline
\\
C  II     & 3920.68 &  82 & 7.77 & 7.81 & 7.81 &  29 & 7.45 & 7.44 & 7.47 &  17 & 7.93 & 7.93 & 7.84  \\
C  II(B)  & 4267.00 & 165 & 7.62 & 7.64 & 7.64 &  95 & 7.52 & 7.50 & 7.52 &  76 & 7.93 & 7.89 & 7.83  \\
C  II     & 6578.05 & 147 & 7.91 & 7.96 & 7.96 &  62 & 7.58 & 7.54 & 7.56 &  37 & 7.81 & 7.80 & 7.71  \\
\end{tabular}
\end{center}
\end{table*}  

\begin{table*}
\setcounter{table}{2}
\begin{center}
\caption[EW's \& Abundances of NGC2004. ]
{\label{ngc2004f}   Contd. STARS:\#61, \#64, \#70}  
\begin{tabular}{llcccccccccccc} \hline \hline\\
          &          & \multicolumn{4}{c}{NGC2004-061} & \multicolumn{4}{c}{NGC2004-064} &	  \multicolumn{4}{c}{NGC2004-070}  \\
ION       & $\lambda$& EW (m\AA) & (1) & (2) & (3) & EW (m\AA) &  (1) & (2) & (3) & EW (m\AA) & (1) & (2) & (3) \\
\\
\hline
\\
C  II     & 3920.68 &  84 & 7.97 & 7.90 & 7.90 &     &      &      &      &     &      &      &       \\
C  II(B)  & 4267.00 & 175 & 7.78 & 7.75 & 7.75 &  94 & 7.50 & 7.48 & 7.49 &  88 & 7.59 & 7.56 & 7.58  \\
C  II     & 6578.05 & 138 & 7.98 & 7.92 & 7.92 &  47 & 7.37 & 7.35 & 7.36 &  67 & 7.75 & 7.67 & 7.68  \\
\end{tabular}
\end{center}
\end{table*}  

\begin{table*}
\setcounter{table}{2}
\begin{center}
\caption[EW's \& Abundances of NGC2004. ]
{\label{ngc2004g}   Contd. STARS:\#84, \#90, \#91}  
\begin{tabular}{llcccccccccccc} \hline \hline\\
          &          & \multicolumn{4}{c}{NGC2004-084} & \multicolumn{4}{c}{NGC2004-090} &	  \multicolumn{4}{c}{NGC2004-091}  \\
ION       & $\lambda$& EW (m\AA) & (1) & (2) & (3) & EW (m\AA) &  (1) & (2) & (3) & EW (m\AA) & (1) & (2) & (3) \\
\\
\hline
\\
C  II     & 3920.68 &  36 & 7.71 & 7.68 & 7.71 &     &     &      &      &  60 & 8.02& 8.04& 8.04  \\
C  II(B)  & 4267.00 &  96 & 7.64 & 7.61 & 7.64 &  41 &7.53 & 7.54 & 7.63 & 125 & 7.80& 7.81& 7.80  \\
C  II     & 6578.05 &  76 & 7.85 & 7.77 & 7.80 &     &     &      &      & 100 & 8.04& 8.09& 8.08  \\
\hline	
\end{tabular}
\end{center}
\end{table*}  

\begin{table*}
\setcounter{table}{2}
\begin{center}
\caption[EW's \& Abundances of NGC2004. ]
{\label{ngc2004h}   Contd. STARS:\#108, \#119}  
\begin{tabular}{llcccccccc} \hline \hline\\
          &          & \multicolumn{4}{c}{NGC2004-108} & \multicolumn{4}{c}{NGC2004-119}  \\
ION       & $\lambda$& EW (m\AA) & (1) & (2) & (3) & EW (m\AA) &  (1) & (2) & (3)  \\
\\
\hline
\\
C  II     & 3920.68 &  50 & 7.50 & 7.50 & 7.50 &  58 & 7.67 & 7.67 & 7.67  \\
C  II(B)  & 4267.00 & 103 & 7.29 & 7.29 & 7.29 & 123 & 7.48 & 7.48 & 7.48  \\
C  II     & 6578.05 &  86 & 7.63 & 7.63 & 7.63 &  95 & 7.71 & 7.71 & 7.71  \\
\end{tabular}
\end{center}
\end{table*}  

\begin{table*}
\setcounter{table}{3}
\begin{center}
\caption[EW's \& Abundances of NGC330. ]
{\label{ngc330a}      Equivalent Widths and line by line absolute abundances of NGC330 for stars \#2, \#3, \& \#4.
Columns are as for Table~\ref{ngc3293a}.}
\begin{tabular}{llcccccccccccc} 
\hline 
\hline\\
          &          & \multicolumn{4}{c}{NGC330-002} & \multicolumn{4}{c}{NGC330-003} & \multicolumn{4}{c}{NGC330-004}  \\
ION       & $\lambda$& EW (m\AA) & (1) & (2) & (3) & EW (m\AA) &  (1) & (2) & (3) & EW (m\AA) & (1) & (2) & (3) \\
\\
\hline
\\
C  II     & 3920.68 &  33 & 7.01 & 7.02& 7.02 &  75 & 7.36 & 7.33 & 7.32 & 31& 6.87 & 6.87 & 6.87 \\   
C  II(B)  & 4267.00 &  82 & 7.08 & 7.09& 7.09 & 134 & 7.28 & 7.25 & 7.25 & 62& 6.84 & 6.84 & 6.84 \\ 
C  II     & 6578.05 &     &      &     &      &     &      &      &      &   &      &      &      \\
\hline																															   
\end{tabular}
\end{center}
\end{table*} 

\begin{table*}
\setcounter{table}{3}
\begin{center}
\caption[EW's \& Abundances of NGC330. ]
{\label{ngc330b}   Contd. STARS:\#5, \#9, \#10 }  
\begin{tabular}{llccccccccccccc} \hline \hline\\
          &          & \multicolumn{4}{c}{NGC330-005} & \multicolumn{4}{c}{NGC330-009} & \multicolumn{4}{c}{NGC330-010}  \\
ION       & $\lambda$& EW (m\AA) & (1) & (2) & (3) & EW (m\AA) &  (1) & (2) & (3) & EW (m\AA) & (1) & (2) & (3) \\
\\
\hline
\\
C  II     & 3920.68 & 17 & 6.88 & 6.88 & 6.88& 27 & 7.14 & 7.22 & 7.24 &  29 & 7.09& 7.20 & 7.21 \\   
C  II(B)  & 4267.00 & 33 & 6.77 & 6.77 & 6.77& 48 & 6.99 & 7.04 & 7.05 &  64 & 7.07& 7.15 & 7.17 \\ 
C  II     & 6578.05 &    &     &      &     &    &     &      &      &     &    &      &      \\
\hline				     																											   
\end{tabular}
\end{center}
\end{table*} 

\begin{table*}
\setcounter{table}{3}
\begin{center}
\caption[EW's \& Abundances of NGC330. ]
{\label{ngc330c}   Contd. STARS:\#14, \#16, \#17  }  
\begin{tabular}{llccccccccccccc} \hline \hline\\
          &          & \multicolumn{4}{c}{NGC330-014} & \multicolumn{4}{c}{NGC330-016} & \multicolumn{4}{c}{NGC330-017}  \\
ION       & $\lambda$& EW (m\AA) & (1) & (2) & (3) & EW (m\AA) &  (1) & (2) & (3) & EW (m\AA) & (1) & (2) & (3) \\
\\
\hline
\\
C  II     & 3920.68 &      &       &     &      & 22 & 7.01 & 7.12 & 7.12 &  34 & 7.20 & 7.20 &  7.20 \\   
C  II(B)  & 4267.00 &   67 & 6.93  & 6.93& 6.94 & 69 & 7.24 & 7.36 & 7.36 &  75 & 7.07 & 7.07 &  7.07 \\ 
C  II     & 6578.05 &      &       &     &      &    &      &      &      &  59 & 7.24 & 7.24 &  7.24 \\
\hline																															   
\end{tabular}
\end{center}
\end{table*} 

\begin{table*}
\setcounter{table}{3}
\begin{center}
\caption[EW's \& Abundances of NGC330. ]
{\label{ngc330d}   Contd. STARS:\#18, \#20, \#22 }  
\begin{tabular}{llccccccccccccc} \hline \hline\\
          &          & \multicolumn{4}{c}{NGC330-018} & \multicolumn{4}{c}{NGC330-020} & \multicolumn{4}{c}{NGC330-022}  \\
ION       & $\lambda$& EW (m\AA) & (1) & (2) & (3) & EW (m\AA) &  (1) & (2) & (3) & EW (m\AA) & (1) & (2) & (3) \\
\\
\hline
\\
C  II     & 3920.68 &  57 & 7.37 & 7.37 & 7.37 &  27 & 7.04 & 6.95 & 6.95 &  37 & 7.03 & 7.09 & 7.12 \\   
C  II(B)  & 4267.00 & 103 & 7.17 & 7.17 & 7.17 &  77 & 7.14 & 7.06 & 7.05 &  95 & 7.10 & 7.13 & 7.14 \\ 
C  II     & 6578.05 &     &      &      &      &     &      &      &      &     &      &      &      \\
\hline																															   
\end{tabular}
\end{center}
\end{table*} 

\begin{table*}
\setcounter{table}{3}
\begin{center}
\caption[EW's \& Abundances of NGC330. ]
{\label{ngc330e}   Contd. STARS:\#26, \#27, \#32    }  
\begin{tabular}{llccccccccccccc} \hline \hline\\
          &          & \multicolumn{4}{c}{NGC330-026} & \multicolumn{4}{c}{NGC330-027} & \multicolumn{4}{c}{NGC330-032}  \\
ION       & $\lambda$& EW (m\AA) & (1) & (2) & (3) & EW (m\AA) &  (1) & (2) & (3) & EW (m\AA) & (1) & (2) & (3) \\
\\
\hline
\\
C  II     & 3920.68 &  39 & 7.33& 7.33 & 7.33 &     &	   &	  &	 &    &      &      &	   \\	
C  II(B)  & 4267.00 &  97 & 7.3 & 7.30 & 7.30 &   82& 7.13 & 7.13 & 7.13 &  34& 7.16 & 7.16 & 7.16 \\ 
C  II     & 6578.05 &  80 & 7.47& 7.47 & 7.47 &     &	   &	  &	 &    &      &      &	   \\
\hline																															   
\end{tabular}
\end{center}
\end{table*} 

\begin{table*}
\setcounter{table}{3}
\begin{center}
\caption[EW's \& Abundances of NGC330. ]
{\label{ngc330f}   Contd. STARS:\#42, \#47, \#74  }  
\begin{tabular}{llccccccccccccc} \hline \hline\\
          &          & \multicolumn{4}{c}{NGC330-042} & \multicolumn{4}{c}{NGC330-047} & \multicolumn{4}{c}{NGC330-074}  \\
ION       & $\lambda$& EW (m\AA) & (1) & (2) & (3) & EW (m\AA) &  (1) & (2) & (3) & EW (m\AA) & (1) & (2) & (3) \\
\\
\hline
\\
C  II     & 3920.68 &     &     &      &      &     &       &      &      &    &      &      &     \\	
C  II(B)  & 4267.00 &  48 & 7.06& 7.06 & 7.04 &   54&  7.20 & 7.20 & 7.20 &  23&  7.29& 7.31 & 7.26\\ 
C  II     & 6578.05 &  31 & 7.09& 7.09 & 7.08 &     &       &      &      &    &      &      &     \\
\hline																															   
\end{tabular}
\end{center}
\end{table*} 

\begin{table*}
\setcounter{table}{3}
\begin{center}
\caption[EW's \& Abundances of NGC330. ]
{\label{ngc330g}   Contd. STARS:\#114, \#124  }  
\begin{tabular}{llccccccccc} \hline \hline\\
          &          & \multicolumn{4}{c}{NGC330-114} & \multicolumn{4}{c}{NGC330-124}   \\
ION       & $\lambda$& EW (m\AA) & (1) & (2) & (3) & EW (m\AA) &  (1) & (2) & (3)  \\
\\
\hline
\\
C  II     & 3920.68 & 39&7.37&7.36 & 7.36 &   &    &      &     \\   
C  II(B)  & 4267.00 & 83&7.24&7.23 & 7.23 & 54&7.57& 7.57 & 7.55\\ 
C  II     & 6578.05 &   &    &     &      &   &    &      &     \\
\hline																															   
\end{tabular}
\end{center}
\end{table*}

%\begin{figure*}
%\begin{center}
%\epsfig{file=figures/ngc2004_ONvsN_LE25.ps, width=75mm, angle=0}
%\epsfig{file=figures/ngc2004_OvN_LE25.ps, width=75mm, angle=0}
%\epsfig{file=figures/ngc2004_CNvsN_LE25.eps, width=130mm, angle=0}
%\epsfig{file=figures/ngc2004_CvN_LE25.ps, width=130mm, angle=0}
%\caption[]
%{{\bf Upper Panel:} [C/N] as a function of [N/H]. {\bf Lower Panel: }[C/H] as a function of 
%[N/H]. Note that the correlation in [C/N] is due to the increase in nitrogen in
%the stars, as can be understood from the plot of [C/H] against [N/H].
%}
%\label{cnoplots}
%\end{center}
%\end{figure*}

\end{appendix}

%------------------------------------------------------------------------------
%Additional tables and plots of interest but not included in paper
%\begin{appendix}%comparison abundance tables
%\input{tables/origver/ngc3293comp.tex}
%\input{tables/origver/ngc4755comp.tex}
%\end{appendix}

%\clearpage
%\begin{appendix}%Vicky LMC results not this paper....
%\begin{figure*}
%\begin{center}
%\epsfig{file=figures/lmcfield_NvsLogg_GT20.eps, width=100mm, angle=0}
%\epsfig{file=figures/all1lmc_NvsLogg_GT20.eps, width=100mm, angle=0}
%\epsfig{file=figures/alllmc1_hist.eps, width=100mm, angle=0}
%\end{center}
%\end{figure*}
%\end{appendix}
%----------------------------------------------------------------------------
\end{document}